\documentclass[11pt,a4paper]{article}
\pdfoutput=1

\usepackage[colorlinks=true, linkcolor=black!50!blue, urlcolor=blue, citecolor=blue, anchorcolor=blue]{hyperref}
\usepackage[font=small,labelfont=bf,margin=0mm,labelsep=period,tableposition=top]{caption}
\usepackage[a4paper,top=1.5cm,bottom=2cm,left=1.5cm,right=1.5cm,bindingoffset=0mm]{geometry}

\usepackage{placeins,cite}
\usepackage{graphicx}
\usepackage{float}
\usepackage{afterpage}
\usepackage{epsfig}
\usepackage{amssymb}
\usepackage{amsmath}
\usepackage{bm}
\usepackage{multirow}
\usepackage{url}
\usepackage{xcolor}
\usepackage{url}
\usepackage{booktabs,multirow}
\usepackage{textcomp}
\graphicspath{{../}{./figures/}}
\usepackage{cleveref}
\usepackage[T1]{fontenc} 
\usepackage{wrapfig}

\makeatletter
\def\thickhline{%
             \noalign{\ifnum0 =`}\fi\hrule \@height \thickarrayrulewidth \futurelet
             \reserved@a\@xthickhline}
\def\@xthickhline{\ifx\reserved@a\thickhline
                \vskip\doublerulesep
                \vskip -\thickarrayrulewidth
                \fi
                \ifnum0 =`{\fi}}
\makeatother
\newlength{\thickarrayrulewidth}
\usepackage{tikz}
\usetikzlibrary{positioning}
\usetikzlibrary{shapes.geometric, arrows}
\usetikzlibrary{arrows.meta}
\usepackage{varwidth}
\usepackage{xcolor}
\definecolor{mtplotlib1}{HTML}{1f77b4}
\definecolor{mtplotlib2}{HTML}{ff7f0e}
\definecolor{mtplotlib3}{HTML}{2ca02c}
\definecolor{mtplotlib4}{HTML}{d62728}

\tikzset{%
  >={Latex[width=2mm,length=2mm]},
            base/.style = {rectangle, rounded corners, draw=black,
                           minimum width=4cm, minimum height=1cm,
                           text centered}, 
            mystyle/.style={rectangle, rounded corners, draw=black,
            minimum width=12cm, minimum height=1cm,
            text centered}, 
    col0/.style = {base, fill=white!30},
    col1/.style = {base, fill=mtplotlib1!30},
    col11/.style = {mystyle, fill=mtplotlib1!30},
    col2/.style = {base, fill=mtplotlib2!30},
    col3/.style = {base, fill=mtplotlib3!30},
    col4/.style = {base, minimum width=2.5cm, fill=mtplotlib4!15,}
}

\newcommand{\be}{\begin{equation}}
\newcommand{\ee}{\end{equation}}
\newcommand{\bea}{\begin{eqnarray}}
\newcommand{\eea}{\end{eqnarray}}
\newcommand{\bi}{\begin{itemize}}
\newcommand{\ei}{\end{itemize}}
\newcommand{\ben}{\begin{enumerate}}
\newcommand{\een}{\end{enumerate}}

\newcommand{\lc}{\left[}
\newcommand{\rc}{\right]}
\newcommand{\lp}{\left(}
\newcommand{\rp}{\right)}

\def\gsim{\mathrel{\rlap{\lower4pt\hbox{\hskip1pt$\sim$}}
    \raise1pt\hbox{$>$}}}       
\def\lsim{\mathrel{\rlap{\lower4pt\hbox{\hskip1pt$\sim$}}
    \raise1pt\hbox{$<$}}}

\newcommand{\rep}{\mathrm{rep}}

\newcommand{\draft}[1]{}

\def\beq{\begin{equation}}
\def\eeq{\end{equation}}

\newcommand{\thetaq}{\theta(\mathcal{Q} \bar{\mathcal{Q}})}

\numberwithin{equation}{section}
\numberwithin{figure}{section}
\numberwithin{table}{section}
\usepackage[normalem]{ulem}

\usepackage{tabularx}
\newcolumntype{C}[1]{>{\centering\arraybackslash}p{#1}}

\definecolor{darkblue}{rgb}{0.0,0,0.5}
\definecolor{darkgreen}{rgb}{0.0,0.3,0.0}
\definecolor{redish}{rgb}{0.675,0,0.2}
\definecolor{red}{rgb}{0.8,0,0}
\definecolor{green}{rgb}{0,0.6,0}
\definecolor{bluish}{rgb}{0.2,0.2,0.675}
\definecolor{mygrey}{rgb}{0.6,0.6,0.6}

\usepackage{tikz} 
\usetikzlibrary{shapes,arrows,positioning,automata,backgrounds,calc,er,patterns,arrows.meta}
\usepackage{tikz-feynman}
\tikzfeynmanset{compat=1.0.0}
\usepackage{varwidth}
\usepackage{xcolor}
\definecolor{mtplotlib1}{HTML}{1f77b4}
\definecolor{mtplotlib2}{HTML}{ff7f0e}
\definecolor{mtplotlib3}{HTML}{2ca02c}
\definecolor{mtplotlib4}{HTML}{d62728}

\tikzset{%
  >={Latex[width=2mm,length=2mm]},
            base/.style = {rectangle, rounded corners, draw=black,
                           minimum width=4cm, minimum height=1cm,
                           text centered}, 
            mystyle/.style={rectangle, rounded corners, draw=black,
            minimum width=12cm, minimum height=1cm,
            text centered}, 
    col0/.style = {base, fill=white!30},
    col1/.style = {base, fill=mtplotlib1!30},
    col11/.style = {mystyle, fill=mtplotlib1!30},
    col2/.style = {base, fill=mtplotlib2!30},
    col3/.style = {base, fill=mtplotlib3!30},
    col4/.style = {base, minimum width=2.5cm, fill=mtplotlib4!15,}
}

\usepackage{tabularx}
\usepackage{subcaption}
\newcolumntype{C}[1]{>{\centering\arraybackslash}p{#1}}

\begin{document}
\newgeometry{top=1.5cm,bottom=1.5cm,left=1.5cm,right=1.5cm,bindingoffset=0mm}

\hfill {\small UCI-TR-2024-13}

$\quad$
\vspace{1.2cm}

\begin{center}
  {\Large \bf FPF@FCC: Neutrino, QCD, and BSM Physics Opportunities with \\[0.3cm] Far-Forward Experiments at a 100 TeV Proton Collider}\\
  \vspace{1.1cm}
  {\small
Roshan Mammen Abraham$^{1}$, Jyotismita Adhikary$^{2}$, Jonathan L.~Feng$^{1}$, Max Fieg$^{1}$, \\[0.13cm]
Felix Kling$^{3}$, Jinmian Li$^{4}$, 
Junle Pei$^{5,6}$, Tanjona R. Rabemananjara$^{7,8}$,\\[0.12cm] Juan Rojo$^{7,8}$, and Sebastian Trojanowski$^{2}$
  }\\
  
\vspace{0.7cm}

{\it \small
	~$^1$Department of Physics and Astronomy, University of California, Irvine, CA 92697 USA\\[0.1cm]
    $^{2}$	National Centre for Nuclear Research, Pasteura 7, Warsaw, PL-02-093, Poland\\[0.1cm]
    ~$^3$Deutsches Elektronen-Synchrotron DESY, Notkestr.~85, 22607 Hamburg, Germany\\[0.1cm]
    ~$^4$College of Physics, Sichuan University, Chengdu 610065, China\\[0.1cm]
    ~$^5$Institute of High Energy Physics, Chinese Academy of Sciences, Beijing 100049, China\\[0.1cm]
    ~$^6$Spallation Neutron Source Science Center, Dongguan 523803, China\\[0.1cm]
    ~$^7$Department of Physics and Astronomy, Vrije Universiteit Amsterdam, NL-1081 HV Amsterdam\\[0.1cm]
    ~$^8$Nikhef Theory Group, Science Park 105, 1098 XG Amsterdam, The Netherlands\\[0.1cm]
 }

\vspace{1.0cm}

{\bf \large Abstract}

\end{center}

Proton-proton collisions at energy-frontier facilities produce an intense flux of high-energy light particles, including neutrinos, in the forward direction.
At the LHC, these particles are currently being studied with the far-forward experiments FASER/FASER$\nu$ and SND@LHC, while new dedicated experiments have been proposed in the context of a Forward Physics Facility (FPF) operating at the HL-LHC.
Here we present a first quantitative exploration of the reach for neutrino, QCD, and BSM physics of far-forward experiments integrated within the proposed Future Circular Collider (FCC) project as part of its proton-proton collision program (FCC-hh) at $\sqrt{s} \simeq 100$ TeV.
We find that $10^9$ electron/muon neutrinos and $10^7$ tau neutrinos could be detected, an increase of several orders of magnitude compared to (HL-)LHC yields. 
We study the impact of neutrino DIS measurements at the FPF@FCC to constrain the unpolarised and spin partonic structure of the nucleon and assess their sensitivity to nuclear dynamics down to $x \sim 10^{-9}$ with neutrinos produced in proton-lead collisions.
We demonstrate that the FPF@FCC could measure the neutrino charge radius for $\nu_{e}$ and $\nu_\mu$ and reach down to five times the SM value for $\nu_\tau$.
We fingerprint the BSM sensitivity of the FPF@FCC for a variety of models, including dark Higgs bosons, relaxion-type scenarios, quirks, and millicharged particles, finding that these experiments would be able to discover LLPs with masses as large as 50 GeV and couplings as small as $10^{-8}$, and  quirks with masses up to 10 TeV.
Our study highlights the remarkable opportunities made possible by integrating far-forward experiments into the FCC project, and it provides new motivation for the FPF at the HL-LHC as an essential precedent to optimize the forward physics experiments that will enable the FCC to achieve its full physics potential.

\clearpage

\tableofcontents
\section{Introduction}

The discovery of collider neutrinos by the FASER~\cite{FASER:2023zcr} and SND@LHC~\cite{SNDLHC:2023pun} experiments, followed by FASER's world-leading constraints on dark photons~\cite{FASER:2023tle} and axion-like particles~\cite{CERN-FASER-CONF-2024-001} and the first laboratory measurement of neutrino cross sections at TeV energies by FASER$\nu$~\cite{FASER:2024hoe}, demonstrate the unique physics potential of the LHC far-forward experiments for beyond-the-Standard-Model (BSM), neutrino, QCD, and astroparticle physics. 
These experiments complement the existing LHC infrastructure with novel avenues to study both the Standard Model (SM) and its possible extensions, with unique capabilities beyond the reach of other LHC detectors.
Several studies have quantified the physics reach of far-forward experiments at the LHC both at the current Run-3 and at the HL-LHC data-taking period starting in 2029, the latter embedded in the context of the proposed Forward Physics Facility (FPF)~\cite{Feng:2022inv,Anchordoqui:2021ghd}.

Beyond the HL-LHC, the particle physics community is considering possible options for future high-energy colliders. 
These proposals include electron-positron colliders, either circular, such as the FCC-ee~\cite{FCC:2018byv} and the CEPC~\cite{CEPCPhysicsStudyGroup:2022uwl}, or linear, such as the ILC~\cite{Behnke:2013xla,ILC:2013jhg}, the C3~\cite{Vernieri:2022fae}, and CLIC~\cite{Linssen:2012hp}; high-energy proton-proton ($pp$) colliders such as the FCC-hh~\cite{FCC:2018byv,FCC:2018vvp} and the SppC~\cite{Tang:2015qga}; muon colliders~\cite{Accettura:2023ked,Aime:2022flm}; and high-energy electron-proton/ion colliders such as the LHeC and the FCC-eh~\cite{LHeC:2020van,FCC:2018byv}. 
Furthermore, with a focus on QCD and hadronic physics, but also with a rich program of electroweak measurements and BSM searches, the Electron Ion Collider (EIC)~\cite{AbdulKhalek:2021gbh} is expected to record first collisions in the early 2030s.

In this landscape, a particularly compelling option is the integrated FCC program at CERN, operating in a new tunnel of around 90 km in circumference, and including two main stages: an electron-positron program (FCC-ee) covering center-of-mass (CoM) energies from $\sqrt{s}=91.2$~GeV (Tera-Z) up to the top-quark pair threshold at $\sqrt{s}=365$~GeV, followed by a hadron-hadron collider (FCC-hh) reaching energies up to $\sqrt{s}=100$~TeV for $pp$ collisions, and enabling proton-ion and heavy-ion collisions at the highest values of $\sqrt{s_{\rm NN}}$ ever achieved.
The same physical mechanisms driving the production of light particles in the forward direction at the LHC~\cite{Kling:2021gos,FASER:2024ykc,Buonocore:2023kna} would lead to unprecedentedly intense neutrino and light particle fluxes at the FCC-hh.
The resulting increase, compared to the HL-LHC, arises 
from combining the 7-fold increase in CoM energy (from $\sqrt{s}=14$~TeV to 100~TeV) with the 10-fold increase in integrated luminosity (from $\mathcal{L}_{\rm int}=3$~ab$^{-1}$ to 30~ab$^{-1}$).

As we demonstrate here, an FPF-like suite of far-forward experiments integrated in the FCC-hh (henceforth denoted as FPF@FCC) could record up to a factor $10^3$ more events as compared to its HL-LHC counterpart, accumulating up to one billion electron and muon neutrino interactions and up to 30 million tau neutrino interactions in charged-current (CC) scattering. 
These enormous event rates enable novel physics opportunities beyond those accessible with far-forward experiments at the LHC.
These include high-precision tests of the universality of neutrino flavour interactions at multi-TeV energies and the measurement of their charge radius; pinning down the proton spin decomposition with neutrino deep-inelastic scattering (DIS) on polarised targets; and the 
detection of the first neutrinos produced in proton-ion collisions, probing cold nuclear matter and non-linear QCD effects in an unexplored region of relevance also for astroparticle physics experiments.

\begin{figure}[t]
\includegraphics[width=1.0\textwidth]{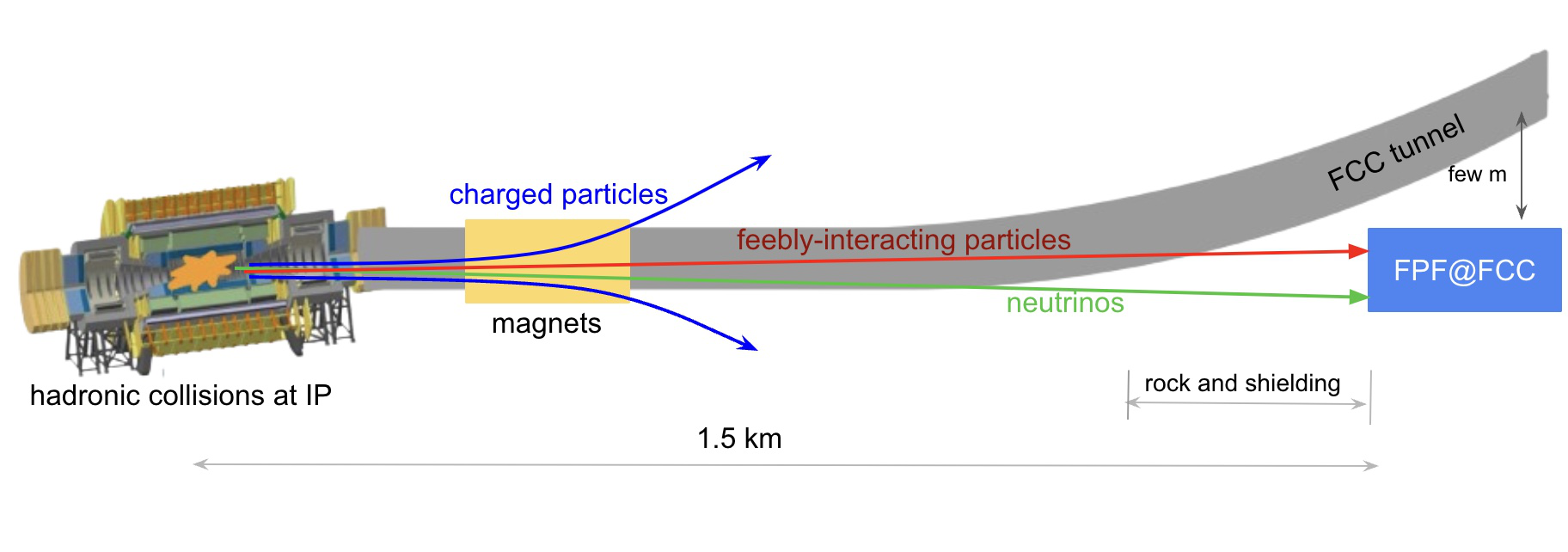}
\includegraphics[width=1.\textwidth]{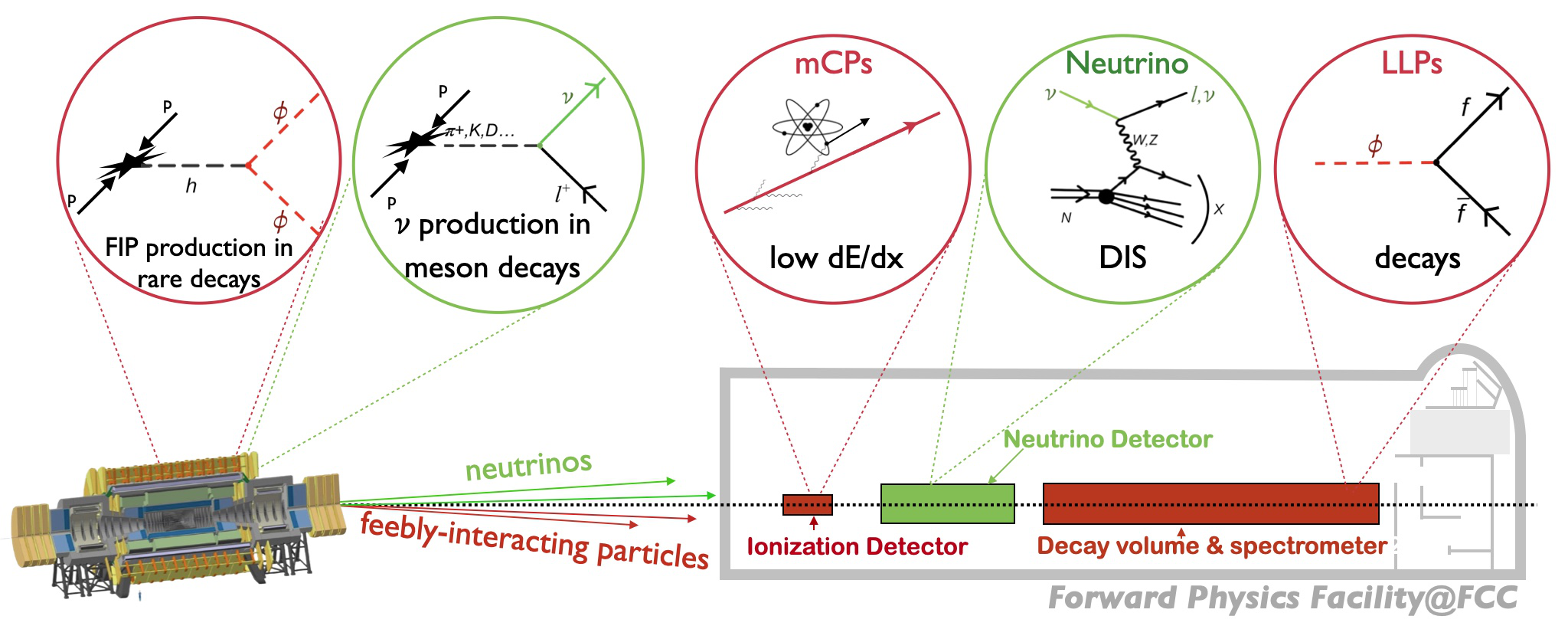}
\caption{Top: High-energy light particles from $pp$ collisions at the FCC-hh with $\sqrt{s}=100$~TeV are produced predominantly in the forward direction. 
With a dedicated sweeper magnet to remove muons, 
mostly neutrinos or LLPs/FIPs from BSM sectors would reach the detectors of the FPF@FCC installed around 1.5~km away from the interaction point (IP).
Bottom: Schematic of representative production, scattering, and decay signatures that could be studied at the FPF@FCC experiments, assuming an ionization detector, a neutrino detector, and a decay volume with spectrometer.
}
\label{fig:schematicphysics}
\end{figure}

Fig.~\ref{fig:schematicphysics} indicates schematically how light particles produced at the interaction point of $pp$ collisions at $\sqrt{s}=100$ TeV from the FCC-hh are produced predominantly in the forward direction. 
By means of a dedicated sweeper magnet, 
the flux of high-energy muons would be significantly reduced\footnote{Although we will not study them here, there are interesting physics opportunities enabled by the detection of such high-energy muons. 
A full FPF@FCC program should therefore also include the capability to make high-precision measurements of the high-energy forward muon flux.} such that mostly neutrinos or long-lived particles (LLPs) and feebly interacting particles (FIPs) from BSM sectors would reach the detectors of the FPF@FCC, installed around 1.5 km away from the interaction point (IP).
The bottom panel displays representative  production, scattering, and decay signatures that could be studied at such a facility, assuming that it is equipped with an ionization detector, a neutrino detector, and a decay volume with spectrometer.
Specifically, we display Higgs invisible decay to LLPs, $h\to \phi\phi$; neutrino production from $D$-meson decays, $D^+ \to \ell^+ \nu_\ell$ (production);
millicharged particles (mCPs), neutrino DIS, $\nu_\ell +N \to \ell^- + X$ (scattering); and finally LLP decays $\phi \to f\bar{f}$ (decay). 

In this paper we present a first quantitative assessment of representative physics opportunities enabled by the FPF@FCC.
Since these experiments would only operate several decades from now, during which detector technologies will certainly advance, we do not assume specific types of detectors, but instead keep our analysis general and technology-agnostic.
For this reason, for the FPF@FCC detectors considered here, we only specify their geometry and the relevant performance requirements such as particle identification capabilities, resolution, and kinematic acceptances. 
Systematic uncertainties are neglected for most case studies considered: our projections hence represent an optimistic scenario, which aims to highlight the best possible sensitivity achievable in these types of experiments in the case of ideal detector response. 
Our study complements and extends previous work on forward physics and beam dump experiments at future colliders, including at the ILC~\cite{Kanemura:2015cxa,Sakaki:2020mqb,Asai:2021xtg}, at FCC-ee (HECATE)~\cite{Chrzaszcz:2020emg}, at the FCC-hh with FASER@FCC~\cite{Kling:2021fwx} and FOREHUNT~\cite{Bhattacherjee:2023plj}, and at a muon collider~\cite{Cesarotti:2022ttv,InternationalMuonCollider:2024jyv}. 

The outline of this paper is the following.
In Sect.~\ref{sec:fluxes}, we evaluate the neutrino fluxes produced by $pp$ and proton-ion collisions at the FCC-hh, study their  kinematic features, and evaluate the associated event yields for different neutrino detectors.
In Sect.~\ref{sec:SM} we study the impact of neutrino-nucleon DIS to pin down the unpolarised and polarised proton structure and assess the constraints on  cold nuclear matter provided by neutrinos from proton-lead collisions.
The impact of FPF@FCC for BSM physics is studied in Sect.~\ref{sec:BSM},
first by demonstrating that the neutrino charge radius can be probed down to SM values for $\nu_e$ and $\nu_\mu$, and then by considering a variety of interesting signatures and models, such as dark decays of the Higgs boson, relaxion-type scenarios, mCPs, and quirks. 
Finally, we conclude and present some ideas for additional developments in Sect.~\ref{sec:summary}.
Technical details are provided in App.~\ref{app:SFs} for neutrino polarised DIS, and in App.~\ref{app:quirk} for information on the timing selection criteria for the quirk signal.
\section{Detectors, fluxes, and event rates}
\label{sec:fluxes}

Here we describe the main features of the detectors we consider for the FPF@FCC, including both those targeting neutrinos and those targeting LLPs, quirks, mCPs, and other BSM states. 
For the proposed neutrino detectors, we provide predictions for the far-forward fluxes and the associated event yields in both $pp$ and in heavy-ion collisions. 

\subsection{Neutrino detectors}
\label{subsec:neutrino_detectors}

As indicated in Fig.~\ref{fig:schematicphysics}, the FPF@FCC would be located around 1.5~km downstream from the IP, where a dedicated cavern would be excavated and aligned with the line-of-sight (LoS) of the primary $pp$ collisions.
This distance from the IP is motivated by the geometry and the bending of the FCC tunnel, and it results in approximately 500~m of shielding from rock and concrete. 
Indeed, the FPF@FCC detectors should be installed on the LoS as close as possible to the IP, but still sufficiently far separated from the FCC beam to reduce radiation to acceptable levels.
Using the geometry presented in Ref.~\cite{Abelleira:2020dkg}, the closest distance that still provides at least 10~m separation from the beam is about 1.2 km downstream from the IP. 
Our chosen configuration, located 1.5~km downstream from the IP, enables a larger facility and also a bigger arm-length for the sweeper magnet to improve the suppression of the high-energy muon flux, though, if needed, the FPF@FCC could be moved closer to the IP.

The far-forward neutrino detectors considered for the FPF@FCC, together with FASER$\nu$ and FASER$\nu$2 as reference, are described in Table~\ref{tab:neutrino_detectors}.
For each detector, we indicate its geometry (transverse $\times$ longitudinal dimensions),
its coverage in neutrino pseudo-rapidity $\eta_\nu$,
the integrated luminosity $\mathcal{L}_{\rm pp}$ of $pp$ collisions at $\sqrt{s}$ at the interaction point, and the acceptance for the final-state charged lepton energy and scattering angle, $E_\ell$ and $\theta_\ell$ respectively, as well as for the energy of the hadronic final state $E_h$. 

\begin{table*}[h]
  \centering
  \small
  \renewcommand{\arraystretch}{1.70}
\begin{tabularx}{\textwidth}{Xccccc}
\toprule
Detector &  Geometry & Rapidity &  $\mathcal{L}_{\rm pp}$  & $\sqrt{s}$ &  Acceptance  \\
\midrule
\midrule
FASER$\nu$  & 25 cm $\times$ 30 cm $\times$ 103 cm &   $\eta_\nu \ge 8.5$  &  250 fb$^{-1}$       &  13.6~TeV & $E_\ell,E_h \gsim 100$ GeV, $\theta_\ell \lsim 0.025 $          \\
\midrule
FASER$\nu$2  & 40 cm $\times$ 40 cm $\times$ 6.6 m  &  $\eta_\nu \ge 8.4$  &  3 ab$^{-1}$       &  14~TeV & $E_\ell,E_h \gsim 100$ GeV, $\theta_\ell \lsim 0.05 $          \\
\midrule
\midrule
FCC$\nu$  &  40 cm $\times$ 40 cm $\times$ 6.6 m  &  $\eta_\nu \ge 9.2$  &  30 ab$^{-1}$       &  100~TeV & $E_\ell,E_h \gsim 100$ GeV, $\theta_\ell \lsim 0.05 $          \\
\midrule
FCC$\nu$(d)  &  40 cm $\times$ 40 cm $\times$ 66 m  &  $\eta_\nu \ge 9.2$  &  30 ab$^{-1}$       &  100~TeV & $E_\ell,E_h \gsim 100$ GeV, $\theta_\ell \lsim 0.05 $          \\
\midrule
FCC$\nu$(w)  &  1.25 m $\times$ 1.25 m $\times$ 6.6 m  &  $\eta_\nu \ge 8.1$  &  30 ab$^{-1}$       &  100~TeV & $E_\ell,E_h \gsim 100$ GeV, $\theta_\ell \lsim 0.05 $          \\
  \bottomrule
\end{tabularx}
\vspace{0.2cm}
\caption{\small 
The far-forward neutrino detectors FCC$\nu$, FCC$\nu$(d), and FCC$\nu$(w), considered in this work.  For reference, we also include the parameters of the  FASER$\nu$ and FASER$\nu$2 detectors.
For each detector, we indicate its geometry (transverse $\times$ longitudinal dimensions),
its coverage in neutrino pseudo-rapidity,
the integrated luminosity, the CoM energy $\sqrt{s}$ from $pp$ collisions at the IP, and the acceptance for the final-state charged lepton energy and scattering angle, $E_\ell$ and $\theta_\ell$, respectively, as well as for the energy of the hadronic final state $E_h$. 
  \label{tab:neutrino_detectors}
}
\end{table*}

The FASER$\nu$ detector~\cite{FASER:2019dxq, FASER:2020gpr, FASER:2022hcn}, operating since the beginning of LHC Run 3, is located 480 m from the ATLAS IP and has dimensions 25~cm $\times$ 30~cm $\times$ 103~cm. It is composed of a tungsten passive target interleaved with emulsion films, adding up to a fiducial mass of 1.1 tonnes.
It provides charged-lepton identification, is sensitive to neutrinos with pseudorapidity $\eta_\nu \ge 8.5$, and its emulsion technology also allows the identification of charm-tagged events. 
Here we assume that FASER$\nu$ is exposed to a luminosity of $\mathcal{L}_{\rm pp}=250$ fb$^{-1}$ at $\sqrt{s}=13.6$~TeV.
The proposed upgrade of FASER$\nu$, FASER$\nu$2~\cite{Feng:2022inv}, would be installed at the FPF during the HL-LHC data-taking period.
The FASER$\nu$2 detector is to be located approximately 620 m from the ATLAS IP, and has dimensions of 40~cm $\times$ 40~cm $\times$ 6.6~m, with a target mass that is approximately 20 times larger than FASER$\nu$.
FASER$\nu$2 is designed to be exposed to the complete HL-LHC luminosity of $\mathcal{L}_{\rm pp}=3$~ab$^{-1}$ at $\sqrt{s}=14$~TeV.

For the FPF@FCC, we consider three different options for the neutrino detector of Fig.~\ref{fig:schematicphysics}, taking the geometry and kinematic acceptance of FASER$\nu$2 as a baseline.
The first is denoted as FCC$\nu$ and is assumed to have the same dimensions as FASER$\nu$2, but now exposed to the FCC-hh luminosity of $\mathcal{L}_{\rm pp}=30$ ab$^{-1}$ at the higher CoM energy of $\sqrt{s}=100$~TeV.
The choice of the same geometry is motivated by the desire to disentangle the effects related to differences in production ($\sqrt{s}$ and $\mathcal{L}_{\rm pp}$) from possible improvements from an enlarged detector.
Two variants of the FCC$\nu$ detector are then considered: FCC$\nu$(d), with a depth 10 times larger than the baseline detector, and a wider variant, FCC$\nu$(w), with a transverse area 10 times the baseline detector.
Although we do not assume any specific detector technology for FCC$\nu$ and its variants, for simplicity we also adopt tungsten as the target material and assume detector acceptances similar to those considered in Ref.~\cite{Cruz-Martinez:2023sdv}.
As mentioned above, the five detectors of Table~\ref{tab:neutrino_detectors} are centered on the beam LoS.
While other detector configurations may be considered, the ones studied here are sufficiently representative to highlight the main physics opportunities enabled by neutrino detection at the FPF@FCC.

\paragraph{Polarised detectors.}
The neutrino detectors in Table~\ref{tab:neutrino_detectors} do not admit target polarisation, a feature that would enable the study of polarised DIS with neutrino beams~\cite{Forte:2001ph}.
With this motivation, here we consider two additional neutrino detectors based on a target that can be polarised, summarised in Table~\ref{tab:neutrino_detectors_polarised}.
The first is inspired by the COMPASS configuration~\cite{COMPASS:2007rjf}, namely, a $^6$LiD target with density $\rho=950$ kg/m$^3$ and an active volume of $V=6\times 10^{-3}$ m$^3$, with a radius of 2.5 cm and a depth of 1.2 m; this detector is therefore denoted COMPASS$\nu$ in the following.
The second polarised detector is dubbed FCC$\nu$-pol and assumes that one can polarise a COMPASS-like target enlarged to transverse dimensions of 40~cm $\times$ 40~cm and longitudinal dimension of 6.6~m (the same geometry as for FASER$\nu$2), corresponding to an active volume of $V=1.05$~m$^3$.

As in the case of the neutrino detectors for unpolarised DIS, one cannot predict which technologies would be available by the time the FCC-hh becomes operative, and hence one should take our projections as an initial estimate of the sensitivity that the FPF@FCC can achieve in polarised neutrino DIS measurements.  

\begin{table*}[t]
  \centering
  \small
  \renewcommand{\arraystretch}{1.90}
\begin{tabularx}{\textwidth}{Xccccc}
\toprule
Detector &  Geometry & Rapidity &  $\mathcal{L}_{\rm pp}$  & $\sqrt{s}$ &  Acceptance  \\
\midrule
\midrule
COMPASS$\nu$  &  $\pi  (2.5~{\rm cm})^2\times$ 1.2 m  &  $\eta_\nu \ge 11.6$  &  30 ab$^{-1}$       &  100~TeV & $E_\ell,E_h \gsim 100$ GeV, $\theta_\ell \lsim 0.05 $          \\
\midrule
FCC$\nu$-pol  &  40 cm $\times$ 40 cm $\times$ 6.6 m  &  $\eta_\nu \ge 9.2$  &  30 ab$^{-1}$       &  100~TeV & $E_\ell,E_h \gsim 100$ GeV, $\theta_\ell \lsim 0.05 $          \\
\midrule
\end{tabularx}
\vspace{0.2cm}
\caption{\small 
Same as Table~\ref{tab:neutrino_detectors}, but for two possible detectors for the FPF@FCC equipped with a polarised target.
While inspired by the features of COMPASS, here we are agnostic about the technology upon which these would be based.
We assume the same material as COMPASS ($^6$LiD) and that 100\% polarisation can be achieved.
  \label{tab:neutrino_detectors_polarised}
}
\end{table*}

\subsection{Neutrino fluxes and event rates}
\label{subsec:fluxes_pp}

\begin{figure}[t]
    \centering
    \includegraphics[width=0.7\linewidth]{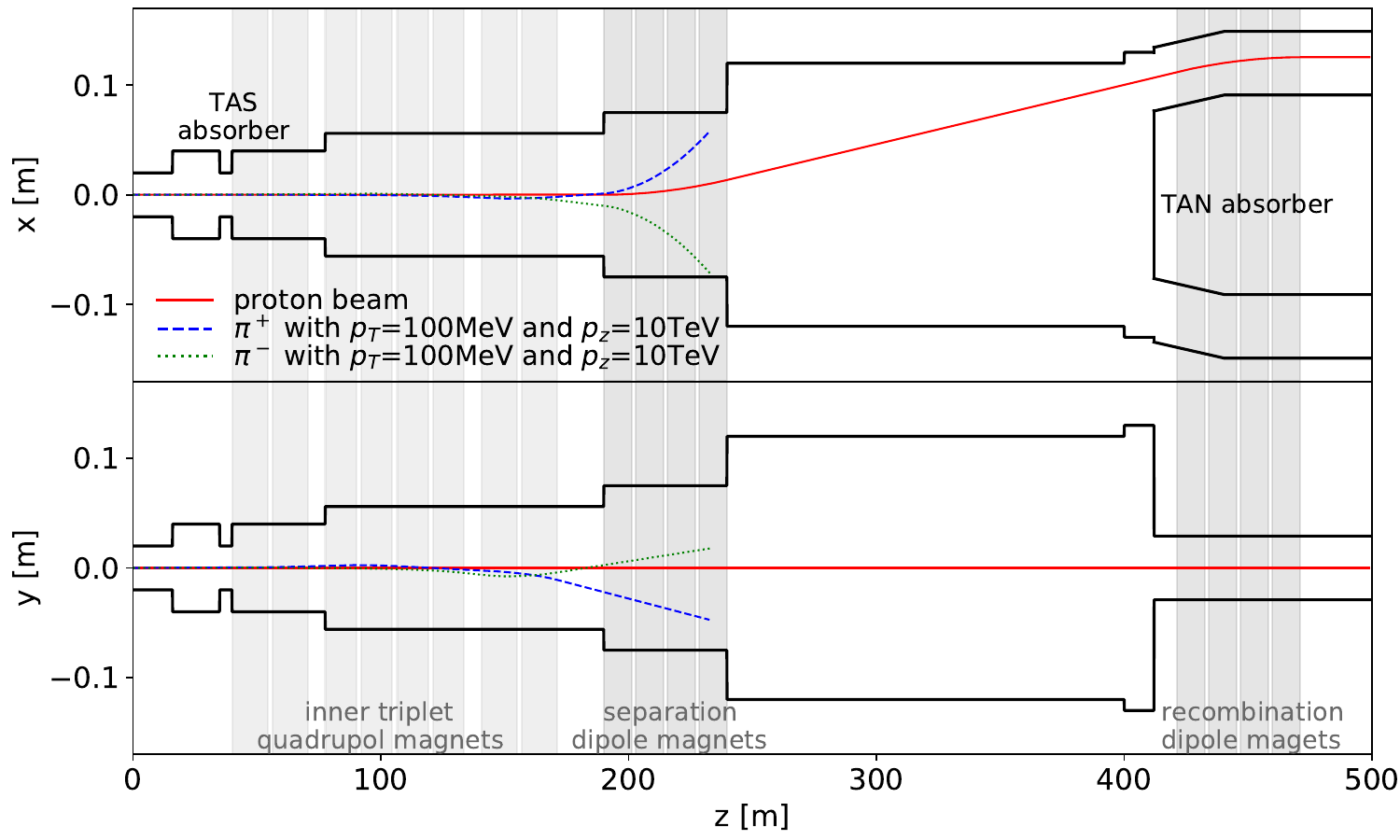}
\caption{The FCC beam pipe geometry, including the magnets, which are assumed in the simulations presented in this work.
We show the boundaries of the FCC’s beam pipe as black lines and the
magnetic fields as gray shaded areas.
The solid red line shows the trajectory of the proton beam with $E_p=50~{\rm TeV
}$, and the other lines show representative trajectories followed by pions with $E_\pi=10$~TeV.}
\label{fig:geometry}
\end{figure}

We now evaluate the forward neutrino fluxes reaching the detectors described in Sect.~\ref{subsec:neutrino_detectors} and the associated event yields, first for $pp$ and then for heavy-ion collisions.
We follow the procedure described in Refs.~\cite{Kling:2021gos,FASER:2024ykc,Buonocore:2023kna} and use the fast neutrino flux simulation introduced in Ref.~\cite{Kling:2021gos}, adjusted to match the FCC-hh configuration, beam pipe geometry, and magnetic fields described in Ref.~\cite{Abelleira:2020dkg} and displayed in Fig.~\ref{fig:geometry}.
We include the neutrino flux from light, charm, and bottom hadron decays, where we generate the light hadrons (mostly pions and kaons) using {\sc\small EPOS-LHC}~\cite{Pierog:2013ria}, and the charm and bottom hadrons using {\sc\small POWHEG}~\cite{Alioli:2010xd} matched with {\sc\small Pythia}8.3~\cite{Bierlich:2022pfr} for parton showering and hadronisation.
Following the work in Ref.~\cite{Buonocore:2023kna}, the latter calculation is accurate at NLO in the QCD coupling and takes as input the NNPDF3.1sx+LHCb NLO+NLL$x$ parton distribution function (PDF) set~\cite{Ball:2017otu,Bertone:2018dse,Gauld:2016kpd}, whose small-$x$ behaviour accounts for BFKL resummation and is tuned to describe the $D$-meson production from LHCb.

For the inclusive neutrino CC interaction cross section, we use the Bodek-Yang model~\cite{Bodek:2002vp,Bodek:2004pc,Bodek:2021bde} as implemented in the {\sc\small GENIE} neutrino event generator~\cite{Andreopoulos:2009rq,Andreopoulos:2015wxa}, which simulates both DIS and non-DIS contributions to the cross section. 
The Bodek-Yang model overestimates the DIS cross section at TeV energies by about 6\% when compared with higher-order QCD calculations based on state-of-the-art PDFs~\cite{Garcia:2020jwr,Candido:2023utz,FASER:2024ykc}, but this difference has a negligible effect in the interpretation of the neutrino flux calculation.
When assessing the impact of unpolarised neutrino DIS measurements at the FPF@FCC in Sect.~\ref{sec:SM}, we will instead use NNLO structure functions based on the PDF4LHC21 combination~\cite{PDF4LHCWorkingGroup:2022cjn} with {\sc\small YADISM}~\cite{Candido:2024rkr}.
We note that the {\sc\small GENIE} calculations could be replaced with the recent implementation of neutrino DIS in {\sc\small POWHEG}~\cite{FerrarioRavasio:2024kem,Buonocore:2024pdv,vanBeekveld:2024ziz}, enabling the exclusive simulation of particle-level final states at NLO accuracy.

In the following we consider in turn the case of $pp$ collisions, first with unpolarised and then with polarised detectors, and then the case of neutrinos from heavy-ion collisions.

\subsubsection{Proton-proton collisions at 100 TeV}

\begin{figure}[t]
    \centering
    \includegraphics[width=\linewidth]{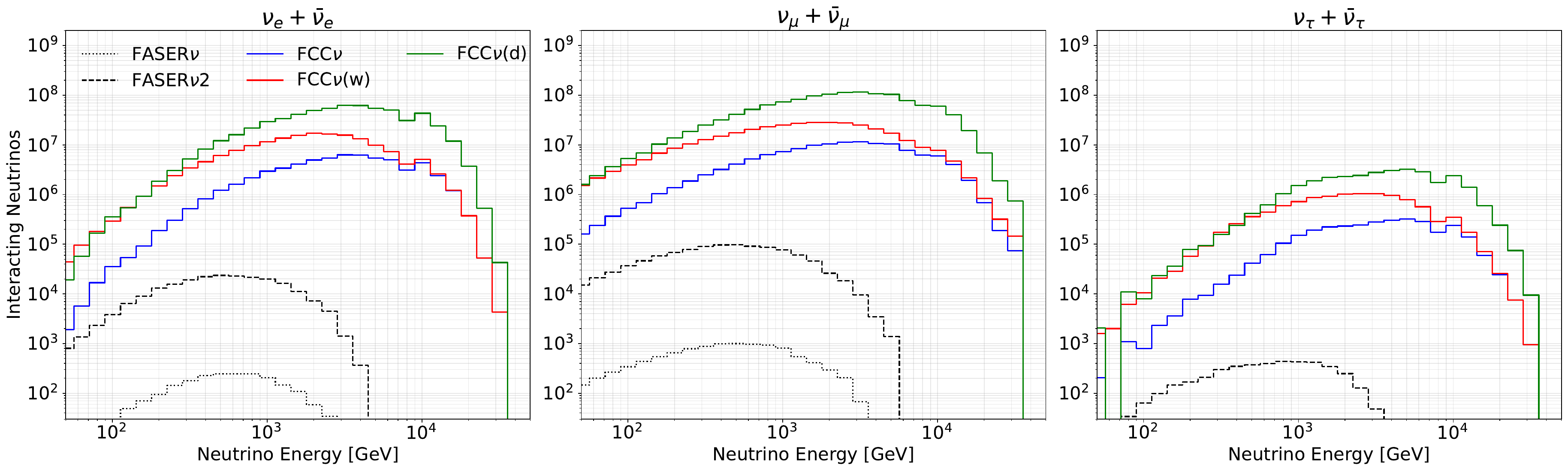}
     \includegraphics[width=\linewidth]{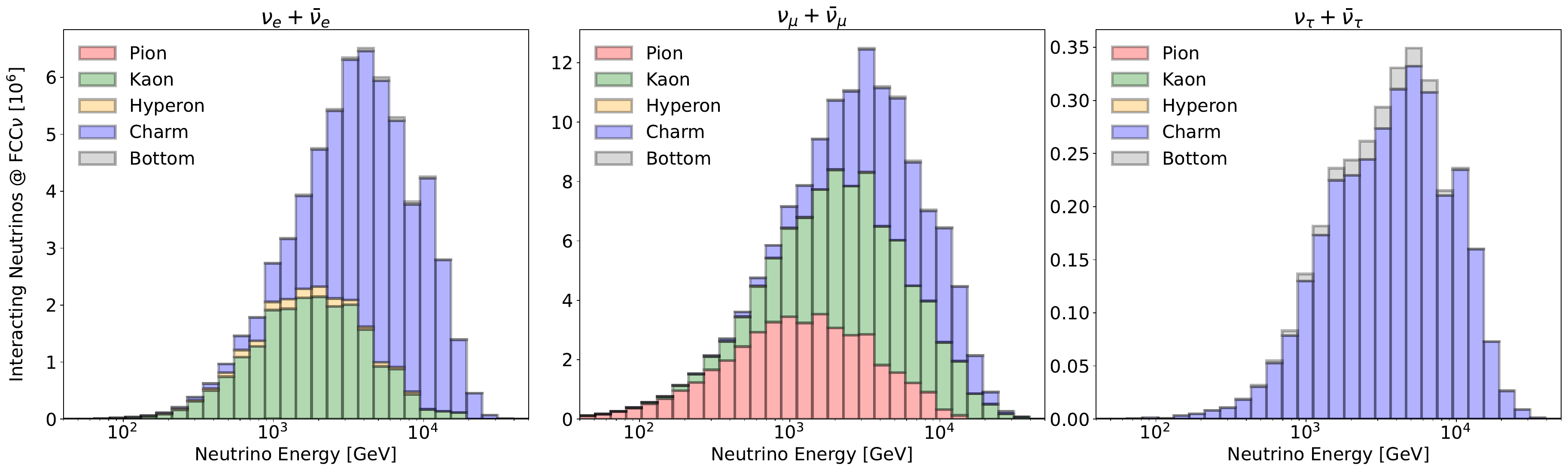}
    \caption{Top: From left to right, the number of interacting electron, muon, and tau neutrinos ($\nu_\ell+\bar{\nu}_{\ell}$) for the  detectors listed in Table~\ref{tab:neutrino_detectors} as a function of $E_\nu$.
    See Table~\ref{tab:integrated_rates} for the associated event yields.
    Bottom: Event yields for the FCC$\nu$ detector, for the different flavours, left to right as above, but now decomposed according to the parent particle whose decays produce the detected neutrinos.
    }
    \label{fig:fluxes_FCC}
\end{figure}

By folding the forward neutrino fluxes from $pp$ collisions with the DIS CC interaction, we determine the number of neutrinos interacting in each of the five detectors described in Table~\ref{tab:neutrino_detectors} as a function of the neutrino energy $E_\nu$, as well as the corresponding inclusive event yields collected in Table~\ref{tab:integrated_rates}.
Fig.~\ref{fig:fluxes_FCC} (top) displays the resulting 
interaction yields for electron, muon, and tau neutrinos as a function of $E_\nu$ for the three considered FCC$\nu$ detectors, with the predictions for FASER$\nu$ and FASER$\nu$2 included for comparison.
The bottom panel of Fig.~\ref{fig:fluxes_FCC} presents the decomposition of the event yields for FCC$\nu$ in terms of the parent particle whose decay leads to the detected neutrinos.
For electron neutrinos, kaon production dominates for $E\lsim 2$ TeV; for higher energies, charm production is the leading mechanism.
For muon neutrinos, the neutrino rates receive sizeable contributions from pion, kaon, and charm decays, with the first two dominant for low energies, and the last two dominant for high energies.
As at the LHC, only heavy meson and baryon decays can produce tau neutrinos. 

\begin{table*}[t]
  \centering
  \renewcommand{\arraystretch}{1.30}
\begin{tabularx}{\textwidth}{XC{3.5cm}C{3.5cm}C{3.5cm}}
\toprule
Detector &  $N_{\nu_e} + N_{\bar{\nu}_e}$ & $N_{\nu_\mu} + N_{\bar{\nu}_\mu}$ 
& $N_{\nu_\tau} + N_{\bar{\nu}_\tau}$ 
\\
\midrule
\midrule
FASER$\nu$   & 2.1k  &   11k  & 36 \\
\midrule
FASER$\nu$2   & 220k  & 1.1M  & 4.3k \\
\midrule
\midrule
FCC$\nu$  &   62M & 130M  & 3.2M \\
\midrule
FCC$\nu$(d)  &  620M & 1.3B  & 32M \\
\midrule
FCC$\nu$(w) &   170M & 370M  & 11M \\
  \bottomrule
\end{tabularx}
\vspace{0.2cm}
\caption{\small The inclusive event yields for the
LHC and FCC-hh neutrino detectors listed in Table~\ref{tab:neutrino_detectors}, separately
for electron, muon, and tau neutrinos. 
See Fig.~\ref{fig:fluxes_FCC} for the differential distributions as a function of  $E_\nu$. 
  \label{tab:integrated_rates}
}
\end{table*}

From Fig.~\ref{fig:fluxes_FCC} and Table~\ref{tab:integrated_rates}, one finds that despite the FPF@FCC neutrino detectors being located further away from the IP than its LHC counterparts (1.5~km compared to 480~m and 620~m for FASER$\nu$ and FASER$\nu$2, respectively), the resulting loss in angular acceptance is more than compensated for by the increase in the luminosity and the CoM energy of the primary $pp$ collision, as well as by the stronger collimation of the neutrino beam and enhanced neutrino-nucleon interaction cross section at the increased neutrino energies.
Indeed, for the same detector geometry, the FCC$\nu$ detector is projected to record 280, 120, and 750 times more electron, muon, and tau neutrino charged current interactions than FASER$\nu$2. 
The higher luminosity accounts for a factor of 10 in the comparison, and the remaining difference can largely be attributed to the higher neutrino energy (and thus larger cross section) and the enhanced forward hadron production rates, in particular for $D$-mesons, at the FCC compared to the LHC. 
Indeed, it is worth noting that the fraction of interacting neutrinos that come from charm decay is roughly 34\%  and 6\% for $\nu_e$ and $\nu_\mu$, respectively, at FASER$\nu$2  (and $\approx$100\% for $\nu_\tau$), while the charm component increases to 65\% for $\nu_e$ and 29\% for $\nu_\mu$ at FCC$\nu$, highlighting the relevance of forward charm production for the FPF@FCC experiments. In contrast, at the FCC, tau neutrino production from bottom hadrons is only increased by a few percent compared to the LHC, with negligible contributions for $\nu_e$ and $\nu_\mu$.

The neutrinos scattering in the FPF@FCC detectors also display a harder $E_\nu$ distribution in comparison with their LHC counterparts.
Although the FASER$\nu$2 neutrino interaction rates peak at energies between $E_\nu\sim 500$ GeV and 1 TeV, the ones for FCC$\nu$ peak at several TeV, and large event rates are expected for neutrinos with energies up to $E_\nu \sim 30$ TeV. 
Note that for the FCC$\nu$ neutrino sample it is unnecessary to apply DIS acceptance cuts on the momentum transfer $Q^2$ and the hadronic invariant mass $W^2$, since non-DIS interactions are negligible at such high energies. 

In terms of the FCC$\nu$ variants considered, FCC$\nu$(d) collects 10 times more neutrinos (for each of the three flavours) than FCC$\nu$, with the same energy spectrum shape, since the total rate grows linearly with the detector length $L$.
In contrast, FCC$\nu$(w) collects 10 times more neutrinos only at low energies, $E_\nu < 1~\text{TeV}$, where the neutrino beam's transverse distribution is spread out sufficiently for it to be considered flat within the detector's spatial extent. 
At high energies, $E_\nu \gsim 10~\text{TeV}$, where the neutrino beam is extremely collimated and essentially contained within the geometric acceptance of FCC$\nu$, the rate observed at FCC$\nu$(w) converges to that observed at FCC$\nu$. 

\begin{figure}[t]
    \centering
    \includegraphics[width=\linewidth]{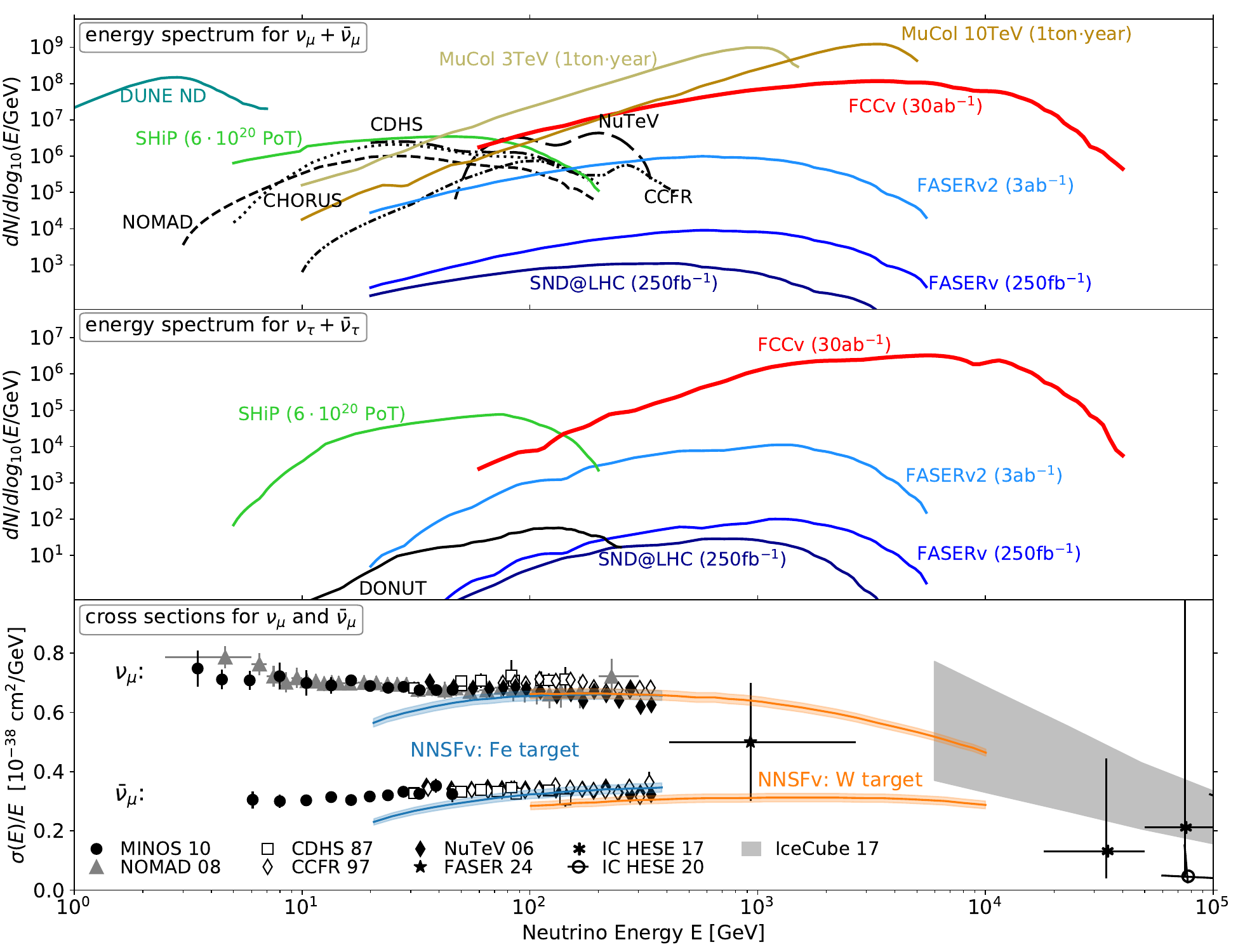}
    \caption{The muon (upper panel) and tau (middle panel) neutrino scattering yields as a function of $E_\nu$ for the FCC$\nu$ detector with $\mathcal{L}_{\rm pp}=30$ ab$^{-1}$.
    These event rates are compared to their counterparts at FASER$\nu$ and SND@LHC (for 250 fb$^{-1}$), FASER$\nu$2 with 3 ab$^{-1}$, the SHiP experiment at ECN3 (with $6\times 10^{20}$ protons on target), the DUNE Near Detector (ND), the muon collider (MuCol), and several previous neutrino experiments: NuTeV, CHORUS, NOMAD, CDHS, and CCFR. 
    The bottom panel shows a comparison of available measurements of the neutrino-scattering cross section with the NNSF$\nu$~\cite{Candido:2023utz} theoretical predictions, considering
    separately Fe and W targets.
    }
    \label{fig:fluxes-global-overview}
\end{figure}

From the inclusive event yields of Table~\ref{tab:integrated_rates}, it is manifest that the unprecedentedly large high-energy neutrino samples that would become available at the FPF@FCC should enable physics opportunities beyond the reach of any other present or future neutrino experiment.
For instance, assuming the FPF$\nu$(d) detector configuration, the $3.2\times 10^{7}$ tau neutrinos collected (combined with the $6.2\times 10^{8}$ and $1.3\times 10^{9}$ samples of electron and muon neutrinos, respectively) would allow testing the (non-)universality of the three neutrino generations at multi-TeV energies with part-per-mille level statistical uncertainties.  Such a prospect is truly tantalizing, given that at present, only roughly 20 tau neutrinos have been directly detected in all particle experiments combined. 
To compare the expected neutrino scattering rates at the FPF@FCC with those provided by previous, ongoing, and future neutrino experiments, we display in Fig.~\ref{fig:fluxes-global-overview} the muon neutrino ($\nu_\mu$+$\bar{\nu}_\mu$, top panel) and tau neutrino ($\nu_\tau$+$\bar{\nu}_\tau$, middle panel) event rates as a function of the neutrino energy for the FCC$\nu$ detector, assuming a total integrated luminosity of $\mathcal{L}_{\rm pp}=30$ ab$^{-1}$, compared to their counterparts at FASER$\nu$ and SND@LHC (250 fb$^{-1}$) and to FASER$\nu$2 operating during the HL-LHC phase (3 ab$^{-1}$).
In addition, we also provide the neutrino event rates projected for the SHiP experiment~\cite{Alekhin:2015byh} recently approved for ECN3~\cite{Ahdida:2023okr}, the DUNE Near Detector (ND)~\cite{DUNE:2021tad}, and several previous neutrino experiments: NuTeV~\cite{NuTeV:2003kth}, CHORUS~\cite{CHORUS:2005cpn}, NOMAD~\cite{NOMAD:2003owt}, DONUT~\cite{DONuT:2007bsg}, CDHS~\cite{Berge:1989hr}, and CCFR~\cite{CCFRNuTeV:1996vbm}.
We also display the muon neutrino event rates that can be obtained at a muon collider (MuCol) operating at either $\sqrt{s}=3$ TeV or 10 TeV using a 1~tonne neutrino detector placed 100~m downstream of the collision point and operating for one year~\cite{InternationalMuonCollider:2024jyv}.
Since the neutrinos at a MuCol are primarily produced in the decay of the beam muons, the neutrino beam does not contain tau neutrinos.
From Fig.~\ref{fig:fluxes-global-overview} one can appreciate both the huge increase in statistics as well as in $E_\nu$ reach offered by the FPF@FCC  detectors as compared to any other previous, ongoing, or future laboratory-based neutrino experiment.

The bottom panel of Fig.~\ref{fig:fluxes-global-overview} shows a comparison of available measurements of the neutrino-nucleon scattering cross section with the theoretical predictions provided by the NNSF$\nu$ calculation~\cite{Candido:2023utz}, considering separately predictions for Fe and W targets and where the  NNSF$\nu$ band indicates the 68\% CL theory uncertainties. 
We include in this overview the cross section data from previous laboratory neutrino experiments (MINOS~\cite{MINOS:2009ugl}, NOMAD~\cite{NOMAD:2007krq}, CDHS~\cite{Berge:1987zw}, CCFR~\cite{Seligman:1997fe}, and NuTeV~\cite{NuTeV:2005wsg}) as well as the recent FASER$\nu$ measurement~\cite{FASER:2024hoe} in the TeV range (averaged over $\nu_\mu$ and $\bar{\nu}_{\mu}$) and the IceCube measurements~\cite{IceCube:2017roe, Bustamante:2017xuy, IceCube:2020rnc} obtained from the analysis of high-energy astrophysical neutrinos.\footnote{For energies $E_\nu\lsim 50$ GeV, neutrino cross sections receive large resonance and quasi-elastic contributions not included in  NNSF$\nu$  (which accounts only for inelastic scattering processes), explaining the disagreement with the data.}
Given the large event rates displayed in Fig.~\ref{fig:fluxes_FCC}, neutrino cross section measurements with 
sub-part-per-mille statistical uncertainties up to $E_\nu\sim 30$ TeV energies and for the three generations would be possible at the FPF@FCC. 

\subsubsection{Event yields with polarised DIS detectors}
\label{subsubsec:polarised}

\begin{figure}[t]
     \centering
     \includegraphics[width = 0.99\textwidth]{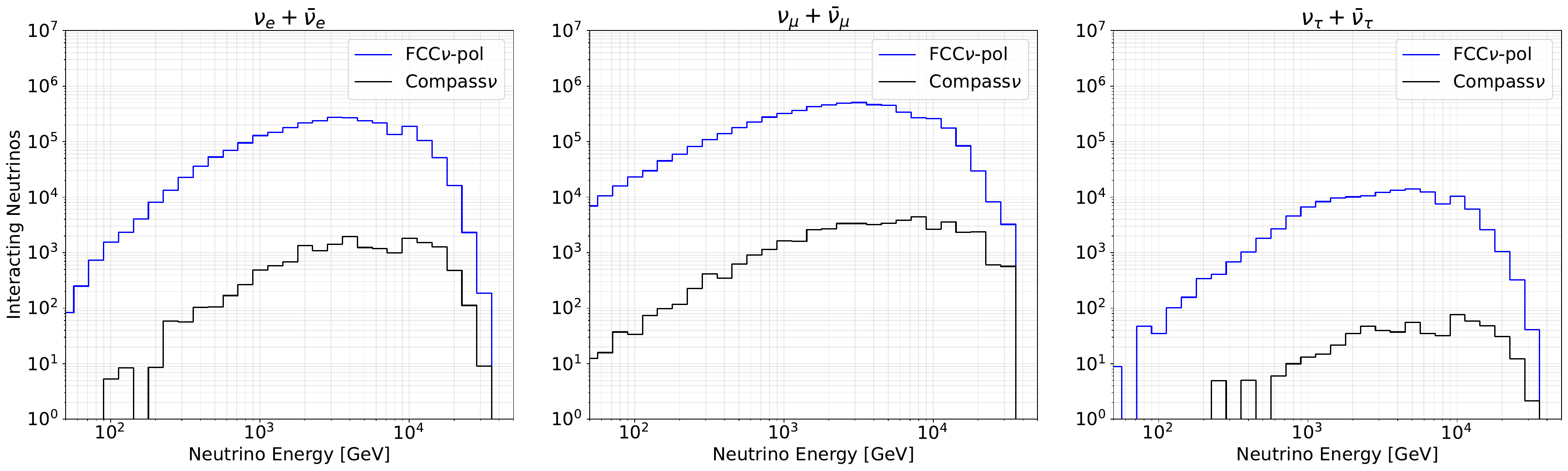}
     \caption{Same as Fig.~\ref{fig:fluxes_FCC}, but for the polarised detectors of Table~\ref{tab:neutrino_detectors_polarised}. 
     See Table~\ref{tab:integrated_rates_polarised} for the associated integrated event yields. 
     }
     \label{fig:6LiD_EventRate}
 \end{figure}

The analogous results for the total (Table~\ref{tab:integrated_rates}) and differential (Fig.~\ref{fig:fluxes_FCC}) event yields for the case of the polarised detectors described in Table~\ref{tab:neutrino_detectors_polarised} are provided in  Table~\ref{tab:integrated_rates_polarised} and Fig.~\ref{fig:6LiD_EventRate}, respectively.
While a small COMPASS$\nu$-like experiment would be sufficient to detect the first neutrinos scattering on a polarised target, meeting the precision targets of polarised DIS structure function measurements would only be possible with a much larger detector, such as FCC$\nu$-pol.
Such a detector could accumulate up to $2.7\times 10^{6}$, $5.9\times 10^{6}$, and $1.4\times 10^{5}$ electron, muon, and tau neutrino scattering events, respectively, when summing over the two target polarisations. 

As opposed to the unpolarised case, it is not possible to directly extract from the binned event yields the statistical precision forecast for the measurement of polarised DIS cross sections.
The reason is that this precision depends on the magnitude of the polarised structure functions, which in turn depends on the proton's polarised PDFs,
as we discuss in Sect.~\ref{sec:protonspin}.
Nevertheless, the event yields evaluated in Table~\ref{tab:integrated_rates_polarised} indicate that a precise measurement of neutrino DIS polarised structure functions should be within reach in the case of the FPF$\nu$-pol detector. 

\begin{table*}[t]
  \centering
  \renewcommand{\arraystretch}{1.30}
\begin{tabularx}{\textwidth}{XC{3.5cm}C{3.5cm}C{3.5cm}}
\toprule
Detector &  $N_{\nu_e} + N_{\bar{\nu}_e}$ & $N_{\nu_\mu} + N_{\bar{\nu}_\mu}$ 
& $N_{\nu_\tau} + N_{\bar{\nu}_\tau}$ 
\\
\midrule
\midrule
COMPASS$\nu$   &  17k & 46k    & 590 \\
\midrule
FCC$\nu$-pol   &  2.7M  &  5.9M    & 140k \\
  \bottomrule
\end{tabularx}
\vspace{0.2cm}
\caption{\small Same as Table~\ref{tab:integrated_rates}, but for the polarised DIS detectors of Table~\ref{tab:neutrino_detectors_polarised}. See Fig.~\ref{fig:6LiD_EventRate} for the associated energy spectra.
  \label{tab:integrated_rates_polarised}
}
\end{table*}

\subsubsection{Neutrinos from proton-lead collisions}
\label{subsec:fluxes_pPb}

As opposed to the FPF operating during the HL-LHC data-taking period, for the FPF@FCC experiments, the combination of increased $\sqrt{s_{\rm NN}}$ and the higher integrated luminosity would enable for the first time the detection of a significant sample of neutrinos produced in both proton-lead and lead-lead collisions.
This feature may also hold in heavy ion collisions involving lighter species, such as with the proton-oxygen and oxygen-oxygen runs foreseen at the HL-LHC~\cite{Brewer:2021kiv}.
Measuring neutrinos produced in high-energy proton-ion and ion-ion collisions would open a new window to extreme cold nuclear mater in unexplored regimes, for instance, those dominated by non-linear effects  modelled by the Color Glass Condensate (CGC)~\cite{Gelis:2010nm}.

For proton-lead (lead-lead) collisions at the FCC-hh, the projected CoM energy per nucleon would be $\sqrt{s_{\rm NN}} = 63~(39)$~TeV.
In the case of proton-lead collisions, protons would have an energy of $E_p=50$~TeV and lead nuclei of $E_{\rm Pb}=A_{\rm Pb}\times 19.7$~TeV, with the lead atomic number being $A_{\rm Pb}=208$. 
This implies that, as is already the case at the LHC, proton-lead collisions at the FCC-hh would be asymmetric with a net boost in the direction of the incoming proton beam, with momentum fractions $x_1 \gg x_2$, where $x_1$ ($x_2$) is associated with the colliding proton (nucleus). 

Here we simulate neutrino production in proton-lead and lead-lead collisions using the {\tt Angantyr} model in {\sc\small Pythia8}~\cite{Bierlich:2018xfw} with the same setup for hadron decay and neutrino propagation as for $pp$ collisions and only modifying the beam.
Therefore, modifications of the initial-state dominating the small-$x$ regime such as those encoded by nuclear PDFs~\cite{Klasen:2023uqj,Ethier:2020way} or non-linear dynamics enhanced in heavy nuclei are neglected; such effects, while potentially large, are currently unconstrained and hence challenging to model.
For this reason, in these heavy-ion simulations we adopt the NNPDF2.3 QED LO proton PDFs~\cite{Carrazza:2013axa,ball:2013hta}  to describe the partonic content of heavy nuclei.

Existing studies~\cite{Dainese:2016gch,FCC:2018vvp} estimate that the FCC-hh could deliver integrated luminosities of up to $\mathcal{L}_{\rm PbPb}=100$ nb$^{-1}$ (30 pb$^{-1}$) per month when operating in the lead-lead (proton-lead) collision mode.
While no detailed operation schedules of the heavy-ion program of the FCC-hh have been put forward, here we assume optimistically a total of 12 months of operation for both proton-lead and lead-lead collisions, to take place distributed over the planned $\mathcal{O}(25)$ years of operation of the FCC-hh.
Therefore, here we generate neutrino fluxes from proton-lead collisions for an integrated luminosity of ${\cal L}_{\rm pPb}=350 ~{\rm pb}^{-1}$, and for lead-lead collisions of ${\cal L}_{\rm PbPb}=$ 1.3 pb$^{-1}$.
In the case of proton-lead collisions, we consider both the configuration where the proton beam is headed in the direction of the FPF@FCC detectors as well as the reversed configuration, with the same integrated luminosity in both cases.\footnote{The physics analysis presented in Sect.~\ref{sec:heavyions} is restricted to neutrinos produced in the proton-going direction.} 
The experimental challenge in separating neutrinos originating from charm hadrons produced in the hard scattering and those produced from the underlying event is left for future studies.

\begin{table*}[t]
  \centering
  \footnotesize
  \renewcommand{\arraystretch}{1.90}
\begin{tabularx}{\textwidth}{Xccccc}
\toprule
\multirow{2}{*}{Detector} & $\sqrt{s_{\rm NN}}$ & \multirow{2}{*}{$\mathcal{L}_{\rm pPb}$~($\mathcal{L}_{\rm PbPb}$)}  & $N_{\nu_e} + N_{\bar{\nu}_e}$ & $N_{\nu_\mu} + N_{\bar{\nu}_\mu}$  & $N_{\nu_\tau} + N_{\bar{\nu}_\tau}$ 
\\
 & p+Pb (Pb+Pb)  &  &  p+Pb, Pb+p, PbPb &  p+Pb, Pb+p, Pb+Pb  &  p+Pb, Pb+p, PbPb
\\
\midrule
\midrule
FASER$\nu$  &  8.6~(5.5) TeV  &  1 pb$^{-1}$~(1.3 nb$^{-1}$) & 0.14, 0.21, 0.006  & 0.7, 1.8, 0.05 & 0.004, 0.002, $<10^{-4}$  \\
\midrule
FASER$\nu$2  & 8.6~(5.5) TeV & 1 pb$^{-1}$~(1.3 nb$^{-1}$) & 1.1, 1.6, 0.051  & 5.6, 14.0, 0.38  &0.03, 0.01, 0.0007  \\
\midrule
\midrule
FCC$\nu$  & 63~(39) TeV &  350 pb$^{-1}$~(1.3 pb$^{-1}$) & 15k, 22k, 1.8k &26k, 72k, 5.0k & 840, 370, 75 \\
\midrule
FCC$\nu$(d)  & 63~(39) TeV & 350 pb$^{-1}$~(1.3 pb$^{-1}$) & 150k, 220k, 18k & 260k, 720k, 50k & 8.4k, 3.7k, 750  \\
\midrule
FCC$\nu$(w)  & 63~(39) TeV & 350 pb$^{-1}$~(1.3 pb$^{-1}$) & 75k, 78k, 8.6k & 120k, 240k, 21k & 5.6k, 2.9k, 520\\
  \bottomrule
\end{tabularx}
\vspace{0.2cm}
\caption{\small Same as Table~\ref{tab:integrated_rates} for the integrated event yields from CC DIS events initiated by neutrinos produced in proton-lead, lead-proton, and lead-lead collisions.
The FPF@FCC detectors are assumed to be installed in the direction of the first particle.  
For simplicity the p+Pb and Pb+p runs are taken to have the same luminosity, and $\sqrt{s_{\rm NN}}$ is the nucleon-nucleon energy for each type of collision.
See Fig.~\ref{fig:IntNu_pPb_direct_FASERnu2} for the associated differential rates.
\label{tab:integrated_rates_pPb}
}
\end{table*}

\begin{figure}[t]
\begin{center}
\includegraphics[width=0.99\textwidth]{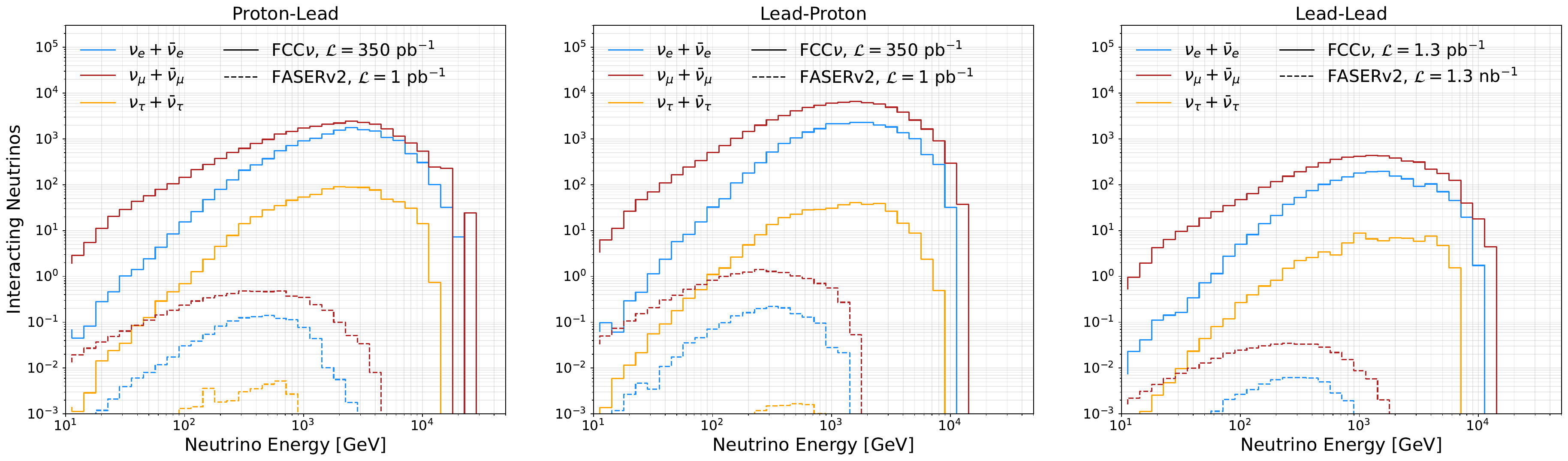}
\caption{Same as Fig.~\ref{fig:fluxes_FCC}, but for the number of interacting neutrinos as a function of $E_\nu$ for
proton-lead, lead-proton, and lead-lead collisions (from left to right) in the FASER$\nu$2 (dashed) and the FCC$\nu$ (solid) detectors. 
We display separately the number of interacting electron, muon, and tau neutrinos.
These results assume
$\mathcal{L}_{\rm pPb}=350~(1.3)$ pb$^{-1}$ for proton-lead and $\mathcal{L}_{\rm PbPb}=1.3$~pb$^{-1}$~(1.3 nb$^{-1}$) for lead-lead collisions at the FCC-hh (HL-LHC) respectively.
See Table~\ref{tab:integrated_rates_pPb} for the associated integrated event rates.
}
\label{fig:IntNu_pPb_direct_FASERnu2}
\end{center}
\end{figure}

Based on these settings, Table~\ref{fig:IntNu_pPb_direct_FASERnu2} provides the integrated event yields from CC DIS events initiated by neutrinos produced in proton-lead, lead-proton, and lead-lead collisions at the FPF@FCC, where the detectors are assumed to be installed in the direction of the first particle.  
For reference, we also indicate the expected number of neutrino events to be detected by FASER$\nu$ and FASER$\nu$2 in proton-lead (lead-lead) collisions at $\sqrt{s_{\rm NN}}=8.6$~(5.5) TeV, corresponding to luminosities of $\mathcal{L}_{\rm pPb}=1$ pb$^{-1}$ ($\mathcal{L}_{\rm pPb}=1.3$ nb$^{-1}$).

As mentioned above, and consistent with previous estimates~\cite{Feng:2022inv}, at the (HL-)LHC the expected event yields associated with neutrinos produced in proton-lead and lead-lead collisions are negligible.
Indeed, even for FASER$\nu$2 operating under the full HL-LHC luminosity, at most $\mathcal{O}(10)$ neutrino events from proton-lead collisions would be expected.
Instead, thanks to the higher $\sqrt{s_{\rm NN}}$ and integrated luminosities, the neutrino event yields from heavy ion collisions become sizable at the FPF@FCC detectors.
For instance, for the baseline FCC$\nu$ detector, $\mathcal{O}(10^4)$ events in proton-lead collisions are predicted for both electron and muon neutrinos, and up to $\mathcal{O}(10^3)$ tau neutrino events are expected.
The expected event yields are increased for the two variants with larger fiducial volume. For example, the FCC$\nu$(d) detector would record up to $\mathcal{O}(10^5)$, $\mathcal{O}(10^6)$, and $\mathcal{O}(10^4)$ electron, muon, and tau neutrino events, respectively.
Interestingly, large samples of neutrinos from lead-lead collisions may also become available at the FPF@FCC, with up to $5\times 10^{4}$  muon-neutrino scattering events recorded in the case of the FCC$\nu$(d) detector.

The total number of electron, muon, and tau neutrino interactions from heavy ion collisions are shown in Table~\ref{fig:IntNu_pPb_direct_FASERnu2}, and the associated energy distributions are displayed in Fig.~\ref{fig:IntNu_pPb_direct_FASERnu2}.
Fig.~\ref{fig:IntNu_pPb_direct_FASERnu2} highlights the substantial increase in the event rates enabled by the higher $\sqrt{s_{\rm NN}}$ and luminosities of the FCC-hh, as well as the shift of the spectra towards higher energies in comparison with the expected FASER$\nu$2 fluxes.
For proton-lead collisions, the $E_\nu$ distribution peaks around 5 TeV with events reaching 20 TeV, while for lead-lead collision the maximum is around 2 TeV and the distribution reaches up to 10 TeV. 
Fig.~\ref{fig:IntNu_pPb_direct_FASERnu2} showcases how the FPF@FCC would be exposed to a sufficiently large flux of forward neutrinos from heavy-ion collisions to scrutinise small-$x$ PDFs and nuclear matter in kinematic regions not accessible by any ongoing or future laboratory experiment, as we quantify in  Sect.~\ref{sec:heavyions}.

\subsection{Detectors targeting LLPs, quirks, and millicharged particles}
\label{subsec:detectors_bsm}

As indicated by the schematic of Fig.~\ref{fig:schematicphysics}, the proposed FPF@FCC would incorporate, in addition to neutrino detectors, other detectors focusing on the direct detection of BSM signatures, such as a detector targeting mCPs and a detector targeting LLPs, FIPs, and similar new physics states produced in the forward direction.
Here we describe the main features and performance targets of the detectors used for the BSM sensitivity studies of Sect.~\ref{sec:BSM}.

We consider two forward FPF@FCC detectors that are dedicated to BSM searches for highly-displaced visible decays of LLPs and can also be used to detect and study quirks. 
We denote them as FCC-LLP1 and FCC-LLP2.
Both detectors are assumed to be centered on the LoS with their front face 1.5~km from the IP.  
The decay volume of FCC-LLP1 has a transverse area that is $5~\text{m} \times 5~\text{m}$ and a length of 50~m, while FCC-LLP2's decay volume has a transverse area that is $20~\text{m} \times 20~\text{m}$ and a length of 400~m.  
The geometry of these detectors is summarized in Table~\ref{tab:BSMdetectors}, where for reference we also indicate the geometry and acceptance of FASER and FASER2, which are 480~m and 620~m from the ATLAS IP, respectively.

\begin{table*}[t]
  \centering
  \small
  \renewcommand{\arraystretch}{1.70}
\begin{tabularx}{\textwidth}{Xcccc}
\toprule
Detector &  Geometry &  $\mathcal{L}_{\rm pp}$  & $\sqrt{s}$ &  Acceptance  \\
\midrule
\midrule
FASER  & $\pi (10~{\rm cm})^2\times 1.5~\textrm{m}$  &  150 fb$^{-1}$ &  14~TeV & $E_{\textrm{vis}} \gsim 100$ GeV \\
\midrule
FASER2  & $\pi (1~{\rm m})^2\times 5~\textrm{m}$  &  3 ab$^{-1}$ &  14~TeV & $E_{\textrm{vis}} \gsim 100$ GeV \\
\midrule
\midrule
FCC-LLP1  & $5~\textrm{m}\times 5~\textrm{m}\times 50~\textrm{m}$  &  30 ab$^{-1}$ &  100~TeV & $E_{\textrm{vis}} \gsim 100$ GeV \\
\midrule
FCC-LLP2  & $20~\textrm{m}\times 20~\textrm{m}\times 400~\textrm{m}$  &  30 ab$^{-1}$ &  100~TeV & $E_{\textrm{vis}} \gsim 100$ GeV \\
\midrule
FCC-mCP  & $5~\textrm{m}\times 5~\textrm{m}\times 4~\textrm{m}$  &  30 ab$^{-1}$ &  100~TeV &  $4\times \lp \bar{N}_{\textrm{PE}}
\ge 1 \rp$\\
\bottomrule
\end{tabularx}
\vspace{0.2cm}
\caption{\small 
The transverse size and length (depth) of the three FPF@FCC detectors considered for the BSM sensitivity analysis of this work.
The front of the detector is taken to be 1.5~km from the IP.
For reference, we also include information on the geometry of the FASER and FASER2 experiments. 
$E_{\rm vis}$ indicates the lower threshold in visible energy required to tag an event. The millicharged particle search relies on four time-coincident detections, each corresponding to at least one photoelectron on average ($\bar{N}_{\textrm{PE}}$).
\label{tab:BSMdetectors}
}
\end{table*}

Although the size of the decay volume of FCC-LLP1 is larger than the one considered for FASER2 in the FPF at the LHC, it is similar to current and near-future beam-dump experiments, in particular to SHiP~\cite{Ahdida:2704147}. 
We consider this as a baseline option for FPF@FCC. 
We also show results for FCC-LLP2, which corresponds to the ultimate size of the forward LLP detector at the considered distance from the FCC interaction point.
This is due to its large length and also because FCC-LLP2's dimensions imply that the nearest part of the detector is only about $15~\text{m}$ from the FCC beamline.
Recall that we require sufficient distance between the two to suppress beam-induced backgrounds and radiation in the FPF@FCC. 
We stress that the baseline FCC-LLP1 detector will be shielded from the beamline by at least a few tens of meters of rock from each direction. Importantly, the shielding from the direction of the FCC IP will be even larger, of the order of $500~\text{m}$.

In our projections we assume that both detectors will be able to resolve charged particle tracks produced in LLP decays and measure their momenta. 
To achieve this goal, given the high energies of at least some such particles, which can exceed $10~\textrm{TeV}$, one requires a strong magnetic field to deflect them in the decay volume and excellent tracking resolution. 
This requirement goes beyond what is currently employed by FASER or in experiments operating at lower energies.
As in the case of the neutrino detectors, we do not specify the experimental technologies to be used, but we note that tracking technologies developed for the FCC-hh~\cite{FCC:2018vvp} could also be used in far-forward experiments.
We also stress that precise momentum measurement might not be required for the most energetic tracks, provided that the BSM signal can be disentangled from backgrounds. 
For this purpose, the FCC-LLP1(2) detectors should veto all the events coincident with muons and other charged particles entering the decay volume from outside. 
The same decay volume should also be kept in a vacuum to suppress neutrino-induced backgrounds. 

In this study we assume that an electromagnetic (EM) calorimeter will be installed downstream of the decay volume. 
This will allow for an improved measurement of energies of electrons and positrons produced in LLP decay chains. Measuring the fraction of LLP decay events with and without substantial EM energy deposition will be crucial for understanding the nature of the LLP in the case of discovery.
The presence of the EM calorimeter will also extend the sensitivity of FPF@FCC detectors to BSM scenarios predicting LLP decays to photons. Developing calorimeter technology relevant to this purpose will greatly benefit from ongoing studies about calorimetry in main FCC detectors; see Ref.~\cite{Aleksa:2019pvl}.

In addition to the FCC-LLP detectors, we also consider the physics reach of the FCC-mCP detector, which targets ionization signals, such as those that could be left by mCPs produced in the forward kinematic region of the FCC-hh. 
To this end, we follow the proposed design of the FORMOSA experiment at the FPF~\cite{Ball:2016zrp,Foroughi-Abari:2020qar,Feng:2022inv}. 
In particular, we assume that the detector will consist of four layers of plastic scintillator bars and that each layer will be $1~\textrm{m}$ long.
The transverse size of the detector depends on the number of scintillator bars in a single detector segment, and we assume an increased size of $5~\text{m} \times 5~\text{m}$, similar to FCC-LLP1. While such a detector setup is sufficient to significantly improve the sensitivity of FCC-mCP, compared with ongoing and near-future experiments, we also note that alternative scintillating materials can be considered in the future with an increased photon yield to improve the discovery potential further. 
For the mCP signal, we require a quadruple coincidence of hits with an average number of photoelectrons $\bar{N}_{\textrm{PE}}$ in each layer satisfying $\bar{N}_{\textrm{PE}}\ge 1$. 
We also assume $10\%$ detector efficiency, similar to the FORMOSA~\cite{Foroughi-Abari:2020qar} and milliQan~\cite{Ball:2016zrp} detectors.

\section{Proton and nuclear structure from FCC-hh neutrinos}
\label{sec:SM}

The results of Sect.~\ref{sec:fluxes} demonstrate that the FPF@FCC experiments would be exposed to unprecedentedly large samples of TeV-scale neutrinos. 
Here we illustrate the impact that neutrino  measurements at these detectors would have on the unpolarised and polarised structure of the proton and of heavy nuclei by means of three representative applications.
These are not meant to provide an exhaustive analysis, but rather to highlight the opportunities provided by FPF@FCC neutrinos for QCD studies.

First, we quantify the constraints that the measurement of high-energy neutrino DIS structure functions would provide on proton PDFs, revisiting the study of Ref.~\cite{Cruz-Martinez:2023sdv} at the LHC, but now for the FPF@FCC detectors.
Second, we assess the sensitivity that neutrino DIS measurements taken on a polarised target would have on the spin structure of the nucleon.
Third, we assess the information that could be extracted on nuclear structure at ultra small-$x$ values by detecting neutrinos originated from proton-lead collisions.

\subsection{Proton structure from high-energy neutrino DIS}
\label{sec:protonstructure}

The impact of measurements of neutrino DIS structure functions at ongoing and future far-forward LHC detectors on proton and nuclear PDFs
has been estimated in Ref.~\cite{Cruz-Martinez:2023sdv}.
For the present study, we have updated the proton PDF projections of Ref.~\cite{Cruz-Martinez:2023sdv} to the case of DIS CC processes at the FPF@FCC detectors listed in Table~\ref{tab:neutrino_detectors}, using the same methodology and analysis settings.

\begin{figure}[t]
\begin{center}
\includegraphics[width=0.49\textwidth]{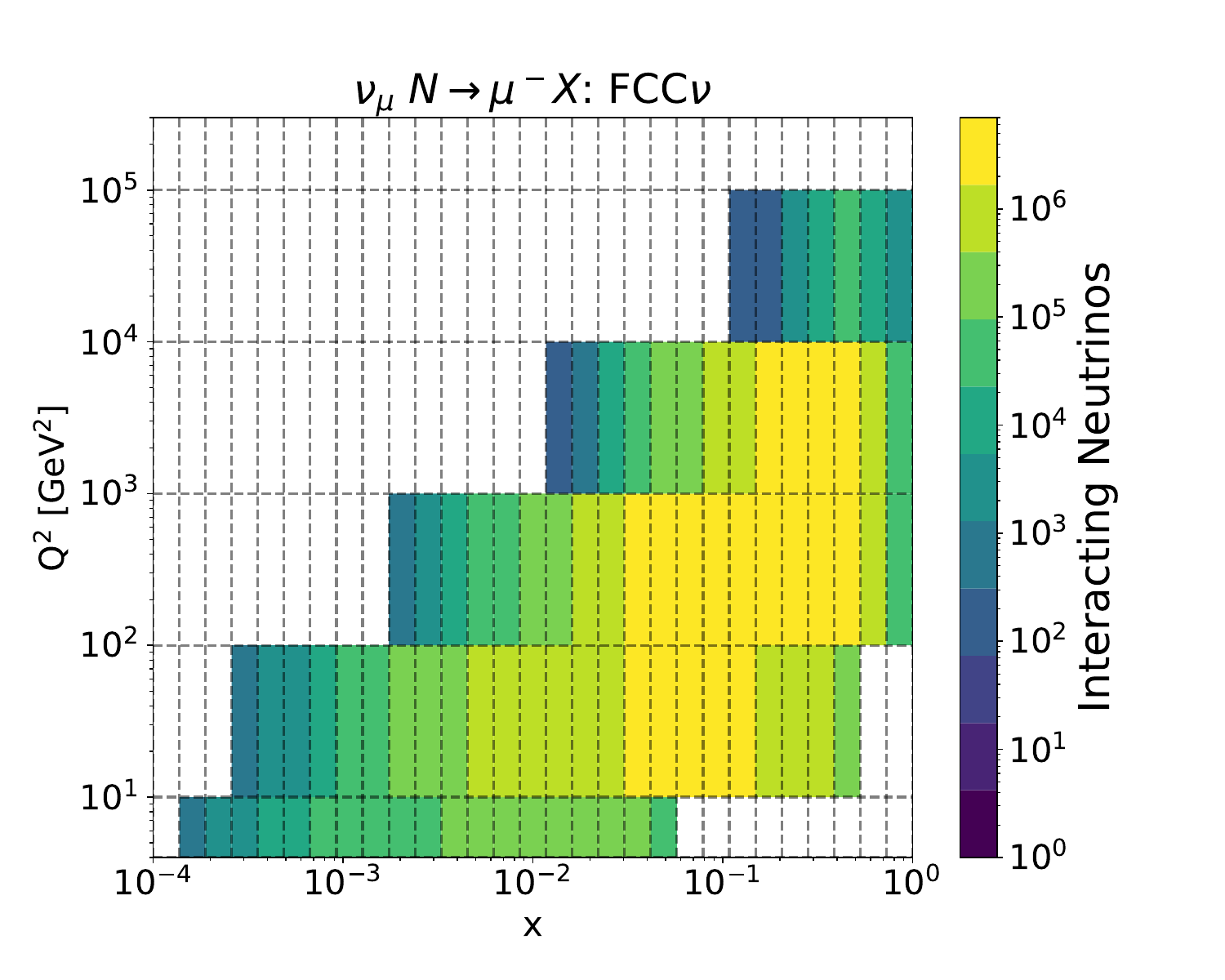}
\includegraphics[width=0.49\textwidth]{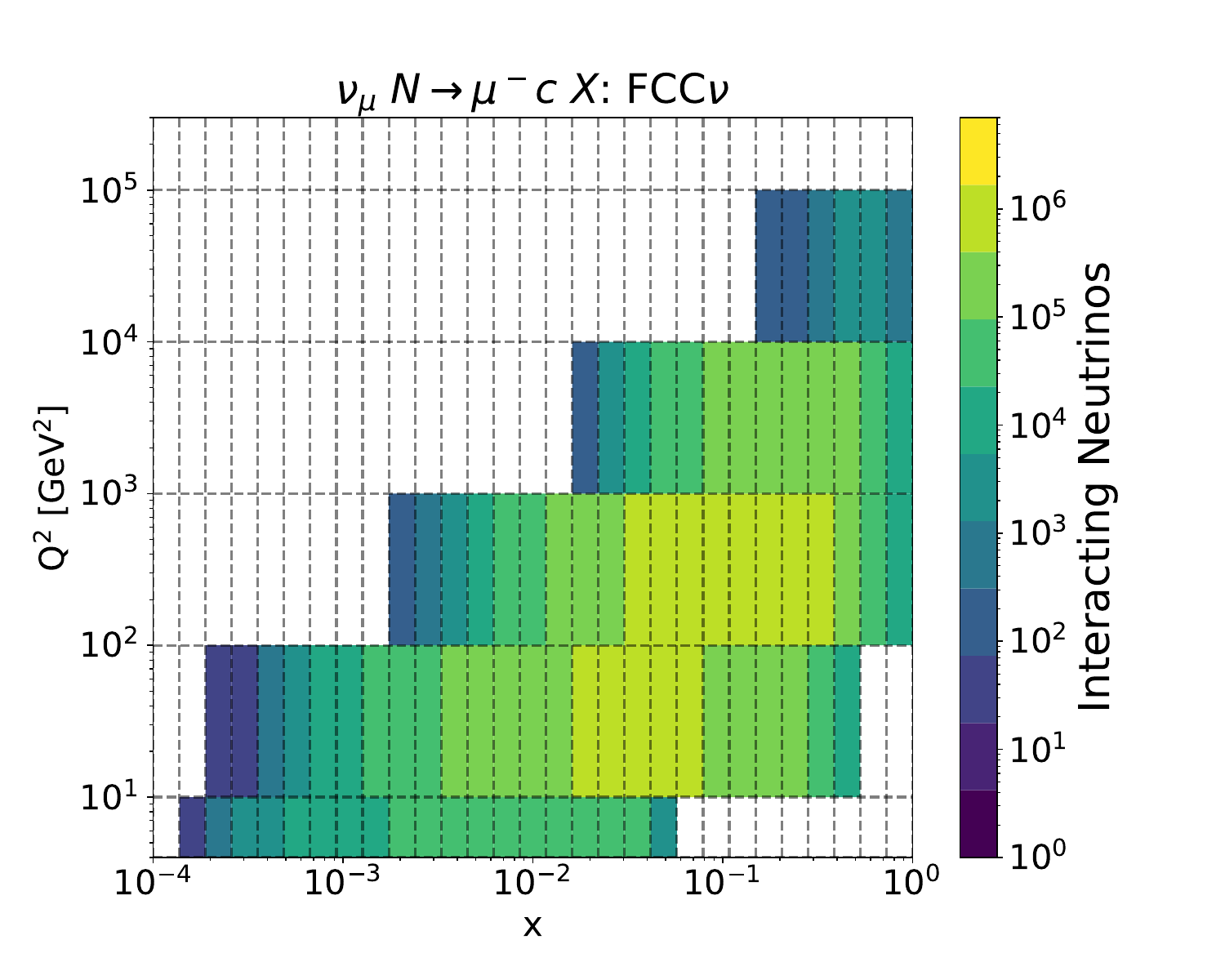}
\caption{The kinematic coverage in the $(x,Q^2)$ plane of muon-neutrino CC DIS measurements at the FCC$\nu$ detector for inclusive scattering (left) and for charm production (right).
For each bin we indicate the event yields to be accumulated for $\mathcal{L}_{\rm pp}=30$ ab$^{-1}$, from which the statistical uncertainties on the double-differential cross section measurement are determined. 
Only events within detector acceptance requirements are retained.
}
\label{fig:kinplane_FASERdeep}
\end{center}
\end{figure}

Fig.~\ref{fig:kinplane_FASERdeep} shows the kinematic coverage of muon-neutrino CC DIS measurements at the FCC$\nu$ detector, both for inclusive scattering and for charm production DIS, in the latter case assuming that final-state charm-tagging is experimentally available, as required to constrain strangeness~\cite{Faura:2020oom}.
Each bin indicates the event yields to be accumulated during the operation of the FCC-hh ($\mathcal{L}_{\rm pp}=30$~ab$^{-1}$), adding up to the total yields of Table~\ref{tab:integrated_rates}.
These binned event yields have been calculated using the NLO structure functions from the NNSF$\nu$ calculation~\cite{Candido:2023utz}, and we impose DIS cuts on the momentum transfer of $Q^2>2~{\rm GeV}^2$ and the hadronic final-state invariant mass of $W^2>4~{\rm GeV}^2$.
Given the multi-TeV energies of the neutrinos reaching the FPF@FCC detectors, DIS cuts have a negligible effect on the event yields. 
For the event yields of Fig.~\ref{fig:kinplane_FASERdeep}, we also impose the acceptance cuts of the FCC$\nu$ detector on the energy and scattering angle of the outgoing muon of $E_\ell>100~{\rm GeV}$ and $\theta_\ell\lsim 0.05$, and we require a total hadronic energy of $E_h>100~{\rm GeV}$, consistent with Ref.~\cite{Cruz-Martinez:2023sdv}. 
As compared to the $(x,Q^2)$ values accessible at the LHC far-forward experiments, the FCC$\nu$ detector extends their coverage in both the small-$x$ and large-$Q^2$ regions by almost an order of magnitude, reaching $x\sim 10^{-4}$ and $Q^2\sim 10^5$ GeV$^2$.

\begin{figure}[!t]
\begin{center}
\includegraphics[width=\textwidth]{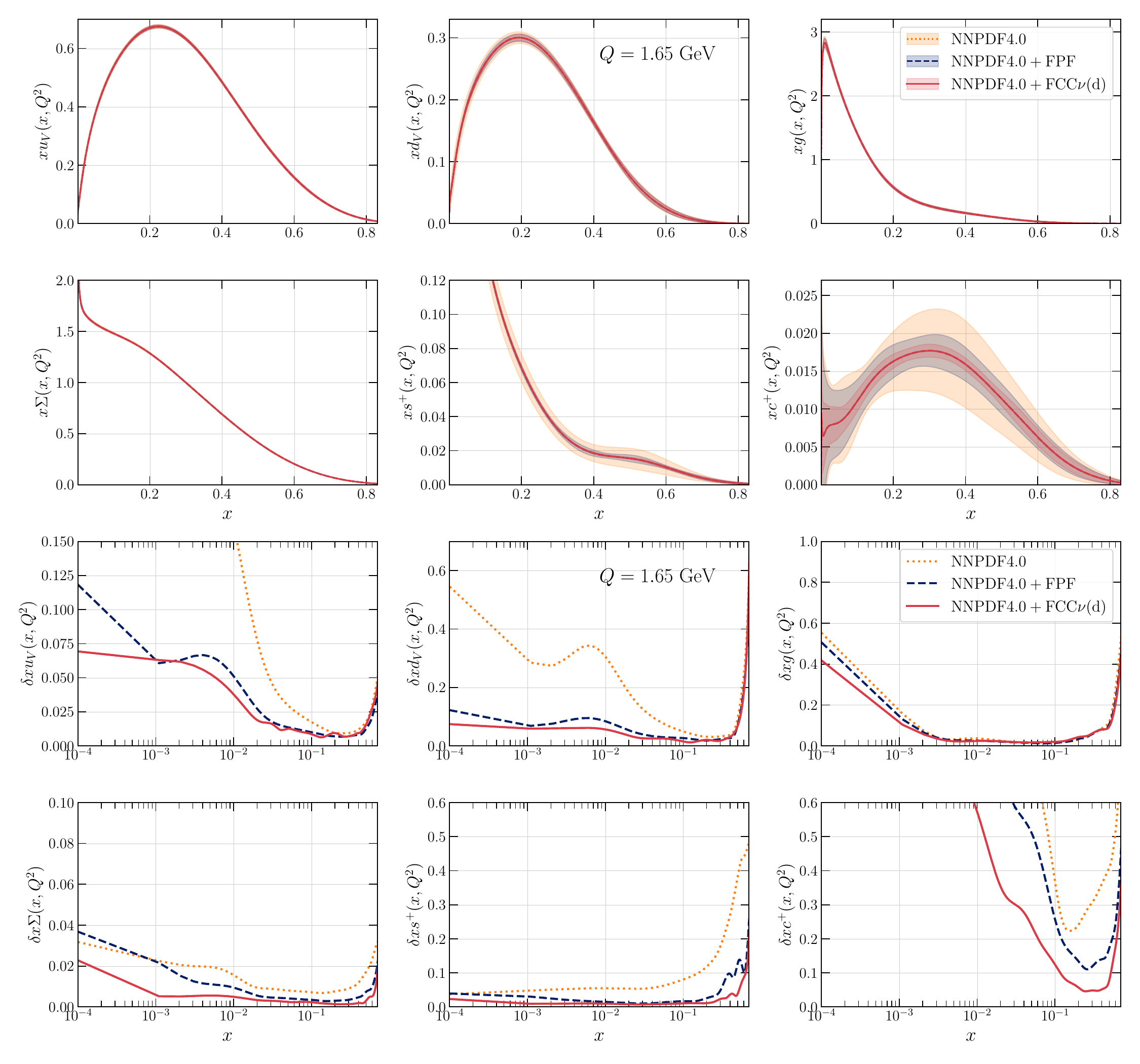}
\caption{The NNPDF4.0 PDFs (top two rows) and their $1\sigma$ PDF uncertainties (bottom two rows) upon the inclusion of neutrino DIS pseudo-data from the FCC$\nu$(d) detector.
For comparison, we also include the corresponding FPF projections from Ref.~\cite{Cruz-Martinez:2023sdv}. 
Results are shown at the input parametrization scale $Q = 1.65~\mathrm{GeV}$. }
\label{fig:proton-pdf-inclusive}
\end{center}
\end{figure}

Following Ref.~\cite{Cruz-Martinez:2023sdv}, we have generated DIS pseudo-data for the FCC$\nu$, FCC$\nu$(w), and FCC$\nu$(d) detectors using NNLO neutrino structure functions computed with {\sc\small YADISM} and the central replica of NNPDF4.0 as input PDF, and subsequently included them in the NNPDF4.0 global fit~\cite{NNPDF:2021njg,NNPDF:2021uiq}.
We limit the analysis to muon-neutrino DIS, which provides the largest event yields and is less sensitive to uncertainties affecting the electron and tau neutrino fluxes from charm decay.
Systematic uncertainties associated with detector performance are neglected, since we are technology-agnostic and want to determine the ultimate sensitivity of these experiments to proton structure. 
Fig.~\ref{fig:proton-pdf-inclusive} displays the results of including the FCC$\nu$(d) projections for neutrino structure functions in NNPDF4.0 at the input parametrisation scale of $Q = 1.65~\mathrm{GeV}$.
For reference, we also show the results of Ref.~\cite{Cruz-Martinez:2023sdv}, where the complete FPF structure function dataset (with only statistical uncertainties) was added to the same NNPDF4.0 prior fit.

The results of Fig.~\ref{fig:proton-pdf-inclusive} demonstrate that, in what concerns unpolarised structure functions, the reach of the FPF@FCC detectors improves that of the FPF at the HL-LHC, as expected due to both the increased event rates and the extended kinematic coverage.
More stringent constraints on the small-$x$ region are obtained from the  FCC$\nu$ measurements as compared to the FPF case, especially for the total quark singlet and for the $u_V$ and $d_V$ PDFs.
The FCC$\nu$ structure functions are particularly powerful to disentangle different quark flavours, as highlighted by the strangeness and the (fitted) charm PDFs, the latter relevant for intrinsic charm studies~\cite{NNPDF:2023tyk,Ball:2022qks}.
The impact of the FCC$\nu$ structure functions would be especially dramatic should the FPF not be realised at the HL-LHC.
These findings should nevertheless be revisited once realistic technological configurations for FCC$\nu$ detectors become available, since, as shown in Ref.~\cite{Cruz-Martinez:2023sdv}, the systematic errors limit the PDF sensitivity of the FPF measurements. 

\begin{figure}[!t]
\begin{center}
\includegraphics[width=0.87\textwidth]{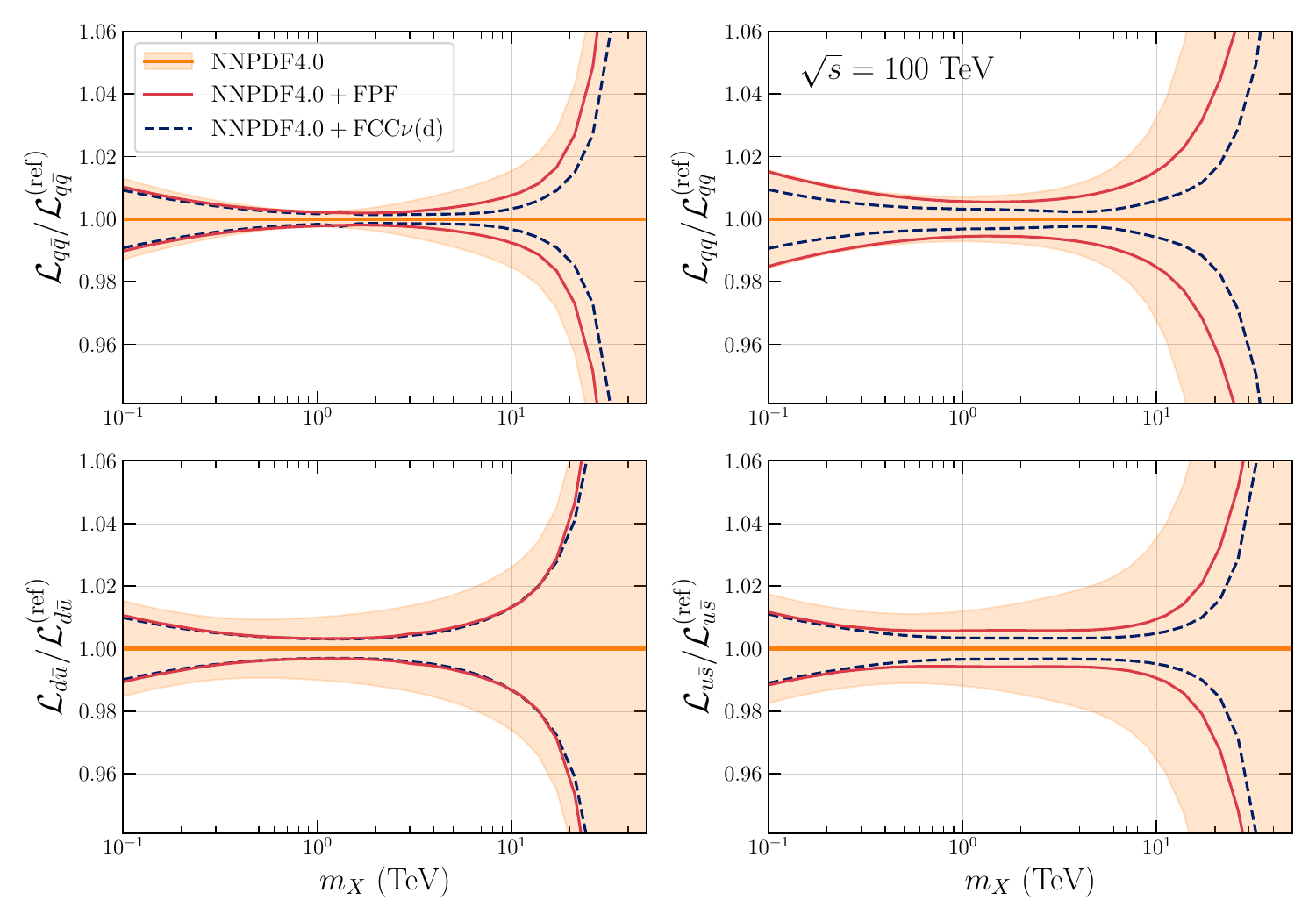}
\caption{Partonic luminosities in $pp$ collisions at $\sqrt{s}=100$~TeV as a function of $m_X$, comparing the 68\% CL uncertainties of the baseline with those from fits including FPF or FCC$\nu$(d) structure function projections.
}
\label{fig:lumi-FCCnu}
\end{center}
\end{figure}

As shown in Ref.~\cite{Cruz-Martinez:2023sdv}, the improved PDFs from FPF neutrino measurements enable precise predictions for key processes at the HL-LHC, from Higgs production to high-mass Drell-Yan distributions. 
The same feature would be present at the FCC-hh, with PDFs constrained by the FPF@FCC neutrino data leading to a reduction of the theory systematics entering $pp$ cross sections at $\sqrt{s}=100$~TeV.
To illustrate this feature, Fig.~\ref{fig:lumi-FCCnu} shows the partonic luminosities for $pp$ collisions at $100$ TeV as a function of the invariant mass of the final state $m_X$ for various quark flavour combinations, comparing NNPDF4.0 with fits including either FPF or FCC$\nu$(d) neutrino projections.
The FCC$\nu$(d) detector provides constraints across the full invariant mass range, including the crucial region for high-mass searches, with $m_X\gsim 10$ TeV, beyond the direct LHC coverage.
This impact is most marked for the quark-quark and quark-antiquark luminosities.

In summary, neutrino structure function measurements at the FPF@FCC would provide a sensitive probe of the partonic content of the nucleon.
They would constrain large-$x$ PDFs from ``low-energy'' measurements, in a manner that prevents the possible entanglement between PDF effects and BSM signals present in the high-$p_T$ tails of the FCC-hh data, in analogy with the situation at the (HL-)LHC~\cite{Greljo:2021kvv,Kassabov:2023hbm,Hammou:2023heg}.

\subsection{The proton spin under the neutrino microscope}
\label{sec:protonspin}

A major open question in Quantum Chromodynamics is explaining how the total spin of the proton ($S=\hbar/2$) arises in terms of the spin and orbital angular momentum of its underlying partonic constituents~\cite{Aidala:2012mv,Bass:2004xa}. 
Since the foundational SMC experiment three decades ago~\cite{SpinMuonSMC:1994met}, it is known that the three valence quarks contribute only a relatively small fraction of the total proton spin, with potentially large contributions from gluons~\cite{Nocera:2014gqa,deFlorian:2014yva}, sea quarks, and orbital angular momentum. 
Inclusive and semi-inclusive polarised DIS with charged leptons is a particularly clean process to extract the polarised PDFs (pPDFs) of the nucleon and connect them with the proton spin puzzle, and hence these processes represent the core of global analysis of pPDFs~\cite{Ball:2013lla,Nocera:2014gqa,deFlorian:2014yva,Borsa:2024mss,Borsa:2022vvp,Bertone:2024taw}.

\begin{figure}[t]
\begin{center}
\includegraphics[width=0.99\textwidth]{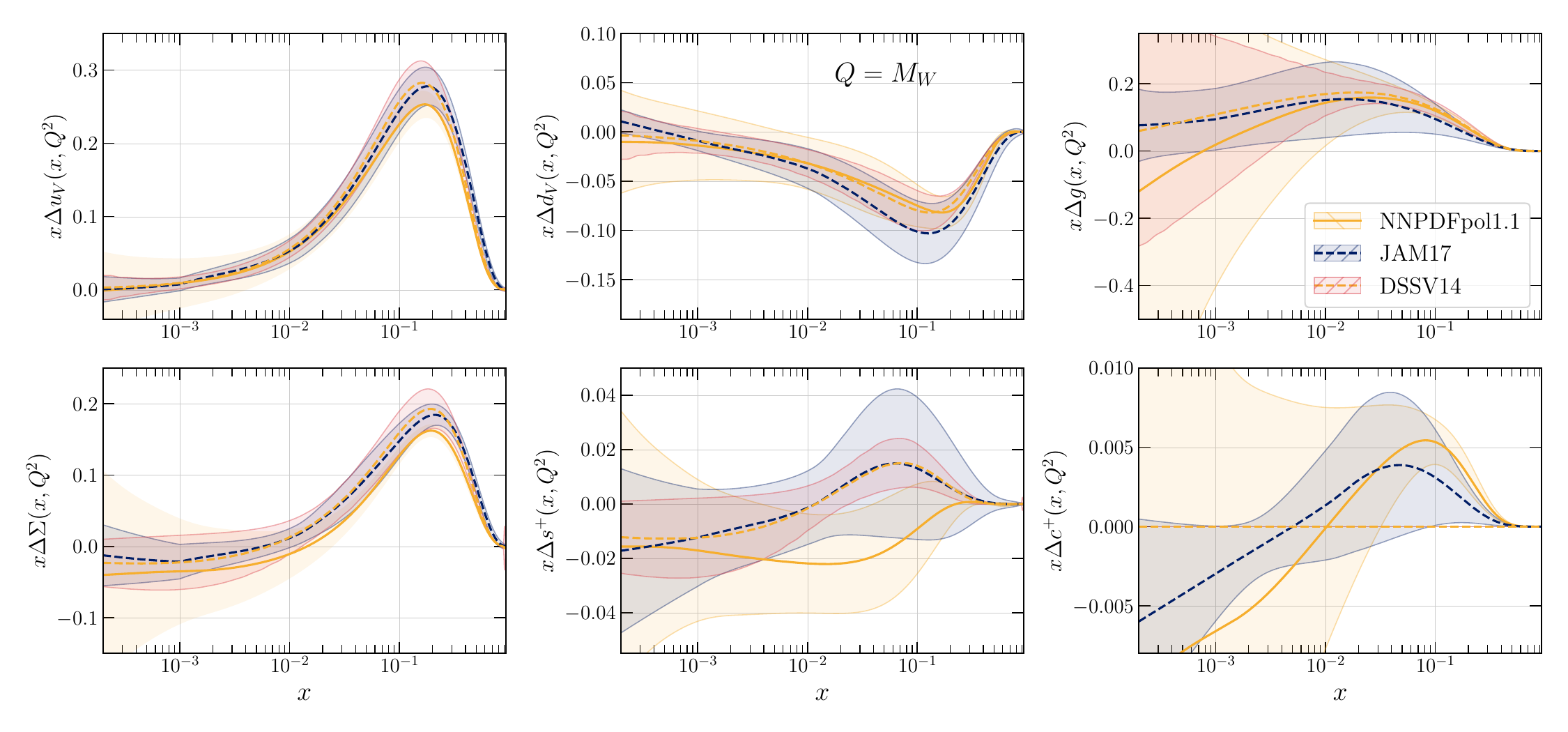}
\caption{The NNPDFpol1.1, DSSV14, and JAM17 sets of NLO polarised PDFs at $Q=m_W$ for the $x$ region accessible at the FCC$\nu$-pol detector (see also Fig.~\ref{fig:kinplane_COMPASS}).
The bands correspond to 68\% CL uncertainties.
}
\label{fig:polarized_PDF}
\end{center}
\end{figure}

Polarised DIS with charged leptons has been extensively studied in past and ongoing experiments, and it is one of the main science drivers of the upcoming EIC~\cite{Khalek:2021ulf}.
To illustrate the current state of the art of pPDFs determinations,
Fig.~\ref{fig:polarized_PDF} displays a comparison of the NNPDFpol1.1, DSSV14, and JAM17 NLO sets at the kinematics accessible with a polarised detector at the FPF@FCC.
Polarised PDF uncertainties are significant given the limited experimental information especially in the medium- and small-$x$ regions.

In the context of impact studies for a  proposed neutrino factory~\cite{Mangano:2001mj} to be installed at the front end of a muon storage ring with $E_\mu=50$~GeV, it was demonstrated~\cite{Forte:2001ph} that neutrino DIS on polarised targets exhibits a unique potential to scrutinise the spin structure of the proton. 
In contrast to polarised DIS using charged leptons, neutrino DIS enables a clean separation between quark and antiquark polarised PDFs of different flavours~\cite{Candido:2023utz}.
On the one hand, being (effectively) massless particles, neutrino beams are naturally polarised (spin aligned with helicity) and therefore carrying out polarised DIS measurements with neutrinos only requires achieving the polarisation of the target.
On the other hand, neutrinos suffer from a weak interaction cross section, which makes their detection with light polarisable targets highly challenging. 
Therefore, any realistic setup for polarised DIS with neutrino beams will result in a significant suppression of the event yields as compared to the charged-lepton case, which can only be compensated by a large enough neutrino flux.
As demonstrated in Sect.~\ref{subsubsec:polarised}, the unprecedented fluxes reaching the FPF@FCC would be intense enough to enable a first measurement of polarised DIS structure functions with neutrino beams. 

\begin{figure}[t]
\begin{center}
\includegraphics[width=0.49\textwidth]{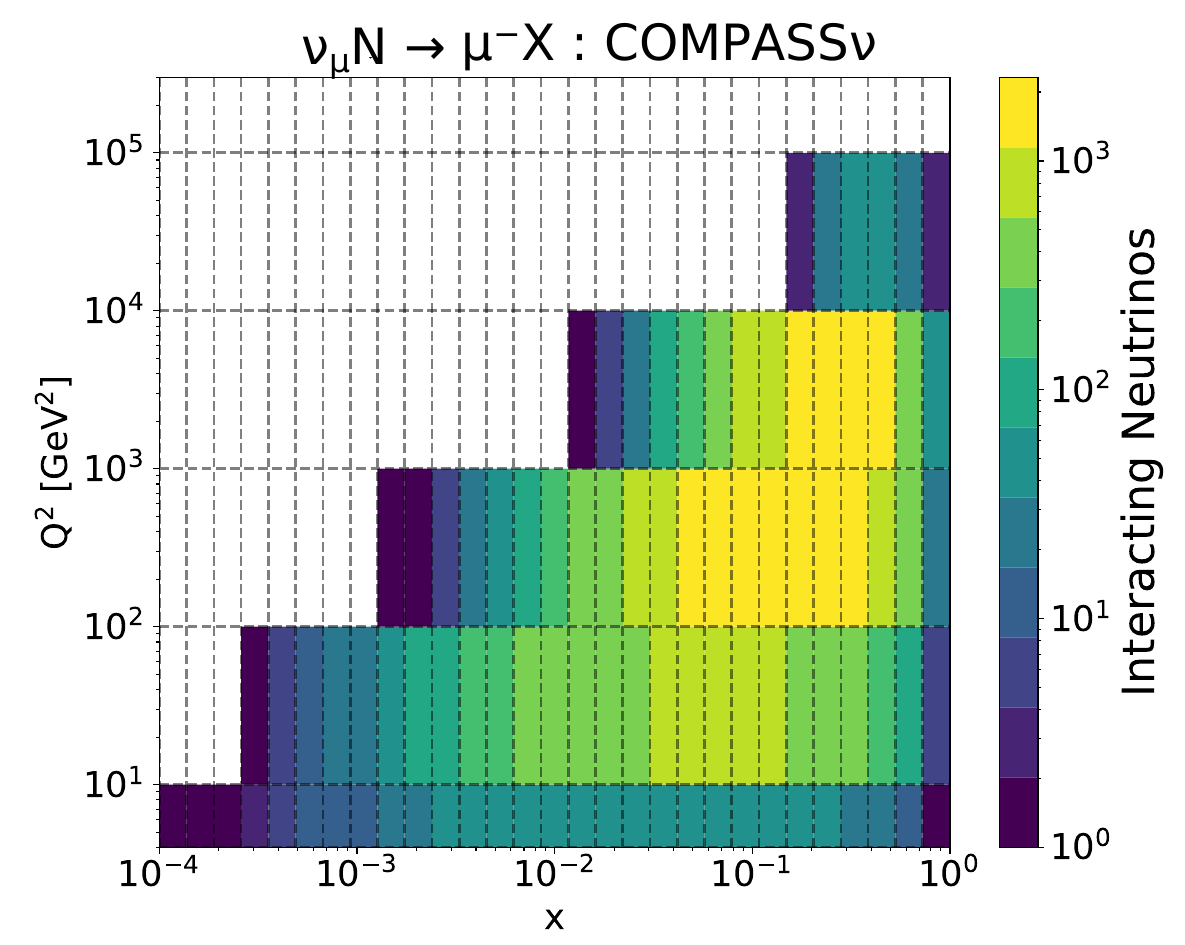}
\includegraphics[width=0.49\textwidth]{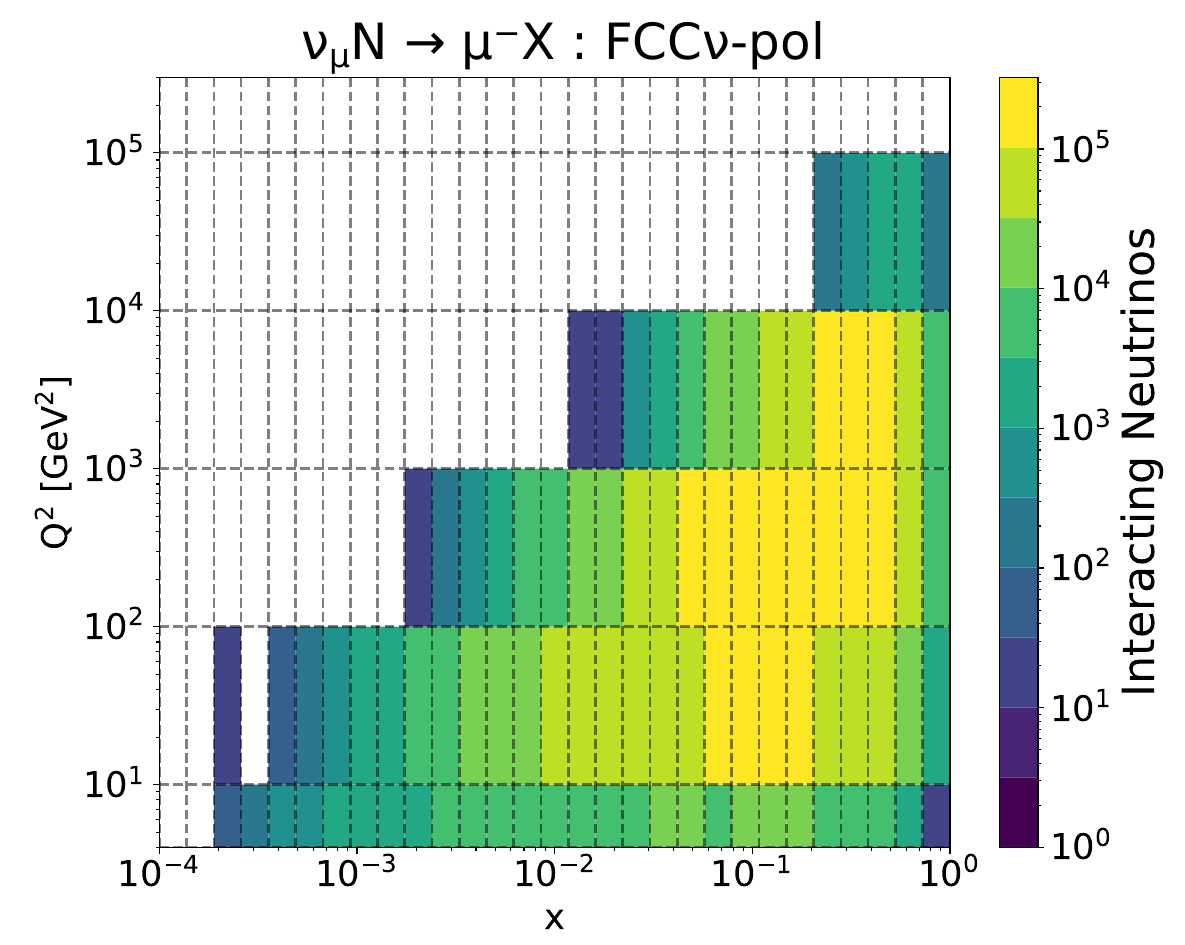}
\caption{Same as Fig.~\ref{fig:kinplane_FASERdeep} now for the two polarised detectors of Table~\ref{tab:neutrino_detectors_polarised}, COMPASS$\nu$ (left) and FCC$\nu$-pol (right), summed over the two target polarisations.
}
\label{fig:kinplane_COMPASS}
\end{center}
\end{figure}

Fig.~\ref{fig:kinplane_COMPASS} displays the
number of expected events binned in the $(x,Q^2)$ plane for the COMPASS$\nu$  and FCC$\nu$-pol polarised detectors of Table~\ref{tab:neutrino_detectors_polarised}, for muon-neutrino scattering and summing over the two polarisations of the target.
Adding up the bin contents results into the total event yields of Table~\ref{tab:integrated_rates_polarised}.
A relatively large polarised detector configuration (FCC$\nu$-pol) is required so that event rates and kinematic coverage become comparable with their counterparts for charged-lepton polarised scattering at the EIC.
Event yields at a more compact detector (COMPASS$\nu$) would be too low to be competitive with other experiments.
 
To translate the event yields shown in Fig.~\ref{fig:kinplane_COMPASS} into the expected precision for the measurement of polarised structure functions at the FPF@FCC detectors, we need to first evaluate the polarised neutrino structure functions~\cite{Forte:2001ph,Mangano:2001mj}.
For an unpolarised proton target, the double differential cross section for (anti-)neutrino scattering are given by
\bea
\label{eq:unpol_2d_xsec}
\frac{d^2\sigma^{\nu p}(x,y,Q^2)}{dxdy}&=&
\frac{G^2_F}{2\pi (1+Q^2/m_W^2)^2}
\frac{Q^2}{xy}
\lc  y \left(1-\frac{y}{2}\right) x F_3^{\nu p}
     + \lp 1-y\rp  F_2^{\nu p} + y^2 x F_1^{\nu p}\rc \, ,
\\
\frac{d^2\sigma^{\bar{\nu} p}(x,y,Q^2)}{dxdy}&=&
\frac{G^2_F}{2\pi (1+Q^2/m_W^2)^2}
\frac{Q^2}{xy}
\lc  -y \left(1-\frac{y}{2}\right) x F_3^{\bar{\nu} p}
     + \lp 1-y\rp  F_2^{\bar{\nu} p} + y^2 x F_1^{\bar{\nu} p}\rc \, , \nonumber
\eea
where $F_i(x,Q^2)$ are the unpolarised structure functions, and target mass effects are neglected.
For polarised targets, one defines the polarised cross section difference between the two target polarisations,
\be
\Delta\sigma\equiv
\sigma(\lambda_p=-1)-\sigma(\lambda_p=+1) \, ,
\ee
where $\lambda_p=\pm 1$ is the proton helicity. 
In this case, the double differential cross sections are given by~\cite{Forte:2001ph}
\bea
\label{eq:pol_2d_xsec}
\frac{d^2\Delta\sigma^{\nu p}(x,y,Q^2)}{dxdy}&=&
\frac{G^2_F}{\pi (1+Q^2/m_W^2)^2}
\frac{Q^2}{xy}
\lc  y \left(2-y\right) xg_1^{\nu p} - \lp 1-y\rp  g_4^{\nu p} - y^2 x g_5^{\nu p}\rc \, ,
\\
\frac{d^2\Delta\sigma^{\bar{\nu} p}(x,y,Q^2)}{dxdy}&=&
\frac{G^2_F}{\pi (1+Q^2/m_W^2)^2}
\frac{Q^2}{xy}
\lc-y \left(2-y\right) xg_1^{\bar{\nu} p} - \lp 1-y\rp  g_4^{\bar{\nu} p} - y^2 x g_5^{\bar{\nu} p}\rc \, , \nonumber
\eea
where $g_i^{\nu(\bar{\nu}) p}(x,Q^2)$ are CC scattering neutrino structure functions, and again target mass effects are neglected. 

To determine the statistical uncertainties associated with the measurements of polarised neutrino structure functions with FCC$\nu$-pol, we evaluate the ratio between the polarised cross section of Eq.~(\ref{eq:pol_2d_xsec}) and the unpolarised one of Eq.~(\ref{eq:unpol_2d_xsec}) for a given set of (un)polarised PDFs,
\be
\label{eq:asymmetry_polarised_cross-sections}
\mathcal{R}^{\nu p}(x,Q^2,y) \equiv \frac{\sigma^{\nu p}(\lambda_p=-1)-\sigma^{\nu p}(\lambda_p=+1)}{\sigma^{\nu p}(\lambda_p=-1)+\sigma^{\nu p}(\lambda_p=+1)} \, ,
\ee
and likewise for antineutrino scattering.
Assuming that  polarised asymmetries are not too large, the {\it absolute} statistical uncertainty on the polarised structure functions, $g_i^{\nu p}$, is the same as that of their unpolarised counterparts, $F_i^{\nu p}$, up to $\mathcal{O}(1)$ factors~\cite{Forte:2001ph}.
Therefore, the relative statistical precision of a measurement of $g_i^{\nu p}$, or what is the same, the absolute statistical precision on Eq.~(\ref{eq:asymmetry_polarised_cross-sections}), can be approximated by
\be
\label{eq:polarised_asy_uncertainties}
\delta \mathcal{R}^{\nu p}(x,Q^2,y) \simeq 
 \Big| \mathcal{R}^{\nu p}(x,Q^2,y)\Big|^{-1}\times \lp N_{\rm ev}^{\rm (bin)}(x,Q^2,y)\rp^{-1/2} \, ,
\ee
where $N_{\rm ev}^{\rm (bin)}$ is the number of expected events for each bin from Fig.~\ref{fig:kinplane_COMPASS}. 
Given a calculation of the asymmetry Eq.~(\ref{eq:asymmetry_polarised_cross-sections}) and the expected binned event rates from Fig.~\ref{fig:kinplane_COMPASS}, one can then determine the relative statistical uncertainty associated with a measurement of the double-differential polarised cross sections Eq.~(\ref{eq:pol_2d_xsec}) at the FPF@FCC detectors.

For the FCC$\nu$-pol projections used in this work, we evaluate the polarised asymmetries  Eq.~(\ref{eq:asymmetry_polarised_cross-sections}) using the central replica of the NNPDFpol1.1 NLO set~\cite{Nocera:2014gqa}.
It suffices to use LO expressions for the unpolarised and polarised structure functions (provided for completeness in App.~\ref{app:SFs}):
higher-order QCD corrections to polarised structure functions are moderate~\cite{Hekhorn:2024tqm} and their inclusion, for example, with {\sc\small YADISM}~\cite{Candido:2024rkr}, would not modify our estimates.
We thus determine this way the value of the statistical precision $\delta R^{\nu p}$, Eq.~(\ref{eq:polarised_asy_uncertainties}), in  the bins accessible for polarised DIS at FCC$\nu$-pol.
This estimate of $\delta R^{\nu p}$ would vary were we to use others pPDF to evaluate the ratio of Eq.~(\ref{eq:asymmetry_polarised_cross-sections}), although Fig.~\ref{fig:PolarisedPDFs-ratioasy} suggests that differences should be moderate.
The projected values of $\delta R^{\nu p}$ are shown in Fig.~\ref{fig:nuPol-StatErrors}, achieving few-percent statistical precision in the $x\gsim 10^{-2}$ region which then quickly worsens for smaller values of $x$.
While here we are detector-agnostic, realistically any polarised detector would have systematic errors at least at the $\mathcal{O}(10\%)$ level due to e.g. finite target polarisation and dilution effects; see, for example, the COMPASS analysis in Ref.~\cite{COMPASS:2006mhr}.
For this reason, we conservatively assume  a point-to-point uncorrelated experimental systematic error of $\delta_{\rm sys}\mathcal{R}^{\nu p}=30\%$.

\begin{figure}[t]
\begin{center}
\includegraphics[width=0.99\textwidth]{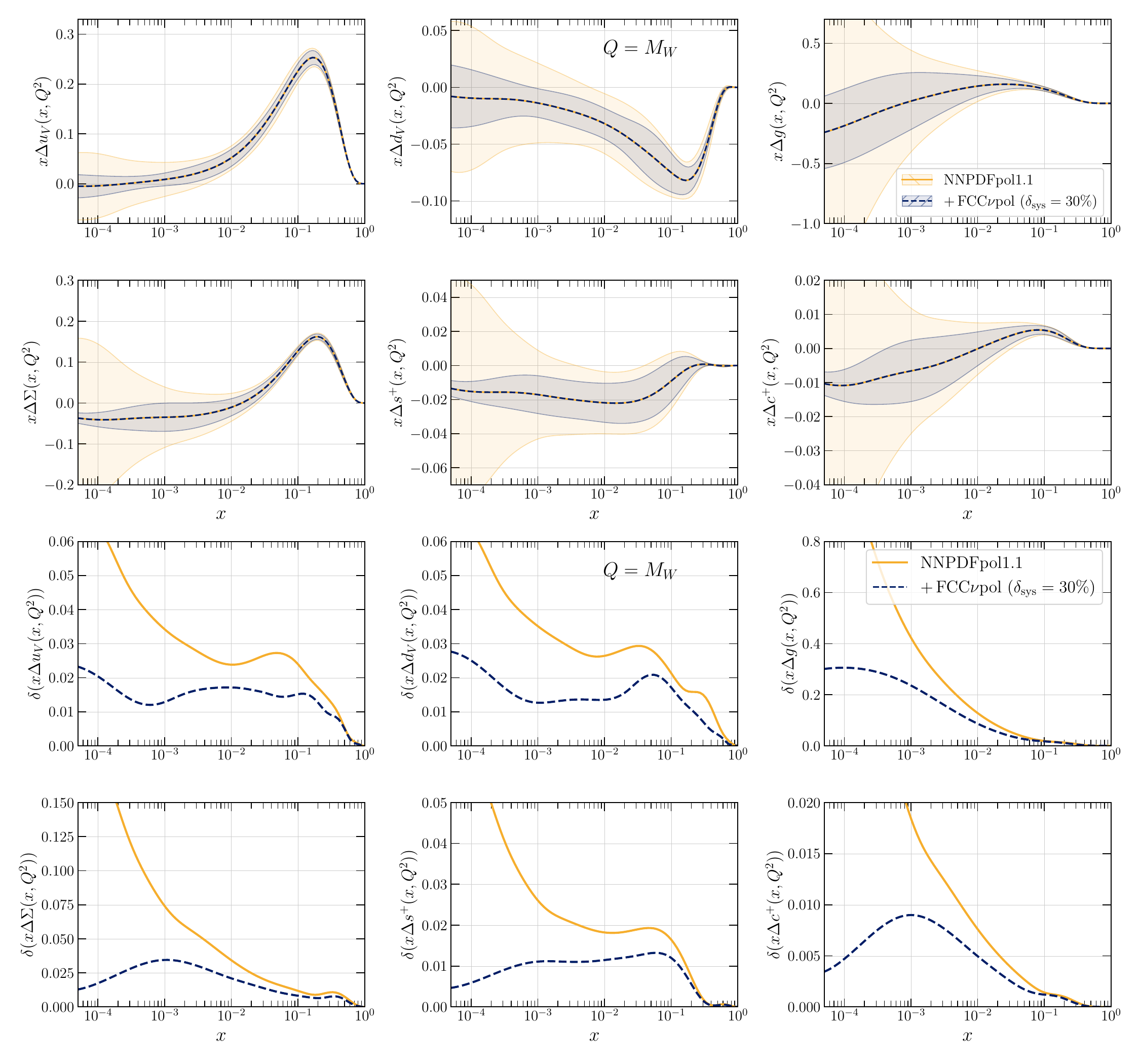}
\vspace{-0.5cm}
\caption{Top: Same as in Fig.~\ref{fig:polarized_PDF}, now comparing the prior NNPDFpol1.1 NLO set with the outcome of its reweighting with the FCC$\nu$-pol projections.
Bottom two rows: The associated absolute 68\% CL PDF uncertainties.
}
\label{fig:PolarisedPDFs-RW}
\end{center}
\end{figure}

To assess the constraints provided by the FCC$\nu$-pol structure function measurements on the spin structure of the proton, we include these projections in the NNPDFpol1.1 NLO fit by means of the  Bayesian reweighting procedure~\cite{Ball:2010gb,Ball:2011gg}.
Fig.~\ref{fig:PolarisedPDFs-RW} compares the prior pPDFs with the results of including the FCC$\nu$-pol structure functions via reweighting.
We find a significant impact on the quark and antiquark pPDFs of all flavours extending down to $x\sim 10^{-4}$ and below, consistently with the kinematic coverage of the FCC$\nu$-pol detector from Fig.~\ref{fig:kinplane_COMPASS}. 
Especially remarkable is the information gain in the region $x\lsim 10^{-2}$, for which there are currently limited experimental constraints available.
The FCC$\nu$-pol structure functions measurements appear to also be sensitive to the gluon and charm  polarised PDFs at small-$x$ via their mixing with the light quark PDFs through DGLAP evolution. 

The encouraging results reported in Fig.~\ref{fig:PolarisedPDFs-RW} motivate a more refined analysis that does not rely on the Bayesian reweighting approximation, includes also higher-order QCD and mass corrections to the neutrino structure functions using {\sc\small YADISM}, and comes accompanied by a realistic estimate of systematic errors associated with specific technologies for polarised DIS detectors.
This updated study would be possible once the upcoming NNPDFpol2.0 global analysis~\cite{Hekhorn:2024jrj} of pPDFs is released, which would allow the FPF@FCC projections to be included on the same footing as both available data on charged-lepton polarised asymmetries and projections for future EIC measurements, as done in Ref.~\cite{Ball:2013tyh}.

\subsection{Mapping cold nuclear matter at ultra-small-$x$}
\label{sec:heavyions}

As demonstrated in Sect.~\ref{subsec:fluxes_pPb}, proton-lead and lead-lead collisions at the FCC-hh would generate a sufficiently large flux of forward neutrinos to record sizeable DIS event samples at the FPF@FCC detectors, up to $\mathcal{O}(10^5)$ with electron-neutrinos
and $\mathcal{O}(10^6)$ for muon-neutrinos;  see Table~\ref{tab:integrated_rates_pPb}.
Neutrinos from proton-lead collisions are of particular interest to study cold nuclear matter effects, such as those encoded by the nuclear PDFs (nPDFs)~\cite{Ethier:2020way,Klasen:2023uqj}, in an extreme kinematical regime uncharted by present or future experiments.

As supported by available LHC measurements~\cite{Klasen:2023uqj}, high-$p_T$ processes in proton-ion collisions can be satisfactorily described by the QCD factorisation theorems in terms of collinear nPDFs.
QCD factorisation down to the $x\sim 10^{-9}$ values accessible at the FPF@FCC experiments, however, has never been demonstrated. 
Indeed, in the small-$x$ regime of QCD relevant for forward neutrinos at hadron colliders, departures from linear DGLAP dynamics are generically expected, with the possible onset of phenomena such as BFKL resummation~\cite{Ball:2017otu,Silvetti:2022hyc} or non-linear (saturation) effects~\cite{Mueller:1989st}, the latter  enhanced in heavy nuclei by a factor $A^{1/3}$ as compared to free nucleons.
The quest to unveil new regimes of QCD, such as the Color-Glass Condensate (QGC), is one of the drivers of the current heavy-ion program at the LHC, as well as of the upcoming EIC.
Improving the modelling of forward particle production in proton-ion collisions is also important for the interpretation of  high-energy astroparticle physics, such as in the simulation of extensive air showers initiated by cosmic rays or the determination of the prompt neutrino fluxes at neutrino observatories.

To estimate the sensitivity of the FPF@FCC to small-$x$ nPDFs, we focus on charm production in proton-lead collisions, which can be evaluated in the framework of perturbative QCD factorisation.
We restrict our simulation to charm production through hard scattering and switch off the copious underlying event.
The calculational settings use POWHEG NLO simulations matched to {\sc\small Pythia8.3} with the nNNPDF3.0 NLO set for lead set as input.
To generate p+Pb collisions with {\sc\small POWHEG}, we first simulate asymmetric $pp$ collisions with the appropriate nucleon beam energies  matching p+Pb collisions at the 
FCC-hh.
The resulting charm production event rates are then rescaled by a factor $A=208$.
These  simulations predict $4.1\times 10^{4}$ ($3.2\times 10^{3}$) $\nu_e$ ($\nu_\tau$)  events recorded by the FCC$\nu$ detector coming from charmed-meson decays in p+Pb collisions at $\sqrt{s_{\rm NN}}=63$~TeV; see also Table~\ref{tab:integrated_rates_pPb} (which corresponds, however, to the sum over all channels and is based on {\sc\small Pythia8} LO simulations with the {\tt Angantyr} model).

\begin{figure}[t]
    \centering
    \includegraphics[width = 0.99\textwidth]{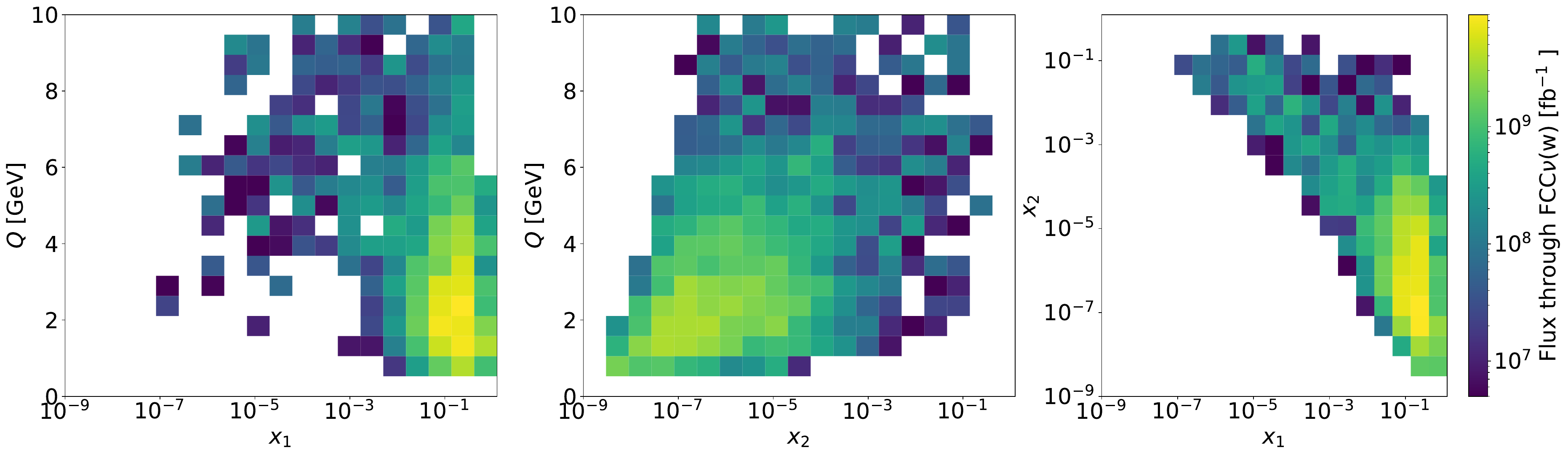}
    \caption{Distributions in the
    $(x_1,Q)$, $(x_2,Q)$, and $(x_1,x_2)$ planes of charm production ($p+p\to c+\bar{c}$) events in $pp$ collisions at $\sqrt{s}=100$ TeV, where $x_1 > x_2$ are the momentum fractions carried by the two colliding partons. 
    We retain only events leading to neutrinos passing through the fiducial volume of the FCC$\nu$(w) detector.
    } 
    \label{fig:x1x2_production}
\end{figure}

%
Fig.~\ref{fig:x1x2_production} displays the distributions in the
$(x_1,Q)$, $(x_2,Q)$, and $(x_1,x_2)$ planes of charm production events in symmetric $pp$ collisions at $\sqrt{s}=100$ TeV, where $x_1 > x_2$ are the momentum fractions carried by the two colliding partons. 
The dominant partonic subprocesses are $gg$ and $gc(\bar{c})$, the latter enhanced for PDFs that account for intrinsic charm effects~\cite{Ball:2022qks,Guzzi:2022rca}.
We only retain events whose neutrinos pass through the fiducial volume of the FCC$\nu$(w) detector.
Fig.~\ref{fig:x1x2_production} highlights how forward charm production at $\sqrt{s}=100$ TeV is dominated by very small-$x$ and large-$x$ values, with $Q\sim {\rm few~GeV}$ due to the charm production cross section peaking at the kinematic threshold $2m_c$, and with large neutrino fluxes down to around $x\sim 10^{-8}$.
This picture is qualitatively unchanged in p+Pb collisions at $\sqrt{s_{\rm NN}}=63$ TeV.
The rightmost panel of Fig.~\ref{fig:x1x2_production} displays a strong anti-correlation between $x_1$ and $x_2$, which follows from the production kinematics and the forward selection requirements, and indicates that the charm neutrino flux from 100 TeV collisions is maximal for $x_1 \sim 0.3$ and $x_2\sim 10^{-7}$.
Since we are interested in probing small-$x$ nPDFs, $x_2 < x_1$ should be the momentum fraction of the colliding nucleus, and the forward detector should be placed in the direction of the proton beam.

\begin{figure}[t]
     \centering
     \includegraphics[width = 0.99\textwidth]{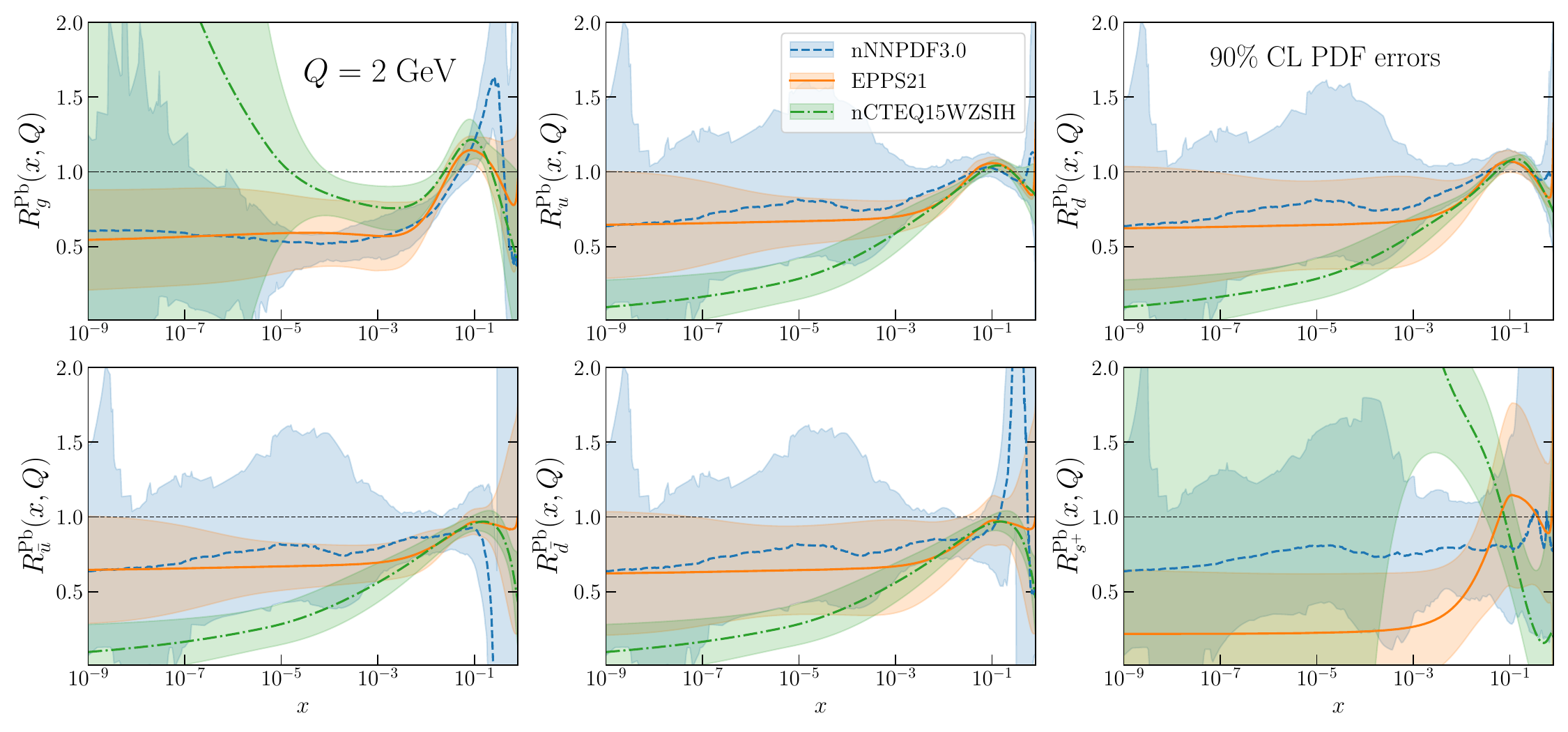}
     \caption{The nPDF modification factors in lead nuclei, Eq.~(\ref{eq:nuclear_modification_factor}),  at $Q=2$ GeV
     for three recent nuclear PDF sets: nNNPDF3.0, EPPS21, and nCTEQ15WZSIH. 
     The bands indicate the  90\% CL nPDF uncertainties, and we display the $x$-region sensitive to forward neutrino production at the FPF@FCC.
     }
     \label{fig:pdfplot-pdg2024_global_q2gev}
 \end{figure}

The state-of-the-art of nPDF determinations is summarised in Fig.~\ref{fig:pdfplot-pdg2024_global_q2gev}, displaying the nPDF modification factors in lead nuclei, defined as
\be
\label{eq:nuclear_modification_factor}
R_i^{\rm (Pb)}(x,Q)=f_i^{\rm (p/Pb)}(x,Q)/f_i^{\rm (p)}(x,Q)\, , \qquad i = q,\bar{q},g \, ,
\ee
where  $f_i^{\rm (p)}$ ($f_i^{\rm (p/Pb)}$) indicates the free proton (bound proton in lead nuclei) PDFs, and $i$ is the flavour index.
These nPDF modification factors are evaluated at $Q=2$ GeV
for three recent global fits of nuclear PDF sets (nNNPDF3.0~\cite{AbdulKhalek:2022fyi}, EPPS21~\cite{Eskola:2021nhw}, and nCTEQ15WZSIH~\cite{Duwentaster:2021ioo, Duwentaster:2022kpv}), and
the bands indicate the associated 90\% CL uncertainties.
The $x$ and $Q^2$ region displayed is the one covered by neutrinos reaching the FPF@FCC detectors and originating from charm production at the FCC-hh; see Fig.~\ref{fig:x1x2_production}.
Despite both nNNPDF3.0 and EPPS21 including the information from LHCb $D$-meson production, in the ultra-small-$x$ region, $x\lsim 10^{-5}$, nPDFs remain  essentially undetermined, motivating the search for new probes to constrain them.

Throughout this section we assume the validity of perturbative QCD to describe charm production in proton-nucleus collisions.
For nucleus-nucleus collisions, this assumption is expected to become invalid due to the presence of collective effects, such as the formation of a quark-gluon plasma (QGP).
For this reason, the ability of the FPF@FCC detectors to observe thousands of neutrinos produced in ion-ion collisions would offer a new window to study forward particle production in heavy ion physics and probe the formation and evolution of particles in hot nuclear matter.  
We note, however, that there is some experimental evidence that collective effects may also be present in proton-ion or even $pp$ collisions.
This evidence includes the observation of multi-particle angular correlations at high particle multiplicities by CMS~\cite{CMS:2016fnw} and of strangeness enhancement at high particle multiplicities by ALICE~\cite{ALICE:2016fzo}, which are often attributed to the formation of a QGP-like medium. 
Nevertheless, collective effects seem to only occur for softer particles in high multiplicity environments, which mainly occur in the central regime.
In the forward direction, where the multiplicity density is typically lower, one would therefore not expect to see these effects. 

Pinning down the small-$x$ nPDFs of lead (or any other nuclear species for which collisions become available at the FCC-hh) using data from the FPF@FCC could be possible by defining tailored observables, such as a nuclear event yield ratio of the form
\be
\label{eq:R_charm_pPb}
\mathcal{R}_{\rm Pb}^{(\nu_\ell)}(E_\nu) \equiv \frac{N_{\nu_\ell + \bar{\nu}_\ell}^{\rm (pPb)}(E_\nu)}{N_{\nu_\ell + \bar{\nu}_\ell}^{\rm (pp)}(E_\nu)} \, ,
\ee
that is, the ratio between the number of measured $\nu_\ell+\bar{\nu}_{\ell}$ CC DIS events originating from proton-lead collisions at $E_\nu$ to the corresponding quantity in $pp$ collisions.
The evaluation of Eq.~(\ref{eq:R_charm_pPb}) should be restricted to the energy region where charm production dominates, namely electron neutrinos with $E_\nu \gsim 3$ TeV and tau neutrinos of any energy (see Fig.~\ref{fig:fluxes_FCC}).
The motivation to define ratio observables such as Eq.~(\ref{eq:R_charm_pPb}) is that theory uncertainties affecting forward charm production, such as higher-order QCD corrections and the modelling of charm hadronization, partially cancel out, while the sensitivity to nPDFs or other types of QCD dynamics at small-$x$ remains.\footnote{For the actual measurement, one should refine this observable by accounting for the different boosts of the final state in $pp$ and proton-lead collisions, such that the same $E_\nu$ values correspond to the same momentum fractions $x_{1,2}$ in production in both types of collisions.
Given that here we work with projected pseudo-data, this difference is immaterial.}  
See Refs.~\cite{Mangano:2012mh,Gauld:2015yia,Cacciari:2015fta,Zenaiev:2019ktw,EuropeanStrategyforParticlePhysicsPreparatoryGroup:2019qin,Kusina:2017gkz} for related approaches.

Using the {\sc\small POWHEG} charm production sample in p+Pb collisions to estimate the projected statistical uncertainties, we generate pseudo-data for the ratio observable  of Eq.~(\ref{eq:R_charm_pPb}) for the FCC$\nu$ detector for electrons and tau neutrinos,
\be
\label{eq:R_charm_pPb_v2}
\mathcal{R}_{\rm Pb}^{(\nu_e)}(E_\nu) =\frac{N_{\nu_e + \bar{\nu}_e}^{\rm pPb}(E_\nu)}{N_{\nu_e + \bar{\nu}_e}^{\rm pp}(E_\nu)} \,\, ~(E_\nu \ge 3~{\rm TeV} )\, ,
\qquad 
\mathcal{R}_{\rm Pb}^{(\nu_\tau)}(E_\nu) =\frac{N_{\nu_\tau + \bar{\nu}_\tau}^{\rm pPb}(E_\nu)}{N_{\nu_\tau + \bar{\nu}_\tau}^{\rm pp}(E_\nu)} \, ,
\ee
for each of the $N_{\rep}=200$ Monte Carlo replicas of nNNPDF3.0 NLO set for lead ($A=208)$,
and freezing the proton PDF to the central value of the nNNPDF3.0 $A=1$ set.
The latter choice is justified by the dominance of nPDF uncertainties in the evaluation of Eq.~(\ref{eq:R_charm_pPb_v2}), especially since nNNPDF3.0 already includes the constraints from LHCb charm production for both $A=1$ and $A=208$.

Pseudo-data for Eq.~(\ref{eq:R_charm_pPb_v2}) is constructed in terms of the central predictions from nNNPDF3.0, with statistical uncertainties determined from the expected yields per energy bin of $N_{\nu_\ell + \bar{\nu}_\ell}^{\rm (pPb)}$ at the FCC$\nu$ detector.
While experimental and theoretical  systematic errors would partially cancel out in the ratio, to be conservative we assume a bin-per-bin uncorrelated systematic error of $\delta_{\rm sys}=25\%$. 
The pseudo-data for $\mathcal{R}_{\rm Pb}^{(\nu_e)}(E_\nu)$ and $\mathcal{R}_{\rm Pb}^{(\nu_\tau)}(E_\nu)$ is then included in the nNNPDF3.0 global fit by means of the Bayesian reweighting method~\cite{Ball:2011gg,Ball:2010gb} in the same manner as in the polarised PDF study of Sect.~\ref{subsubsec:polarised}.

 \begin{figure}[t]
     \centering
     \includegraphics[width = 0.99\textwidth]{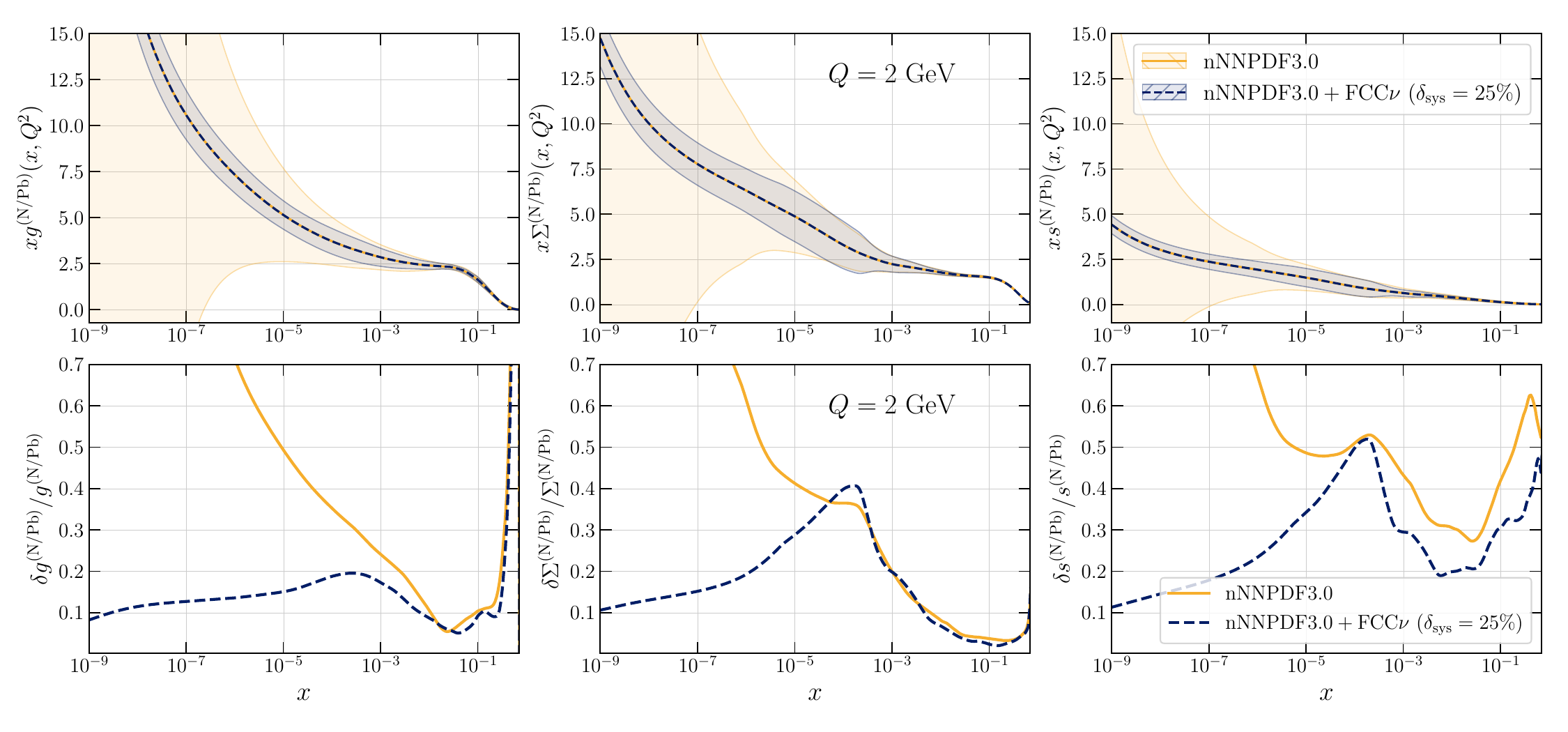}
     \caption{The lead PDFs from nNNPDF3.0 compared with the outcome of their reweighting with the FCC$\nu$ projections for the nuclear ratios $\mathcal{R}_{\rm Pb}^{(\nu_e)}$
     and $\mathcal{R}_{\rm Pb}^{(\nu_\tau)}$ of Eq.~(\ref{eq:R_charm_pPb_v2}).
     Results are shown for the gluon, quark singlet, and total strangeness PDFs at $Q=2$~GeV in the kinematic region sensitive to forward charm production at the FPF@FCC.
     An uncorrelated systematic uncertainty of $\delta_{\rm sys}=25\%$ has been assumed.
     The top panels display the absolute PDFs, with the bands indicating the 68\% CL uncertainties, while the bottom show the relative PDF uncertainties.
     }
     \label{fig:NuclearPDFs-RW}
 \end{figure}

Fig.~\ref{fig:NuclearPDFs-RW} displays
the PDFs of an average nucleon $N$ bound within a lead nucleus, $xf^{\rm (N/Pb)}(x,Q^2)$, from the nNNPDF3.0 determination, compared with the outcome of their reweighting with the FCC$\nu$ projections for the nuclear event yield ratios of Eq.~(\ref{eq:R_charm_pPb_v2}).
Results are shown for the gluon, quark singlet, and total strangeness PDFs in the kinematic region sensitive to forward charm production at the FPF@FCC; see also Fig.~\ref{fig:pdfplot-pdg2024_global_q2gev}.
For the gluon nPDF, a huge reduction of its uncertainties enabled by the FPF@FCC measurements is found for $x\lsim 10^{-4}$ with the constraints being the largest at $x\sim 10^{-8}$, consistent with the kinematics of charm production in Fig.~\ref{fig:x1x2_production}.
A similar qualitative behaviour is obtained for the quark singlet and total strangeness nPDFs, since at small-$x$ the gluon seeds the quark evolution and furthermore the quark sea is flavour-symmetric. 

\begin{figure}[t]
     \centering
     \includegraphics[width = 0.90\textwidth]{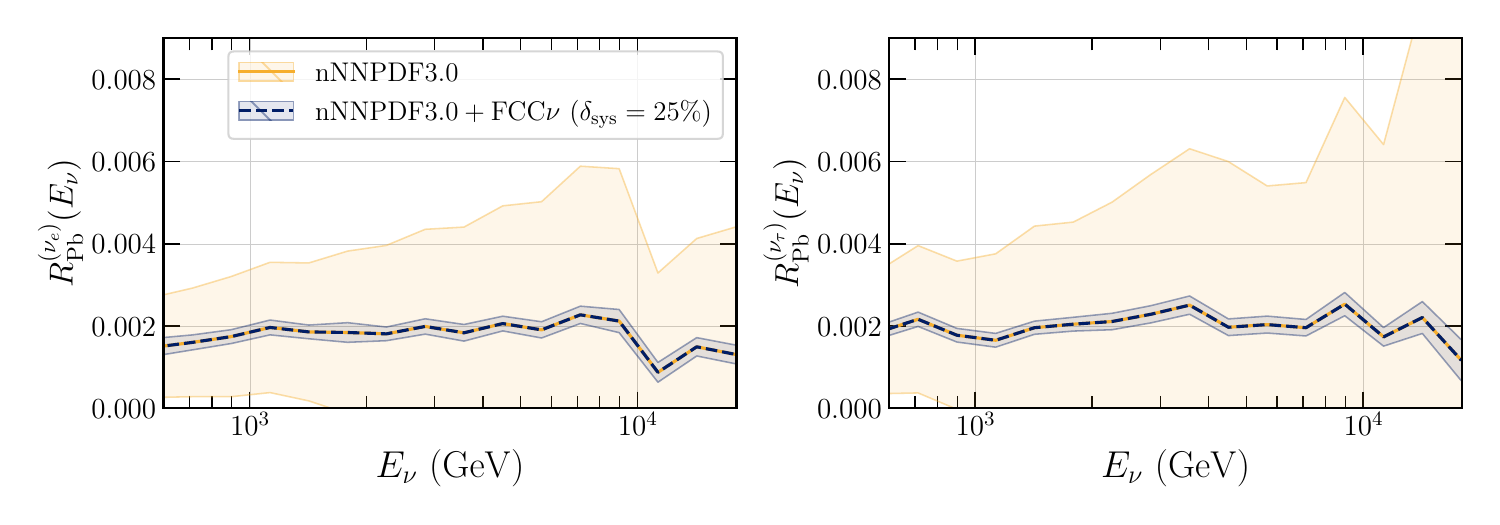}
     \caption{The nuclear ratios $\mathcal{R}_{\rm Pb}^{(\nu_e)}$ (left)
     and $\mathcal{R}_{\rm Pb}^{(\nu_\tau)}$ (right) of Eq.~(\ref{eq:R_charm_pPb_v2}), computed with nNNPDF3.0 before and after the inclusion of the FCC$\nu$ projections in the fit.  See also Fig.~(\ref{fig:NuclearPDFs-RW}).
     }
     \label{fig:NuclearPDFs-RatPb-RW}
 \end{figure}

Finally, Fig.~\ref{fig:NuclearPDFs-RatPb-RW} shows the neutrino event yield nuclear ratios $\mathcal{R}_{\rm Pb}^{(e)}$
and $\mathcal{R}_{\rm Pb}^{(\tau)}$, restricted to neutrinos from charm decays, computed with nNNPDF3.0 before and after the inclusion of the FCC$\nu$ projections shown in Fig.~\ref{fig:NuclearPDFs-RW} in the fit.
While nPDF uncertainties in the prior theory predictions are above 100\%, these are reduced to a few percent following their inclusion in nNNPDF3.0 via reweighting.
We also note that the impact of the FCC$\nu$ projections is approximately constant with $E_\nu$.
The central values of the ratios of Eq.~(\ref{eq:R_charm_pPb_v2}) are primarily determined by the difference in the integrated luminosity, $\mathcal{L}_{\rm pPb}/\mathcal{L}_{\rm pp}\sim 10^{-5}$, partially compensated by a factor $A\sim 200$ in p+Pb collisions plus additional subleading corrections.

All in all, the results of Figs.~\ref{fig:NuclearPDFs-RW} and~\ref{fig:NuclearPDFs-RatPb-RW} demonstrate that, at the FPF@FCC, a precision determination of the gluon and quark PDFs of lead nuclei down to $\sim 10^{-9}$ should be possible at the $\mathcal{O}\lp 10\rp\%$ level.
Such determination would therefore be highly sensitive to eventual deviations from collinear QCD factorisation and to nonlinear evolution effects.
\section{BSM sensitivity of the FPF@FCC}
\label{sec:BSM}

Complementing the QCD and hadron structure studies presented in Sect.~\ref{sec:SM}, precise high-energy neutrino measurements in the forward kinematic region of the FCC-hh will also offer novel opportunities for probing BSM physics and testing the electroweak sector of the SM in unexplored regimes. 
Here we illustrate these capabilities by using the example of the neutrino charge radius measurement.
Furthermore, $pp$ collisions at $\sqrt{s}=100$~TeV could lead to a collimated flux of forward-going BSM states with masses of up to several hundred GeV, giving access to rare and displaced BSM signal events that would otherwise evade detection in traditional detectors. 
We discuss here such BSM physics opportunities for displaced decay and ionization signatures based on proposed searches for dark Higgs bosons, quirks, and mCPs.

\subsection{The neutrino charge radius}
\label{subsec:charge_radius}
 
The electromagnetic properties of neutrinos have long been recognized as a potential window to new physics; see Ref.~\cite{Giunti:2014ixa} for a review.
Already in 1930, Pauli speculated about the existence of a non-zero neutrino magnetic moment~\cite{Pauli:1930pc}. 
Later, it was shown that a non-zero neutrino mass necessarily implies a non-zero magnetic moment~\cite{Fujikawa:1980yx, Shrock:1982sc, Pal:1981rm}. 
In addition, a variety of BSM mechanisms have been proposed that generate an effective electromagnetic current for the neutrino~\cite{Voloshin:1987qy, Barbieri:1988fh, Rajpoot:1990hj, Aboubrahim:2013yfa, Lindner:2017uvt, Babu:2020ivd}. 
Such an effective coupling of the neutrino to the photon can modify the event rates at neutrinos scattering experiments, including the far-forward neutrino detectors at FCC.

\begin{figure}[t]
\centering
\includegraphics[width=0.45\textwidth]{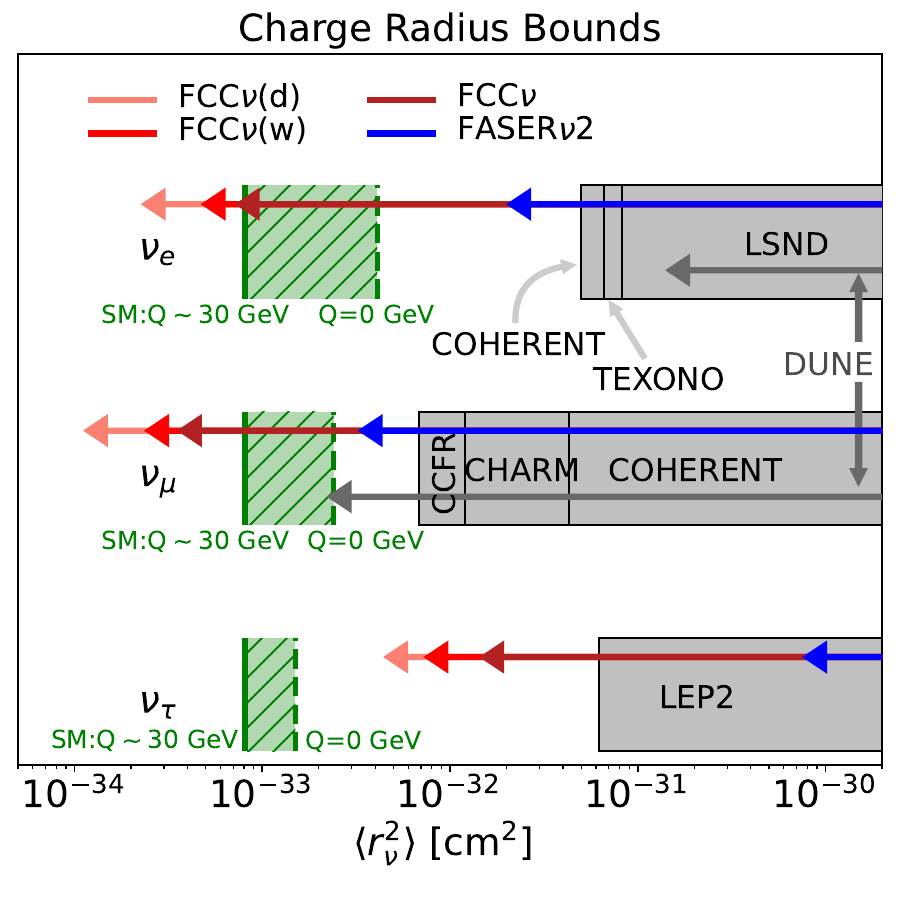}
\hspace{1cm}
\includegraphics[width=0.45\textwidth]{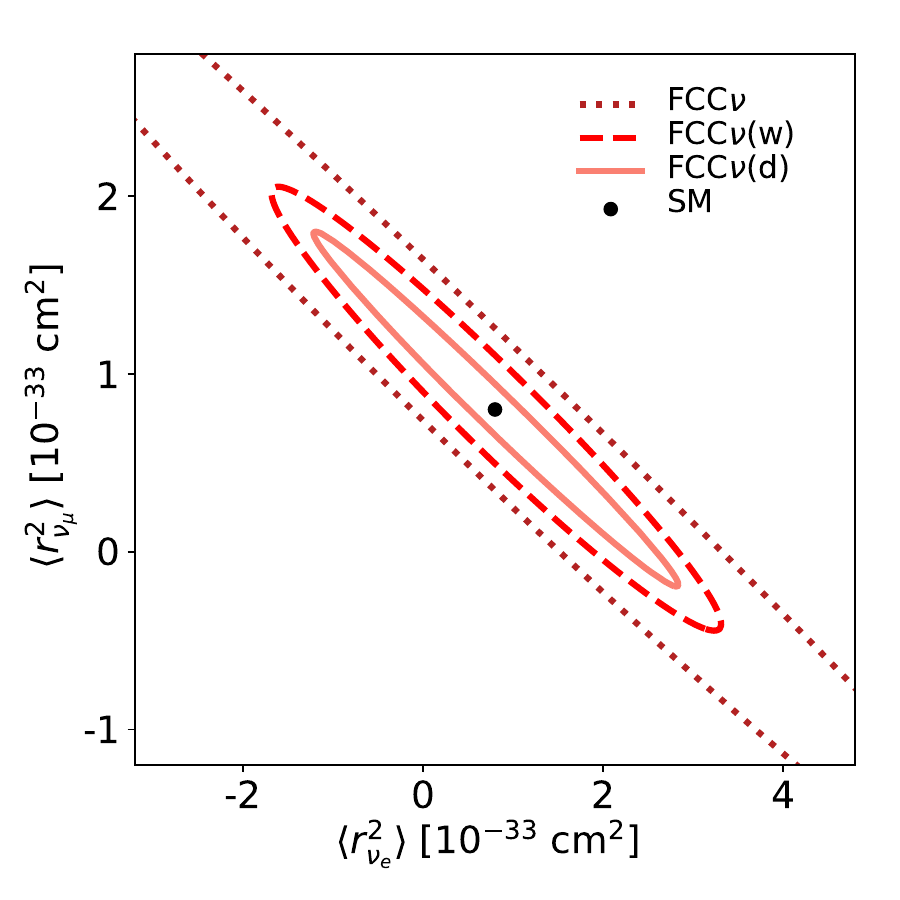}   
\caption{Left: The projected sensitivity to the neutrino charge radius at the far-forward neutrino detectors of \Cref{tab:neutrino_detectors} at 90\% CL. 
The SM predictions are shown in the green hatched regions; they depend on the momentum transfer $Q^2$ and the band shows predicted values ranging from $Q^2$ = 0  to $Q^2$ = 1000 GeV$^2$, the typical values probed at the FPF@FCC (see Fig.~\ref{fig:kinplane_FASERdeep}). 
Also shown are projections from FASER$\nu$2 at the HL-LHC~\cite{MammenAbraham:2023psg} and from DUNE~\cite{Mathur:2021trm} along with other experimental constraints.
Right: The projected sensitivity to the neutrino charge radius (with the SM value as black dot) for $\nu_e$ and $\nu_{\mu}$ at the FPF@FCC considering only statistical uncertainties. 
A potential degeneracy along the diagonal with slope $m=-1$ is broken by the different energy spectra of electron and muon neutrinos.}
\label{fig:NCR}
\end{figure}

The effective interactions between neutrinos and photons can be expressed as $\bar{\nu}\Lambda^\alpha \nu A_\alpha$, where $\Lambda^\alpha$ is the effective electromagnetic current of the neutrino.\footnote{In general, $\Lambda^\alpha$ is a 3 $\times$ 3 matrix in flavour space. Here, we assume it to be a diagonal matrix.} 
In the ultra-relativistic limit relevant for high-energy collider neutrinos, this current can be written as~\cite{Kayser:1982br, Nieves:1981zt} 
\begin{equation}
\label{VrtxSimp}
\Lambda^{\alpha}(Q) = \gamma^{\alpha} \left(q_{\nu}-\frac{Q^2}{6}\left<r^2_\nu\right> \right)-i\sigma^{\alpha\beta}Q_\beta \mu_\nu ~, 
\end{equation} 
where $Q$ is the momentum transfer.
The first term gives the neutrino an electric charge $q_{\nu}$, and the last term gives it a magnetic moment $\mu_\nu$. 
While both $q_{\nu}$ and $\mu_{\nu}$ are predicted to vanish in the SM (for massless neutrinos), the observation of non-zero neutrino masses implies a small magnetic moment $\mu_\nu \sim 10^{-19} \mu_B \, (m_\nu/\text{eV})$, for Dirac neutrinos.
Such tiny values of the magnetic moment are, however, far beyond the scope of ongoing and future experiments. 

The situation is different for the neutrino charge radius $\left<r^2_\nu\right>$, which also enters Eq.~(\ref{VrtxSimp}).
This charge radius receives a non-zero contribution from a SM loop diagram containing the $W$ boson~\cite{Bernabeu:2000hf, Bernabeu:2002pd}. 
The SM contribution can be computed in the limit $Q^2\rightarrow 0$ to be
\begin{equation}
\label{eq:nu_charge_radii}
\left<r^2_{\nu_{\ell}}\right>_{Q^2\rightarrow 0}= \frac{G_F}{4 \sqrt{2}\, \pi^2}\left[3-2 \log\frac{m_{\ell}^2}{m_W^2}\right] \ ,
\end{equation}
where $m_\ell$ is the corresponding charged lepton mass, and $m_W$ is the $W$ boson mass. 
Eq.~(\ref{eq:nu_charge_radii}) evaluates to $4.1 \times 10^{-33}~{\rm cm}^2$ for $\nu_e$, $2.4\times 10^{-33}~{\rm cm}^2$ for $\nu_\mu$, and $1.5 \times 10^{-33}~{\rm cm}^2$ for $\nu_\tau$.
However, the neutrino charge radius depends on momentum transfer~\cite{AtzoriCorona:2024rtv}, and, especially at large values of $Q^2$, it is essential to take this into consideration. As can be seen from Fig.~\ref{fig:kinplane_FASERdeep}, for neutrino DIS interactions at the FCC, the typical momentum transfer is $Q^2 \sim 10^3$~GeV$^2$. For such momentum transfers, $Q^2\gg m_{\ell}^2$, the effects of $m_{\ell}$ are negligible, and $\left<r^2_{\nu_{\ell}}\right>$ becomes approximately $8\times 10^{-34}\text{~cm}^2$ for all three neutrino flavours.  The ranges of SM values of $\left<r^2_{\nu_{\ell}}\right>$ are shown in Fig.~\ref{fig:NCR}, with the band covering the predictions from $Q^2=0$ (rightmost edge) to $Q^2=10^3$ GeV$^2$ (leftmost edge).

Proposed experiments are sensitive to values of $\left<r^2_{\nu_{\ell}}\right>$ that are a factor of a few away from the SM value for $\nu_\mu$,
with FASER$\nu$2 sensitive to values of $\langle r^2_{\nu_{\mu}}\rangle \sim 3\times 10^{-33}\text{~cm}^2$ 
and DUNE expected to reach the SM value. 
These considerations motivate the neutrino charge radius as an important target for future neutrino experiments, especially in the case of $\nu_\tau$, where current constraints are very poor. 

The huge neutrino event rates expected at the FPF@FCC detectors evaluated in Sect.~\ref{sec:fluxes} suggest that these experiments should offer excellent sensitivity to the  neutrino charge radius.  
The effect of a non-zero neutrino charge radius can be accounted for by shifting the vector coupling $g_V^{q}$ entering the neutrino neutral-current (NC) DIS cross section~\cite{Vogel:1989iv}.
The shift in $g_V^{q}$ is
\begin{equation}
g^{q}_{V}\rightarrow g^{q}_{V} - \frac{2}{3}q_q m^2_W\langle r_{\nu_{\ell}}^2\rangle \sin^2 \theta_W \ ,
\end{equation}
where $\theta_W$ is the weak mixing angle, and $q_q$ is the electric charge of the quark. 
This shift induces a change in the inclusive NC DIS cross section for neutrinos, which is approximately given by
\be
\label{eq:nu_charge_radii_XS_scaling}
\sigma_{\rm NC} \to  \sigma_{\rm NC} \times \left[1 - \frac{\langle r_\nu^2\rangle }  { 4\cdot 10^{-31} \text{cm}^2 } + \left(\frac{\langle r_\nu^2\rangle }  { 5\cdot 10^{-31} \text{cm}^2 }\right)^2 \right]\, .
\ee
An excess or deficit of NC DIS events observed in neutrino detectors~\cite{Ismail:2020yqc, MammenAbraham:2023psg} can then be used to constrain and eventually measure the neutrino charge radius. 

Observing a possible difference in the NC DIS event rates due to the neutrino charge radius requires a precise prediction for the SM event rates in the baseline case of $\langle r_\nu^2\rangle=0$.  
Since there are {\em a priori} no sufficiently precise estimates of the neutrino flux that would allow a percent-level prediction of the NC DIS event rate, the neutrino flux has to be constrained directly from the experimental data.  
This can be done using the CC DIS event rate measurements. 
Conceptually, the measurement of the neutrino charge radius therefore corresponds to a search for deviations of the NC DIS event rate compared to the prediction obtained from the CC DIS event rate.  To estimate the sensitivity to $\langle r_\nu^2\rangle$ at the FCC$\nu$ detectors, then, we simulate the expected CC and NC DIS distributions and then perform a $\chi^2$ template fit to obtain the charge radius and the associated uncertainty. 
In this fit, following the reasoning of Sect.~\ref{sec:SM}, we include the statistical uncertainties on the CC and NC DIS event rates, but not systematic errors. 

The projected charge radius sensitivities are shown in the left panel of Fig.~\ref{fig:NCR} for the three detector layouts considered in Table~\ref{tab:neutrino_detectors}, and for all three neutrino flavours. 
For an integrated luminosity of 30~ab$^{-1}$ and $\sqrt{s}\sim 100$~TeV, the FCC$\nu$ detector can probe values down to $\langle r_{\nu_{e}}^2 \rangle \sim 7\times 10^{-34}~{\rm cm}^2,~\langle r_{\nu_{\mu}}^2 \rangle \sim 3\times 10^{-34}~{\rm cm}^2$, and $\langle r_{\nu_{\tau}}^2 \rangle \sim 1\times 10^{-32}~{\rm cm}^2$ at 90\% CL. 
For all three flavours, the FPF@FCC can provide world-leading bounds. For comparison, we also show projections from FASER$\nu$2 at HL-LHC~\cite{MammenAbraham:2023psg} and DUNE~\cite{Mathur:2021trm}, along with existing constraints from COHERENT~\cite{AtzoriCorona:2022qrf, Khan:2022akj}, CHARM-II~\cite{CHARM-II:1994aeb}, LSND~\cite{LSND:2001akn}, CCFR~\cite{CCFR:1997zzq, Hirsch:2002uv}, LEP2~\cite{Hirsch:2002uv}, and TEXONO~\cite{TEXONO:2009knm}. 
Notably, the detectors at FPF@FCC are sensitive to the SM predictions for the $\nu_e$ and $\nu_\mu$ charge radius, and they would be sensitive to within a factor of 5 of the SM value for $\nu_\tau$.

In the right panel of Fig.~\ref{fig:NCR} we show the sensitivity to the SM predictions of the neutrino charge radius in the
($\langle r_{\nu_{e}}^2 \rangle, \langle r_{\nu_{\mu}}^2 \rangle)$ plane for the different FCC neutrino detectors.
We assume $Q^2 = 1000$~GeV$^2$, the typical value at the FCC$\nu$ detectors.
We note that there is a well-constrained direction, in which $\langle r_{\nu_{e}}^2 \rangle$ and $\langle r_{\nu_{\mu}}^2 \rangle$ have the same sign, and a less well-constrained direction, in which the charge radii have opposite signs and their effects on the NC event rate partially cancel. This degeneracy is, however, broken by the different energy spectra of electron and muon neutrinos. 
In the non-degenerate direction, we see that FCC$\nu$ will be able to measure both $\langle r_{\nu_{e}}^2 \rangle$ and $\langle r_{\nu_{\mu}}^2 \rangle$ with $\mathcal{O}(50\%)$ precision and FCC$\nu$(d) with $\mathcal{O}(10\%)$ precision. 

These measurements of the neutrino charge radius require high precision and therefore also extremely good control of all contributing sources of experimental and theoretical systematic uncertainties. 
In particular, it requires a precise modelling of the CC and NC DIS neutrino interaction cross sections, including possible nuclear effects; the ability to measure the flux, separately, for neutrinos and anti-neutrinos of all three flavours; reliable identifications of NC and CC events, even when the final state leptons are relatively soft; good energy resolution; and vanishing backgrounds, for example, from neutral hadrons. 
The ability to measure the neutrino charge radius therefore imposes a variety of detector performance requirements that will guide the ultimate detector design.
In this context, the modelling of the CC and NC neutrino DIS interactions will also benefit from recent progress in higher-order QCD calculations matched to event generators for exclusive event simulation; see  e.g. Refs.~\cite{FerrarioRavasio:2024kem,Buonocore:2024pdv,vanBeekveld:2024ziz} and references therein.

\subsection{Dark Higgs boson and relaxion-type models} 

Complementing the scattering signatures, the FPF@FCC experiments would also be sensitive to the decays of LLPs. 
LLPs may relate to outstanding problems in cosmology and particle theory, and could address some of the persisting experimental anomalies; see Ref.~\cite{Beacham:2019nyx} for a review. 
There exist only a few types of renormalizable portals, which therefore provide a select few motivated targets for experimental searches~\cite{Patt:2006fw,Batell:2009di}.
In particular, the relevant coupling between the SM and the BSM hidden sector could be via a new scalar field that mixes with the SM Higgs boson $h$. 
Indeed, better measurements of SM Higgs properties and searches for extended scalar sectors are among the primary BSM targets of the FCC~\cite{FCC:2018byv}.
The far-forward physics program at the FCC-hh will contribute to these efforts by probing new scalars with very weak couplings and with masses up to tens of GeV.

Here we evaluate the expected sensitivity of the FPF@FCC in the search for the dark Higgs boson $\phi$ coupled to the SM  via the following Lagrangian obtained after electroweak symmetry breaking,
\begin{equation}
\mathcal{L} = -m_\phi^2\phi^2 - \sin{\theta}\frac{m_f}{v}\phi \bar{f}f - \lambda v h \phi\phi \ ,
\end{equation}
where $f$ denotes SM fermions, $\sin{\theta}$ is the $\phi$-$h$ mixing angle between the SM Higgs and its dark sector counterpart, and $\lambda$ is an additional trilinear coupling between the two scalar fields. 
We assume that other couplings between $\phi$ and $h$, which could arise in the most general case, are suppressed by invoking additional symmetries. 
The considered dark Higgs scenario is then among the most popular benchmark models discussed in the literature~\cite{Beacham:2019nyx}. 

The trilinear coupling $\lambda$ implies $\phi$ production via off-shell and on-shell SM Higgs boson decays in the forward kinematic region of the collider~\cite{Feng:2017vli,Boiarska:2019vid}.
The characteristic transverse momentum of SM Higgs bosons produced in primary collisions is of the order of $p_T\sim m_h\ll p_h$, where $p_h$ is the total momentum of $h$.
In the following, we present the projected exclusion bounds in the dark Higgs search for the FCC-LLP1 and FCC-LLP2 detectors discussed in Sect.~\ref{subsec:detectors_bsm}. 
For illustration, the former corresponds to less than $0.0002\%$ of the total forward hemisphere due to its considerable distance from the IP. Still, we find that $\mathcal{O}(2.6\%)$ of the SM Higgs bosons will be forward-boosted towards FCC-LLP1, and this number rises to almost $50\%$ for Higgs bosons with energies above $10~\textrm{TeV}$. Rare decays of the SM Higgs bosons into a pair of dark Higgs bosons, $h\to \phi\phi$, therefore often produce a collimated flux of $\phi$'s that can subsequently decay in forward detectors. 
For $m_\phi\gtrsim \textrm{several GeV}$, the dominant $\phi$ decay modes are into charm and beauty quarks that hadronize into various final states~\cite{Ferber:2023iso}.
The decay length of the boosted dark Higgs boson becomes sufficiently large for small values of the mixing angle $\sin{\theta}$ such that they can reach a distant forward detector before decaying.

\begin{figure}[t]
\includegraphics[width=0.505\textwidth]{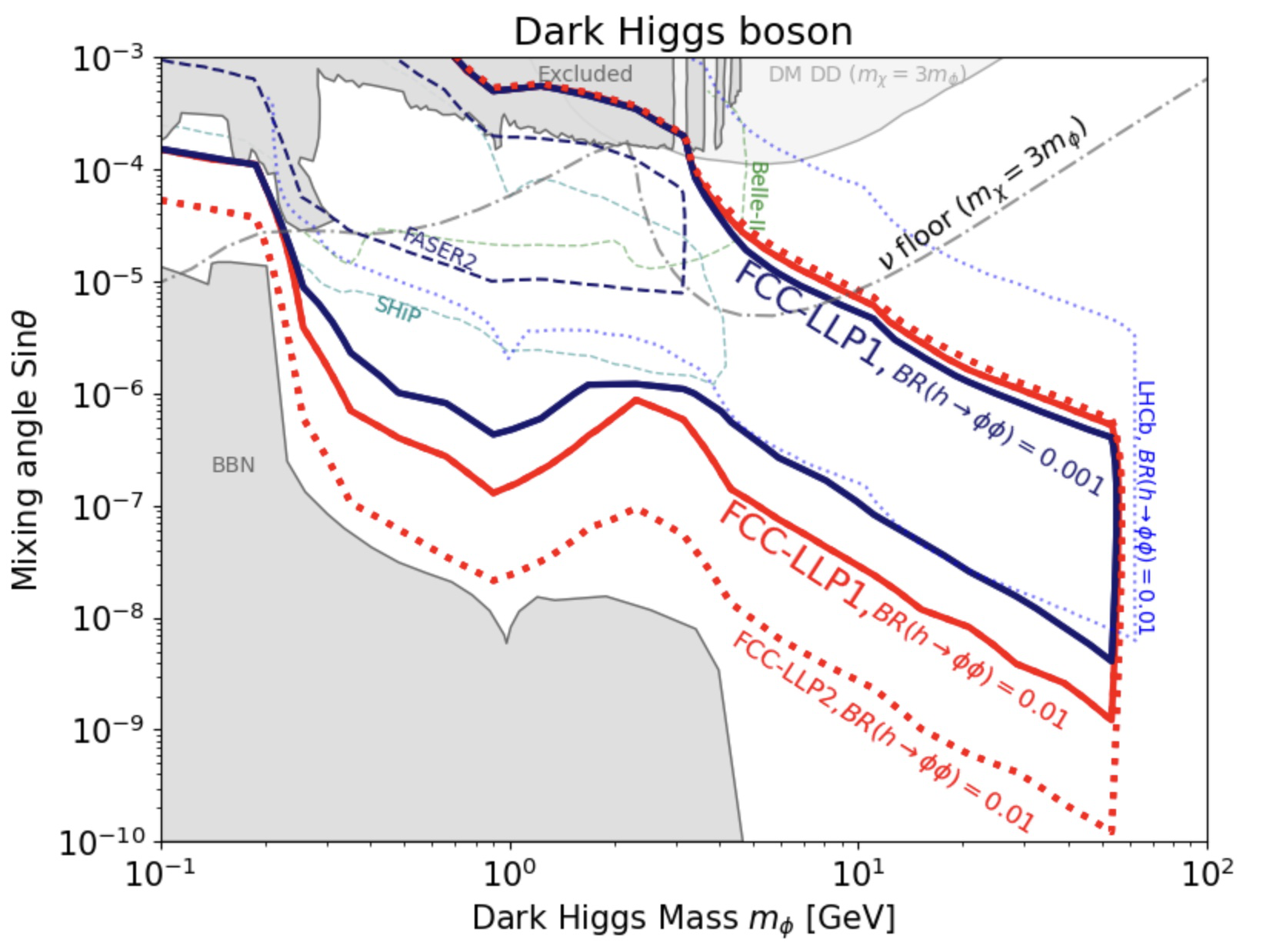}
\includegraphics[width=0.5\textwidth]{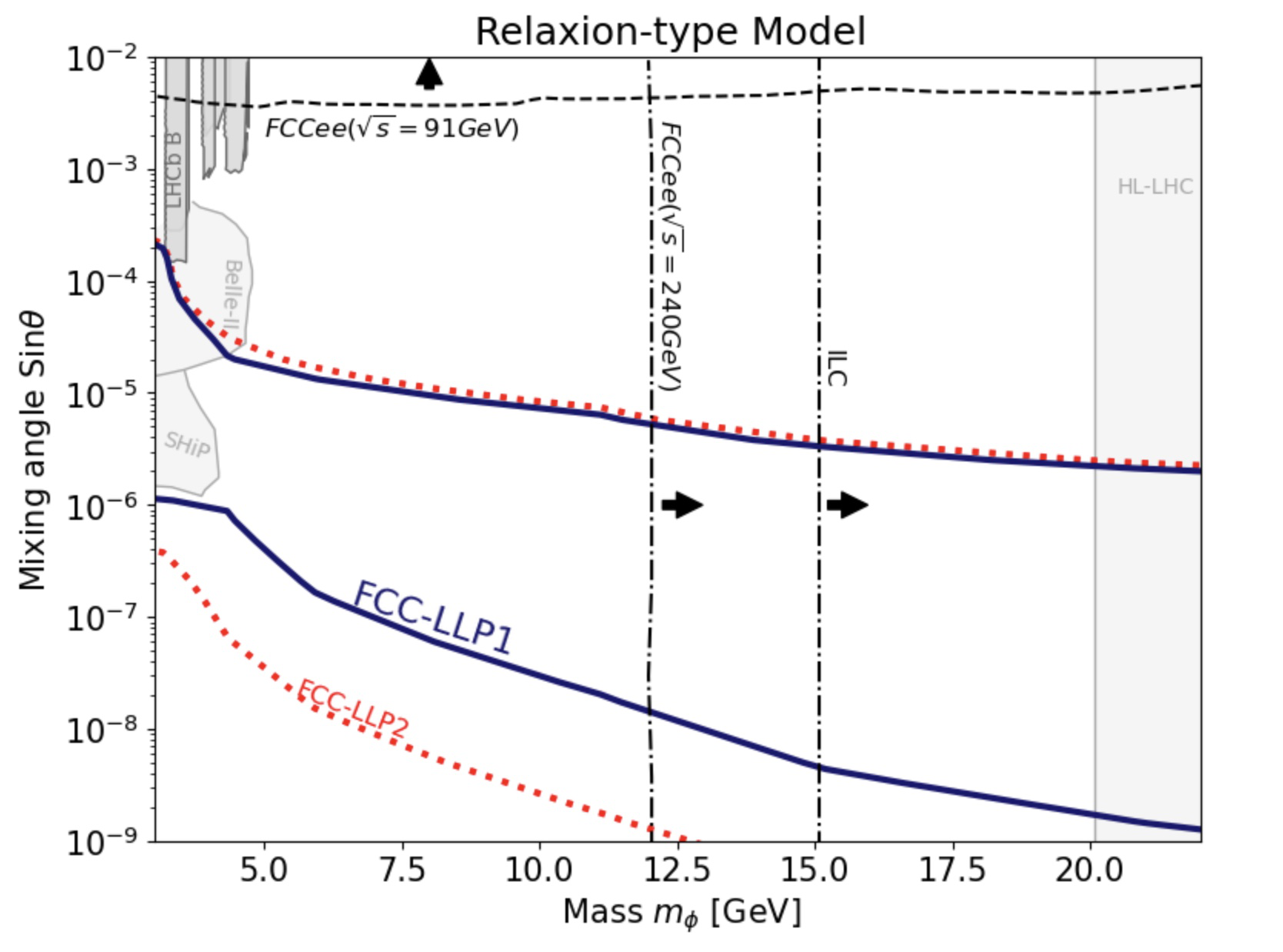}
\caption{Left:
The sensitivity reach for a dark Higgs boson $\phi$ at the FCC-LLP1 detector is shown with blue and red solid lines for two different branching fractions, $\textrm{BR}(h\to\phi\phi) =0.1\%$ and 1\%, respectively. 
We also show the FCC-LLP2 sensitivity with a dotted red line. 
The dark gray shaded region includes bounds from current accelerator and collider experiments; see Ref.~\cite{Ferber:2023iso} for a review. 
The parameter space is constrained by the BBN bound for lower values of the mixing angle $\theta$~\cite{Fradette:2017sdd,Fradette:2018hhl}.
The light gray shaded region shows current dark matter direct detection bounds and the dot-dashed contour is the neutrino floor, assuming $m_{\chi} = 3\,m_{\phi}$, where $m_{\chi}$ is the mass of the dark matter particle that interacts with SM particles via the $\phi$ portal. 
Projected exclusion bounds from Belle-II~\cite{Belle-II:2018jsg}, FASER2 at the FPF@LHC~\cite{Feng:2022inv}, SHiP~\cite{Ahdida:2023okr}, and LHCb for a luminosity of $300\,\textrm{fb}^{-1}$ and $\textrm{BR}(h\to\phi\phi) = 1\%$ are also shown.
Right: The sensitivity reach for relaxion-type models at FCC-LLP1 and FCC-LLP2, with the dark gray shaded region depicting the current LHCb bound. 
Projected exclusion bounds from Belle-II, SHiP, and from the search for invisible Higgs decays at HL-LHC are given by light-shaded regions. Expected constraints from FCC-ee, ILC, and TeraZ (denoted as FCC-ee at $\sqrt{s} = 91~\textrm{GeV}$ in the plot) are shown with arrows pointing towards the region of exclusion~\cite{Frugiuele:2018coc}. 
} 
\label{fig:darkHiggsrelaxion}
\end{figure}

We illustrate the expected sensitivity of searches for dark Higgs bosons at the FPF@FCC in the left panel of Fig.~\ref{fig:darkHiggsrelaxion} in the $(m_\phi,\sin{\theta})$ plane.
We fix the $\lambda$ coupling such that the invisible SM Higgs branching fraction is equal to  either $\textrm{BR}(h\to\phi\phi) = 0.1\%$ or $1\%$. 
The former value is beyond both the expected sensitivities of the HL-LHC, which is approximately $2.5\%$~\cite{Cepeda:2019klc}, and the most promising FCC-ee predictions, which are at the level of $0.2\%$~\cite{deBlas:2019rxi}.   A proposed search at the Muon Collider may probe invisible branching fractions as low as $0.1\%$, assuming perfect muon direction identification~\cite{Ruhdorfer:2023uea}.

Here we assume that muon-induced backgrounds can be reduced to negligible levels. 
The search for LLP decays might suffer from substantial neutrino-induced backgrounds if the decay volume is filled with air. 
We estimate the neutrino-air interaction rates to be about $2.5\textrm{M}$ and $320\textrm{M}$ for the smaller and larger decay volumes, respectively. This should be considered when designing the experiment, and vacuum decay vessels will be required to reduce these backgrounds to manageable levels; see the design of SHiP~\cite{Ahdida:2704147}.

The estimated reach is dominated by rare kaon and $B$-meson decays, $K\to \pi\phi$ and $B\to X_s\phi$, when the $\phi$ mass is below the relevant kinematic thresholds. For heavier dark Higgs bosons the dominant production mechanism is  the aforementioned on-shell decays of the SM Higgs boson. 
We obtain the limits using {\sc\small FORESEE}~\cite{Kling:2021fwx}. We employ the kaon production spectrum obtained with EPOS-LHC~\cite{Pierog:2013ria} implemented in the \texttt{CRMC} package~\cite{CRMC}, and the $B$-meson and SM Higgs $h$ spectra from {\sc\small Pythia8}~\cite{Sjostrand:2006za,Sjostrand:2014zea}.  See also Ref.~\cite{Buonocore:2023kna} for updated predictions for heavy meson production in the forward region.

We also show in Fig.~\ref{fig:darkHiggsrelaxion} other expected exclusion bounds obtained from searches at Belle-II~\cite{Belle-II:2018jsg}, FASER2~\cite{Feng:2022inv}, and SHiP~\cite{Ahdida:2023okr}.
These searches could constrain dark Higgs bosons produced in rare meson decays down to mixing angles $\sin{\theta}\sim 10^{-6}$. 
As can be seen, the proposed FPF@FCC detectors could improve this reach by another two orders of magnitude. For the larger decay volume and $\textrm{BR}(h\to\phi\phi)=1\%$, the projected exclusion bounds are close to the cosmological Big Bang Nucleosynthesis (BBN) bound~\cite{Fradette:2017sdd,Fradette:2018hhl}. 
Thanks to on-shell SM Higgs boson decays, the FPF@FCC limits extend to much larger masses, up to the kinematic limit $m_\phi\lesssim m_h/2$. 
We also show the expected sensitivity of the LHCb detector in this mass regime, assuming that it operates in the HL-LHC era with upgraded track reconstruction algorithms~\cite{Gorkavenko:2023nbk} and $\textrm{BR}(h\to\phi\phi)=1\%$.
The region of the parameter space of this model with larger values of $\sin{\theta}$, i.e., above the expected FPF@FCC exclusion bounds, can additionally be constrained by future FCC central detectors in their search for displaced vertices.  See, for example, Refs.~\cite{ATLAS:2019qrr,ATLAS:2019jcm,ATLAS:2018tup,ATLAS:2021jig,CMS:2024bvl} for current such searches in the ATLAS and CMS detectors at the LHC. 

It is also interesting to consider the cosmological implications of the dark Higgs boson search at the FPF@FCC. 
The dark Higgs boson could mediate interactions between the SM and DM sectors. We illustrate this in the plot for the coupling given by $\mathcal{L} \supset -(1/2)\,\kappa\,\phi\,\bar{\chi}\chi$,
with $\chi$ being the DM particle.
Provided that $\chi$ is heavier than the dark Higgs boson, and we assume specifically $m_\chi = 3m_\phi$, their thermal abundance can be set via the $\bar{\chi}\chi\to\phi\phi$ annihilation process governed by the $\kappa$ coupling, independently of the mixing angle $\sin{\theta}$. 
Fig.~\ref{fig:darkHiggsrelaxion}  shows the current direct detection (DD) bounds on such DM species with a light gray color~\cite{Billard:2021uyg}. Future DD searches will further constrain this scenario. 
However, as shown in the plot, FPF@FCC will remain complementary to these searches and can probe the parameter space beyond the neutrino floor, where DD experiments suffer from significant backgrounds.

Another family of BSM models that contain scalar LLPs that could be probed at far-forward experiments arise in the context of the relaxion solution to the hierarchy problem, which relies on stabilizing the Higgs mass dynamically, instead of using additional symmetry arguments~\cite{Graham:2015cka}. 
The phenomenology of the relaxion scalar field resembles closely the dark Higgs boson with additional couplings to gauge bosons~\cite{Flacke:2016szy}.
Importantly, in this case, the trilinear coupling between the SM Higgs and two $\phi$ fields is not independent, but is instead determined by other parameters of the model. 
In the regime of low mixing angle, it is given by $\lambda \simeq m_\phi^2/v^2$. Therefore, the relaxion-type models predict an increasing invisible SM Higgs branching fraction with the relaxion mass, $\textrm{BR}(h\to\phi\phi)\sim m_\phi^4/v^4$.

This feature results in strong bounds on $m_\phi\sim \textrm{tens of GeV}$ obtained from measuring SM Higgs decays, independent of the mixing angle $\sin{\theta}$. 
The expected bounds from future experiments will improve these upper limits on the relaxion mass. We illustrate this in the right panel of Fig.~\ref{fig:darkHiggsrelaxion}, following Ref.~\cite{Frugiuele:2018coc}. 
The projected bounds from HL-LHC are shown as the light gray shaded region above $20~\textrm{GeV}$ mass. 
We also include the ILC and FCC-ee limits. 
As can be seen, even optimistic assumptions about future colliders will leave a sensitivity gap between their projected bounds and the intensity frontier searches, constraining $m_\phi\lesssim 5~\textrm{GeV}$. 
This gap could only partially be covered at the FCC-ee running at the $Z$-peak (Tera-Z). 
The dedicated search at Tera-Z would probe the relaxion couplings to the SM gauge bosons and exclude $\sin{\theta}\gtrsim \textrm{a few}\times 10^{-3}$. 
Instead, the FPF@FCC experiments will be able to bridge the aforementioned gap for mixing angles in the range $10^{-9}\lesssim \sin\theta\lesssim 10^{-5}$. 
This corresponds to probing the invisible Higgs branching fraction for values as low as $\textrm{BR}(h\to\phi\phi)\sim 10^{-4}$ for $m_\phi \sim \text{few GeV}$.  

\subsection{Quirks} 
\label{subsec:quirks}

A generic possibility for dark sectors is that they contain a non-Abelian gauge force.
Such an interaction would induce confinement of dark sector particles at some scale $\Lambda$, analogous to the scale $\Lambda_{\text{QCD}}$ of the SM strong interactions. 
Quirks $\cal Q$~\cite{Kang:2008ea} are matter particles that are charged under such a hidden gauge group, as well as under a SM gauge group, with the additional condition that $\Lambda$ is much smaller than the mass of the lightest quirk. 
Strongly interacting hidden sectors are motivated by neutral naturalness solutions to the gauge hierarchy problem; for a recent review, see Ref.~\cite{Batell:2022pzc}. 
With this motivation, quirks, if they exist, may naturally be expected to have masses at the TeV scale.  
Such massive quirks cannot be produced in fixed-target or beam-dump experiments and can only be discovered at high-energy colliders.

In particle collisions, quirks and anti-quirks are produced through their SM interactions, but once produced, they do not hadronize with respect to the dark strong force.  Rather, a $\cal Q \bar{\cal Q}$ pair is bound together by a hidden color string with a typical oscillation scale of 
\begin{align}
\ell \sim \frac{m_{\mathcal{Q}}}{\Lambda^2} \sim 1~\text{cm} \ \bigg[ \frac{1~\text{keV}}{\Lambda} \bigg]^2 \ \bigg[\frac{m_{\mathcal{Q}}}{100~\text{GeV}} \bigg] \ .
\label{eq:amplitude}
\end{align} 
For large swaths of the viable $(m_{\cal Q}, \Lambda)$ parameter space, the $\cal Q \bar{\cal Q}$ system is tightly bound, and the bound state has low $p_T$ and so propagates down the beam pipe, making forward detectors well located to discover them~\cite{Li:2021tsy, Li:2023jrt, Feng:2024zgp}. 
Current bounds on quirks are rather weak, given their unusual signature. 
For example, for uncolored quirks and $\Lambda$ above 100~eV, quirk masses as low as 50 GeV are allowed. 
The currently running FASER experiment can extend the sensitivity to quirks to hundreds of GeV, and FASER2 at the HL-LHC can extend this further to $\sim 1$~TeV~\cite{Feng:2024zgp}.  

\begin{figure}[tbp]
	\begin{center}
    \includegraphics[width=0.32\textwidth]{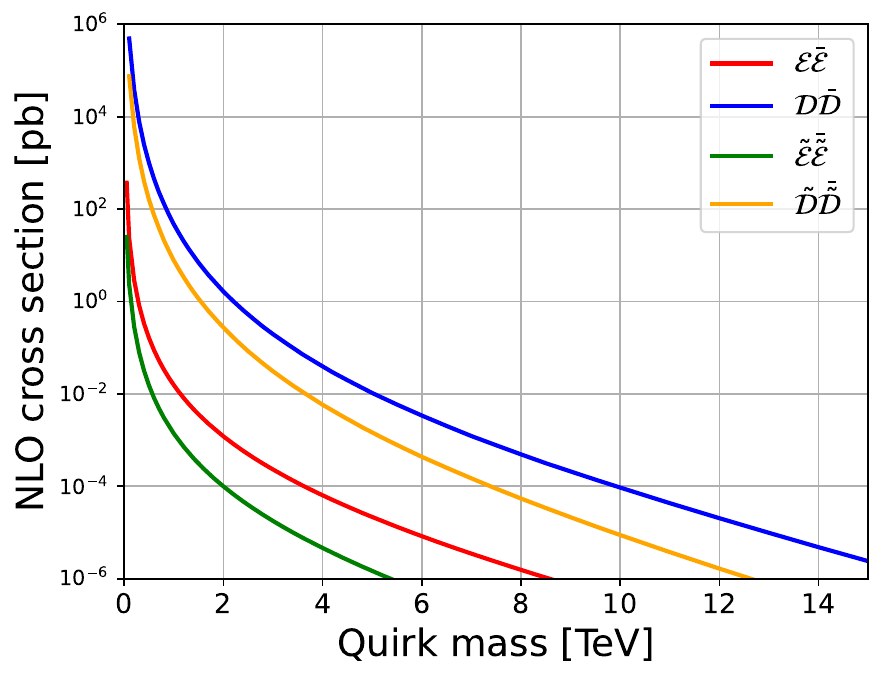}
\includegraphics[width=0.32\textwidth]{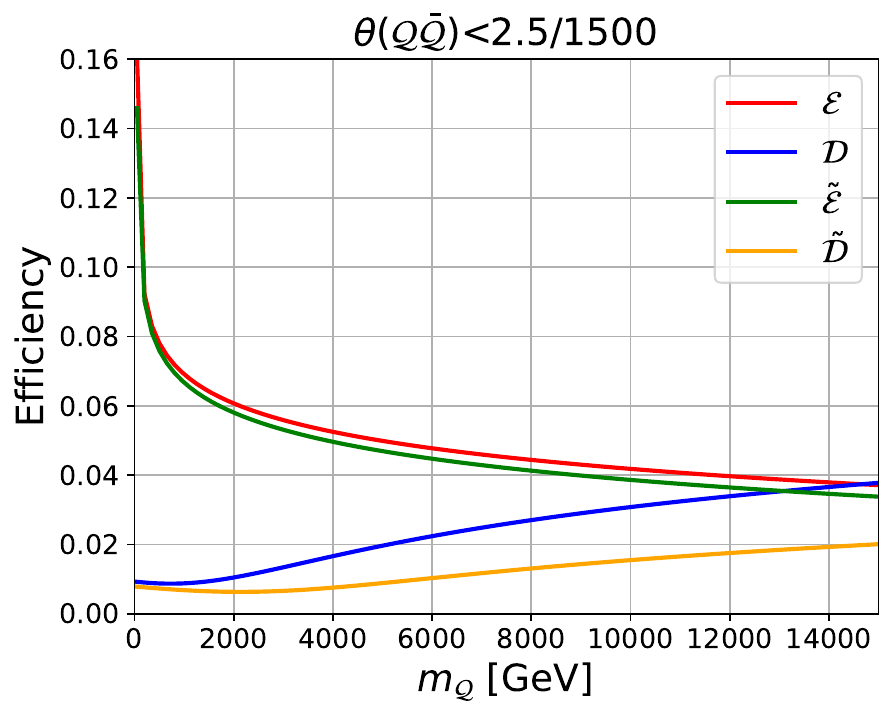}
\includegraphics[width=0.32\textwidth]{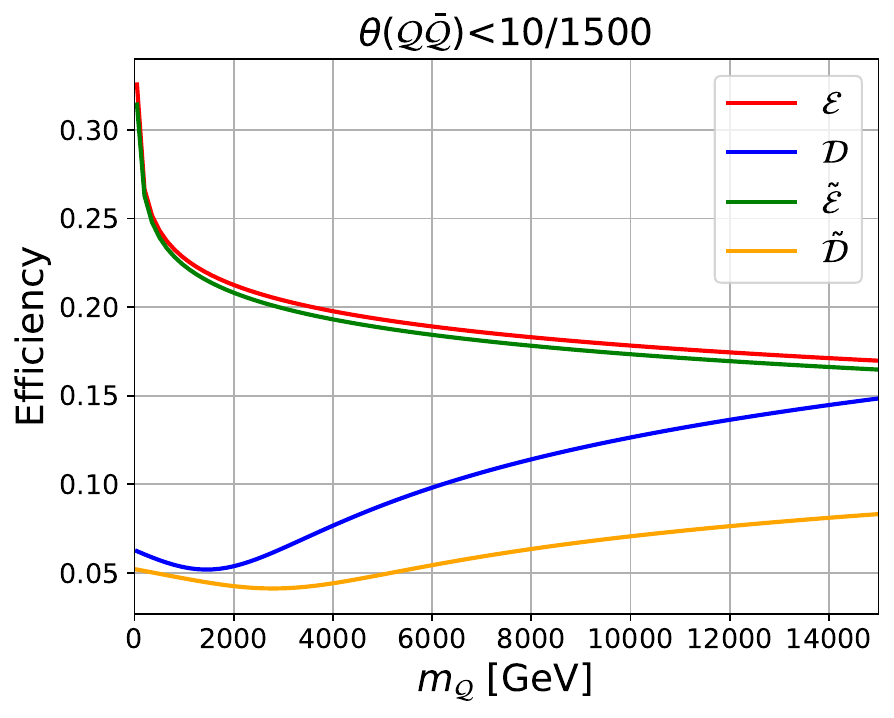}
	\end{center}
 \vspace*{-0.1in}
\caption{Left: NLO cross sections for quirk pair production at the FCC-hh for the quirk models considered. Center and right: The fraction of quirk pair production events produced with $\theta(\mathcal{Q} \bar{\mathcal{Q}}) < 2.5/1500$ (center) and 10/1500 (right), where $\theta(\mathcal{Q} \bar{\mathcal{Q}})$ is the polar angle relative to the beamline of the quirk pair at production.  }
\label{fig::effskins}
\end{figure}

Our goal here is to explore the additional reach in the quirk parameter space accessible at the FPF@FCC experiments.
With this motivation, we examine four simplified quirk models. 
The quirk particles in these models, under the $\text{SU}(N_{\text{IC}}) \times \text{SU}_C(3) \times \text{SU}_L(2) \times \text{U}_Y(1)$ gauge symmetry, are defined as
\begin{align}
    \mathcal{E} &= \left( N_{\text{IC}}, 1, 1, -1 \right) \ , \label{eq::qn4} \\
    \mathcal{D} &= \left( N_{\text{IC}}, 3, 1, -1/3 \right) \ , \label{eq::qn3} \\
    \tilde{\mathcal{E}} &= \left( N_{\text{IC}}, 1, 1, -1 \right) \ , \label{eq::qn2} \\
    \tilde{\mathcal{D}} &= \left( N_{\text{IC}}, 3, 1, -1/3 \right) \ ,
\end{align}
where $\mathcal{E}$ and $\mathcal{D}$ are fermionic quirks, and $\tilde{\mathcal{E}}$ and $\tilde{\mathcal{D}}$ are scalar quirks. The quirk production cross sections are proportional to $N_{\text{IC}}$; for this analysis, we conservatively set $N_{\text{IC}}=2$.  
Each model is thus fully specified by the quirk mass $m_{\mathcal{Q}}$ and the hidden color confinement scale $\Lambda$.
In the left panel of Fig.~\ref{fig::effskins} we show the NLO cross sections for the production of quirk pairs from $pp$ collisions at CoM energy $\sqrt{s} = 100~\text{TeV}$, which are obtained by using \texttt{mg5\_aMC@NLO}~\cite{Alwall:2014hca}. Given their SM charges, the color-neutral quirks $\mathcal{E}$ and $\tilde{\mathcal{E}}$ are mainly produced through the Drell-Yan process, whereas the colored quirks $\mathcal{D}$ and $\tilde{\mathcal{D}}$ are primarily produced via QCD processes involving an $s$-channel gluon.

We will consider the prospects for quirk searches with the FCC-LLP1 and FCC-LLP2 detectors described in Table~\ref{tab:BSMdetectors}. 
As described below, the signal will be based on timing measured in the detectors' scintillators, which are assumed to be located at the front and back faces of these detectors.  
Due to the limited transverse size of these forward detectors, $\thetaq$, the polar angle relative to the beamline of the quirk-pair system's momentum when produced, including the effects of initial state radiation (ISR) and final state radiation (FSR), influences the overall signal efficiency.
In the center and right panels of Fig.~\ref{fig::effskins}, we present the fractions of quirk pair events with $\thetaq < 2.5/1500$ and $\thetaq < 10/1500$ for the four quirk models as functions of quirk mass. 
The condition $\thetaq < 2.5/1500~(10/1500)$ roughly corresponds to the scenario where the quirk system's momentum is directed into the FCC-LLP1 (FCC-LLP2) detector. 
Despite the small solid angle coverage of these two detectors, we see that the fraction of quirk events produced in the direction of these forward detectors is remarkably large, ranging from a few to tens of percent. 
For colored quirks, the fraction of events within the detector's acceptance increases with larger masses because FSR is more dominant than ISR, and FSR deflects the momentum direction less for heavier quirks. In contrast, for color-neutral quirks, only ISR is significant, and the angular cut efficiencies are higher for lighter color-neutral quirks.

To be detected at the FPF@FCC detectors, the quirk--antiquirk system must survive long enough to pass through them. 
At the FCC-hh, quirks are produced in pairs with kinetic energy in their CoM frame.
This kinetic energy can be reduced through the emission of hidden color glueballs, QCD hadrons, and SM photons as the quirks oscillate around their CoM. 
These radiation processes may eventually cause the quirk pair to transition into the quirkonium ground state, potentially leading to the annihilation of the constituent quirks.
The mechanisms of radiative energy loss and quirk annihilation have been previously explored in Refs.~\cite{Evans:2018jmd, Li:2023jrt, Feng:2024zgp}. 
These studies assume that during each period of quirk oscillation, the quirk pair has a probability $\epsilon$ of emitting an infracolor glueball with energy $\Lambda$, and, if the quirks are colored, there is an additional probability $\epsilon^\prime$ of emitting a QCD hadron.

Because quirks are expected to have masses ranging from a few 100s of GeV up to the TeV scale, they are often produced with velocities significantly below the speed of light. 
At the FCC-LLP1(2) detector, the front scintillator can be used to measure the arrival time of particles originating from the IP.
The delayed arrival of quirks at the front scintillator, which is discussed in greater detail in App.~\ref{app:quirk}, can provide compelling evidence for the characteristic quirk signature.
To define a Delayed Track (DT) quirk signal, we apply the same criteria used in Ref.~\cite{Feng:2024zgp}:
\begin{enumerate}
\item Two simultaneous charged tracks are detected that pass through both the front scintillator and the trackers.
\item The signal in the front scintillator is outside the $[-3\,\text{ns}, 3\,\text{ns}]$ muon timing window.
\item The signal in the trackers is two tracks that are separated by more than $16~\mu\text{m}$ in the vertical direction.
\item The momentum of each track is greater than 100 GeV, that is, is consistent with a fairly straight track, as measured by its curvature in the magnetic field.  
\end{enumerate}
The DT signal does not rely on a decay volume or multiple detector components, but does require precise knowledge of bunch crossing times at the IP.

Given that the FCC-LLP1(2) detector can be expected to be equipped with multiple scintillators, it is also possible to search for slow-moving tracks. 
These are tracks that pass through both the front and back scintillators with a time delay greater than what is expected for a particle traveling at the speed of light. 
The impact of such a time delay cut on the quirk signal is discussed in App.~\ref{app:quirk}, where we show the distribution of time differences for quirks at both FCC-LLP1 and FCC-LLP2.
We apply the same criteria used in Ref.~\cite{Feng:2024zgp} for a Slow Track (ST) analysis:
\begin{enumerate}
\item Two simultaneous charged tracks are detected that pass through both the front and back scintillators and the trackers.
\item The time difference of the hits in the front and back scintillators is more than 2~ns greater than what it would be for particles traveling at the speed of light.
\item The signal in the trackers is two tracks that are separated by more than $16~\mu\text{m}$ in the vertical direction.
\item The track momentum is greater than 100 GeV, as measured by its curvature in the magnetic field.   
\end{enumerate}

\begin{figure}[H]
	\begin{center}
		\includegraphics[width=0.45\textwidth]{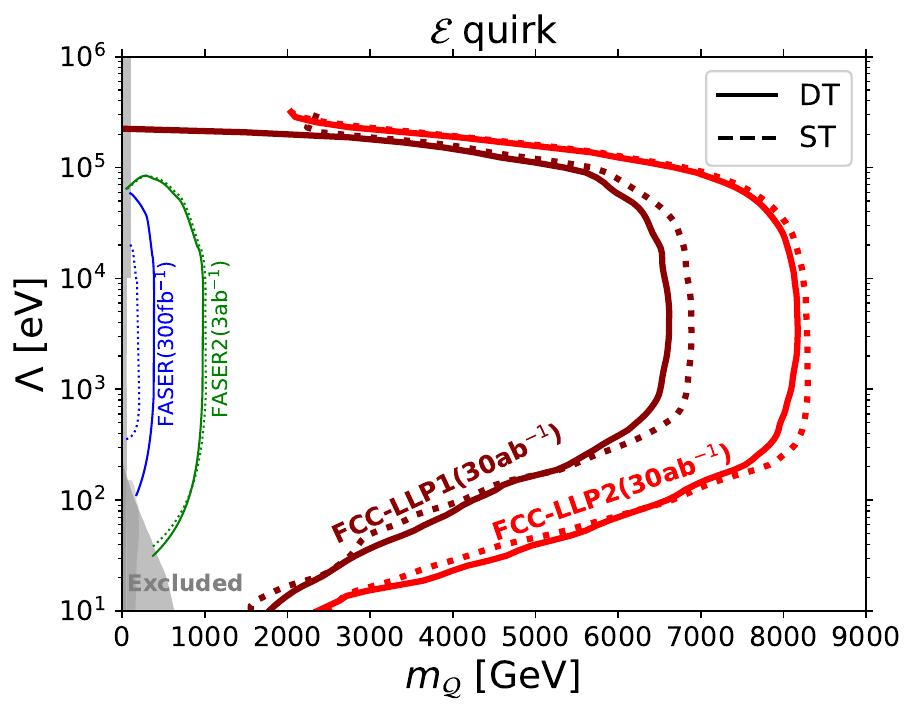}
		\includegraphics[width=0.45\textwidth]{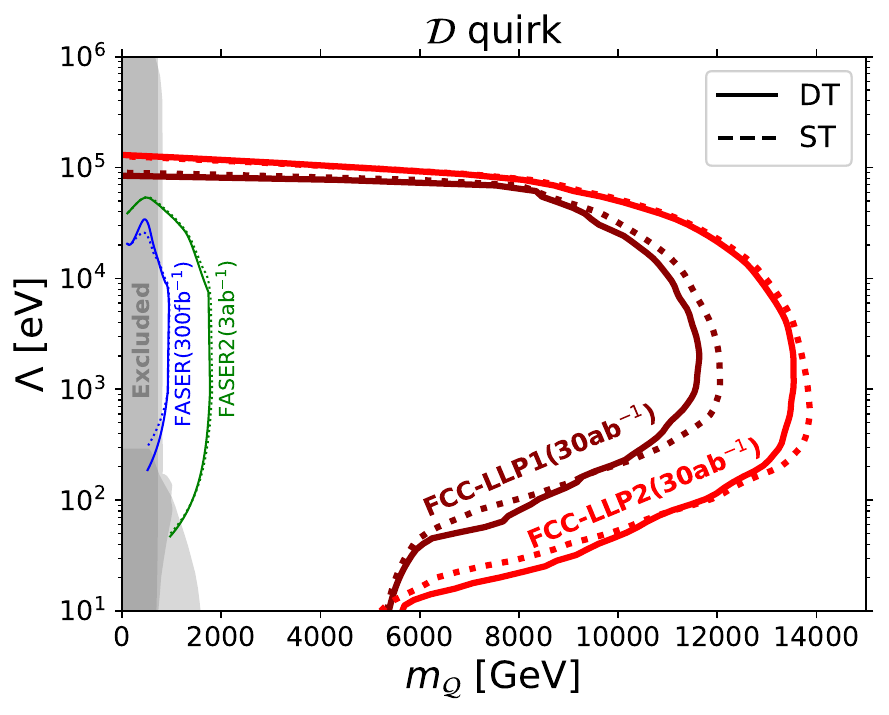}
	\end{center}
 \vspace*{-0.15in}
\caption{The $N=3$ sensitivity contours for FCC-LLP1 and FCC-LLP2 for fermion quirk models ($\mathcal{E}$ and $\mathcal{D}$) and the DT and ST analyses, as indicated. The gray shaded area indicates existing LHC bounds from mono-jet~\cite{Farina:2017cts}, heavy stable charged particle~\cite{Farina:2017cts} and out-of-time decay~\cite{Evans:2018jmd} searches. Projected sensitivities of FASER and FASER2 are given by the blue and green contours, respectively~\cite{Feng:2024zgp}. The energy loss radiation parameters are set to $\epsilon = 0.1$ and $\epsilon' = 0.01$. }
	\label{fig::exclusionquirk1}
\end{figure}

By combining the production cross sections with the selection efficiencies, we can determine the signal rate and discovery potential for each quirk scenario. 
The DT and ST requirements are expected to reduce the background to negligible levels for FASER(2)~\cite{Feng:2024zgp}, and here we assume that this is also the case for the FCC-LPP  detectors and hence three signal events are sufficient to claim a discovery. 
In Fig.~\ref{fig::exclusionquirk1}, we present the $N=3$ signal event contours for the DT and ST analyses of fermion quirk models ($\mathcal{E}$ and $\mathcal{D}$) at FCC-LLP1 and FCC-LLP2, with the energy loss radiation parameters set to $\epsilon = 0.1$ and $\epsilon' = 0.01$.  
For both of the quirk models shown in Fig.~\ref{fig::exclusionquirk1}, for $\Lambda \sim 10-300$~keV, the signal rate is suppressed due to the requirement that the quirk-pair system survives long enough to reach the detector. The lifetime is highly sensitive to the probabilities of infracolor glueball ($\epsilon$) and QCD hadron ($\epsilon^\prime$) radiation. The exact values of $\epsilon$ and $\epsilon^\prime$ are determined by non-perturbative processes and are not known, but larger values of $\epsilon$ and $\epsilon'$ shift the sensitivity contours to smaller values of $\Lambda$.

The FPF@FCC detectors probe multi-TeV quirk masses for $\Lambda \sim 10~{\rm eV} - 100~{\rm keV}$, with the greatest reach for $\Lambda$ in the range of [500~\text{eV}, 30~\text{keV}]. 
For 30~ab$^{-1}$, FCC-LLP1(2) can discover quirks with masses up to 6.9 (8.3) TeV in the $\mathcal{E}$ scenario and 12.1 (13.9) TeV in the $\mathcal{D}$ scenario. The sensitivities for the scalar quirk models ($\tilde{\mathcal{E}}$ and $\tilde{\mathcal{D}}$) are slightly weaker.
All in all, for the the quirk models studied and for a range of four orders of magnitude in $\Lambda$, the reach of the FPF@FCC extends to the multi-TeV region, improving current mass bounds by as much as two orders of magnitude, and effectively covering the range of masses most motivated by neutral naturalness solutions to the gauge hierarchy problem.

\subsection{Millicharged particles}

The FPF@FCC experiments would also be sensitive to ionization or scattering signatures of new feebly-interacting particles. Different detection technologies become relevant for proposed searches depending on the typical visible-energy deposition in such interactions.

The ionization signals can be signatures of new particles with a fractional electric charge. 
These could arise in the presence of massless dark vector bosons manifesting new unbroken gauge symmetries that extend the SM gauge sector. The massless dark photons $A^\prime$ could gain couplings to the SM gauge bosons via the kinetic mixing term $(\epsilon^\prime/2\cos{\theta_W})B^{\mu\nu}X_{\mu\nu}$, where $X^{\mu\nu}$ ($B^{\mu\nu}$) is the dark photon (hypercharge) field strength tensor, and we have included the factor $\cos{\theta_W}$ as a convention~\cite{Holdom:1985ag}. 
If, additionally, a new dark fermion $\chi$ exists, which is coupled to $A^\prime$, a new interaction term between $\chi$ and the hypercharge gauge boson appears after field redefinition to the canonical basis, $(\epsilon^\prime e^\prime/\cos{\theta_W}) \bar{\chi}\gamma^\mu\chi B_\mu$, where we denote the dark coupling constant by $e^\prime$.
After electroweak symmetry breaking, this gives rise to $\chi$ couplings to the SM photon and the $Z$ boson. The former is described by a tiny electromagnetic charge of order $Q = \epsilon e = \epsilon^\prime e^\prime$. 

\begin{figure}[t]
\includegraphics[width=0.50\textwidth]{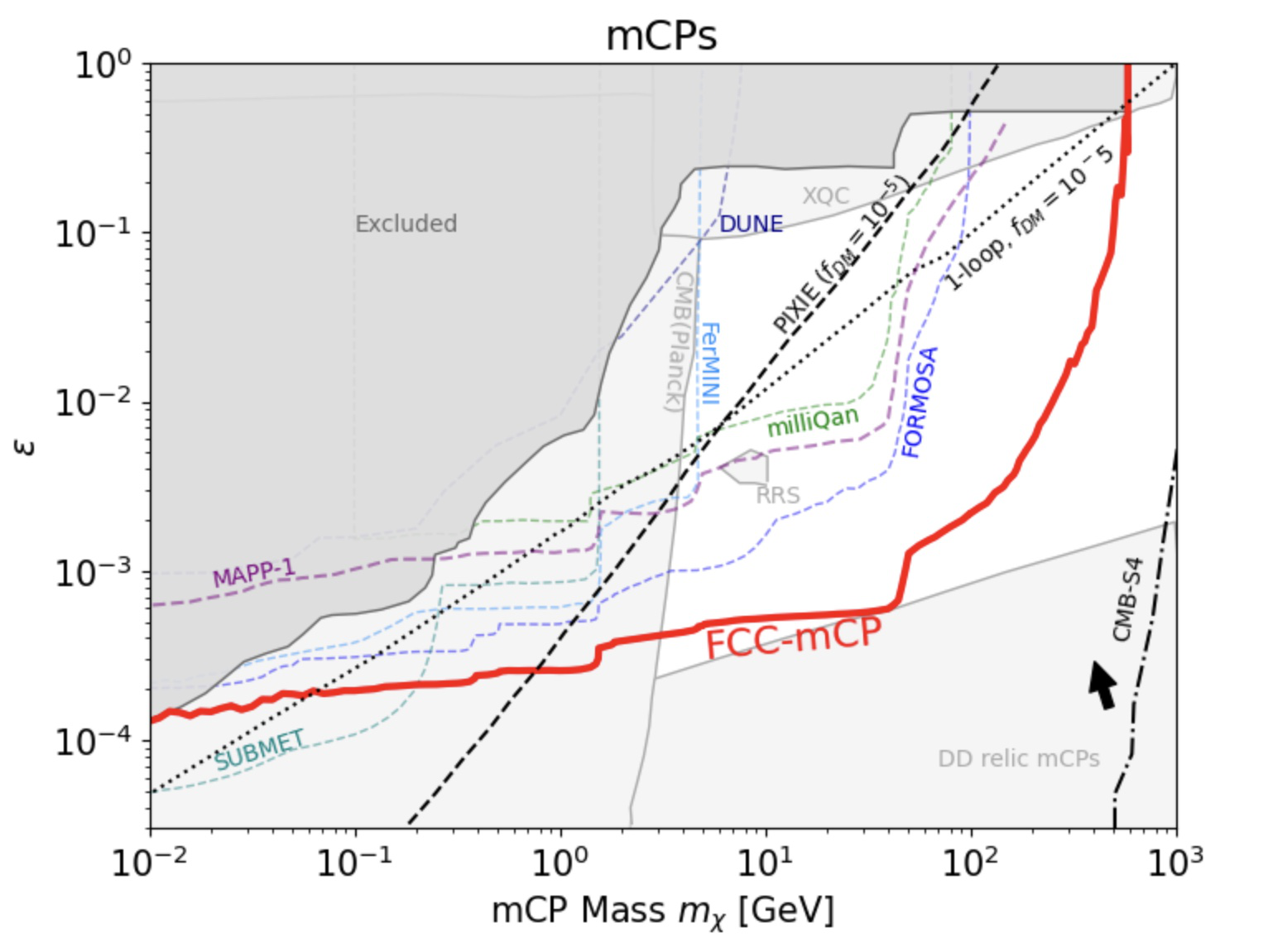}
\includegraphics[width=0.50\textwidth]{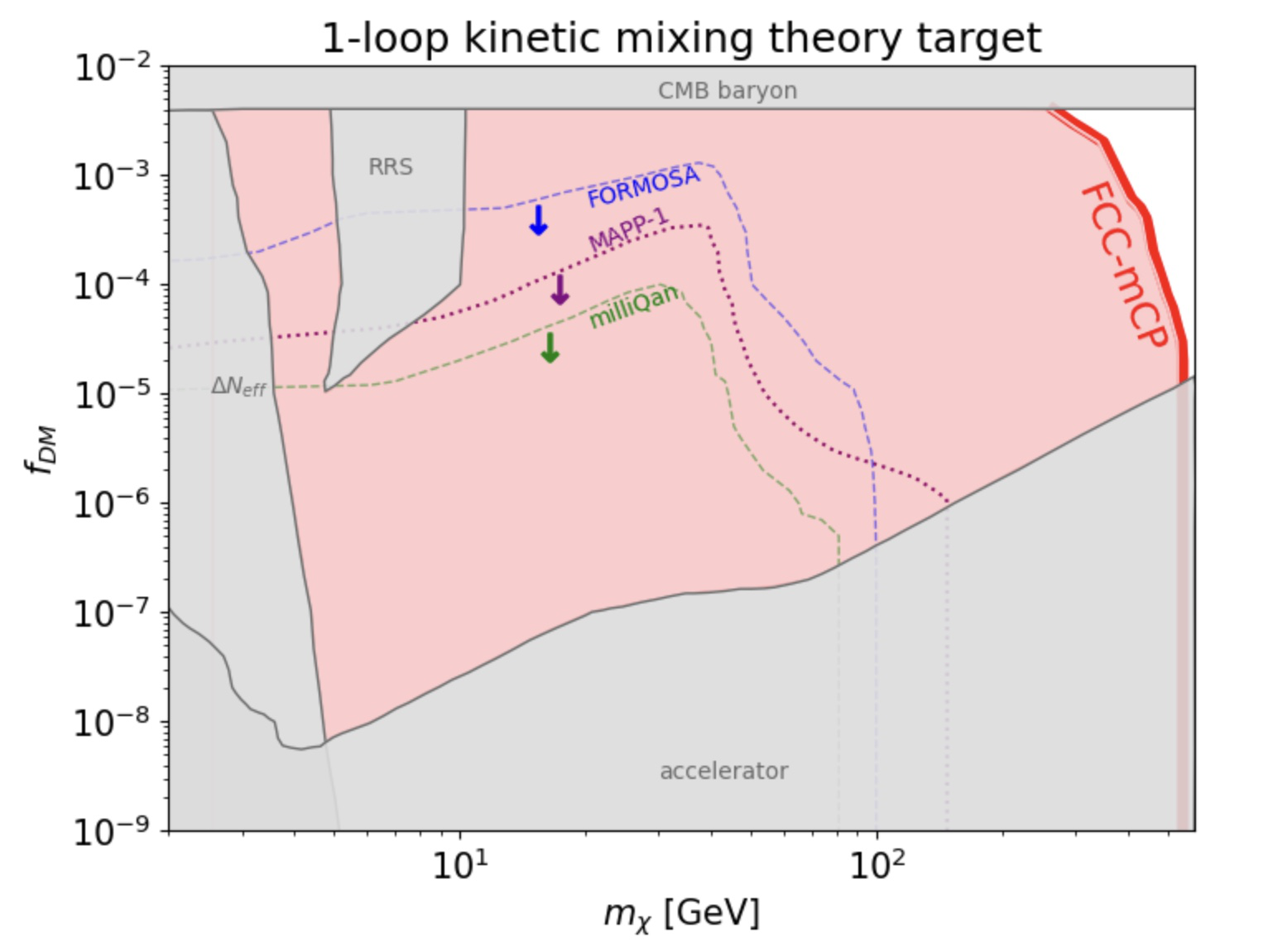}
\caption{Left: The sensitivity reach for mCPs at the FCC-mCP detector. 
The dark gray shaded area depicts the current exclusion bounds, whereas the dashed lines correspond to projected constraints from accelerator-based searches in milliQan~\cite{Haas:2014dda,Ball:2016zrp}, DUNE~\cite{Magill:2018tbb}, FerMINI~\cite{Kelly:2018brz}, FORMOSA~\cite{Foroughi-Abari:2020qar}, MAPP-1~\cite{Mitsou:2023dkg}, SUBMET~\cite{Choi:2020mbk,Kim:2021eix}, and CMB measurements in PIXIE~\cite{Berlin:2022hmt}. Bounds from direct detection experiments, Planck CMB observations, as well as balloon and satellite searches are shown as light gray shaded areas. The latter two assume that mCPs constitute a fraction $f_{\textrm{\rm DM}} = 10^{-5}$ of the total DM abundance.
The projected reach of the upcoming CMB-S4 survey is shown with a dash-dotted black line with the arrow pointing into the exclusion region. A theoretical target line, along which the kinetic mixing parameter is generated at the one-loop level and $f_{\rm DM} = 10^{-5}$, is depicted with a black dotted line. 
Right: The FCC-mCP reach for the theory target with the kinetic mixing parameter generated at the one-loop level and with the thermally-produced fraction of millicharged DM $f_{\rm DM}$ as a function of the mCP mass $m_\chi$. The gray shaded areas indicate the current bounds, while the colored lines show the projected sensitivity regions of other future experiments with arrows pointing toward the excluded region.}
\label{fig:mCP}
\end{figure}

The search for such mCPs in accelerator-based experiments has been the subject of considerable interest. 
We indicate relevant bounds in the left panel of Fig.~\ref{fig:mCP} as a dark gray shaded region, following Refs.~\cite{Prinz:1998ua,Davidson:2000hf,ArgoNeuT:2019ckq,Marocco:2020dqu,milliQan:2021lne,CMS:2022mfm,SENSEI:2023gie}. We show the projected future exclusion bounds for the milliQan detector at the LHC~\cite{Haas:2014dda,Ball:2016zrp} and proposed searches in DUNE~\cite{Magill:2018tbb}, FerMINI~\cite{Kelly:2018brz}, FORMOSA at the FPF@LHC~\cite{Foroughi-Abari:2020qar}, MAPP-1~\cite{Mitsou:2023dkg}, and SUBMET~\cite{Choi:2020mbk,Kim:2021eix}. 
On top of this, constraints from the effective number of relativistic degrees of freedom, $\Delta N_{\textrm{eff}}$, determined by CMB observations, have been shown to exclude mCPs with sub-GeV masses and $\epsilon \gtrsim 10^{-7}$ in the standard cosmological scenario with a high reheating temperature~\cite{Adshead:2022ovo,Gan:2023jbs}. These bounds correspond to the massless dark photon mediator scenario, and we illustrate them with a light gray shaded region labeled as CMB(Planck). We also show prospects for probing this model in the proposed CMB-S4 survey.

As shown in Fig.~\ref{fig:mCP}, the FPF@FCC could further extend the accelerator-based searches up to $m_\chi\sim 500~\textrm{GeV}$.
This provides an important direct probe of this scenario, which could only be independently tested by indirect cosmological constraints. 
The projected exclusion bound has been obtained for the milliQan type detector dubbed FCC-mCP discussed in Sect.~\ref{subsec:detectors_bsm}. 
A detailed discussion of the mCP production modes and the ionization signature at the FPF@LHC can be found in Refs.~\cite{Foroughi-Abari:2020qar,Kling:2022ykt}.
For the purpose of this study, the relevant analysis has been implemented in the \texttt{FORESEE} package~\cite{Kling:2021fwx}. We find that the Drell-Yan process is the dominant production mode of massive mCPs with $m_\chi\gtrsim 5~\textrm{GeV}$.

mCPs produced in the early Universe could naturally remain stable and constitute a fraction of DM. Such a strongly interacting DM component is additionally bounded by direct detection (DD) searches and cosmology. We show the DD bound based on underground experiments as a light gray shaded region at the bottom of the plot~\cite{Emken:2019tni}. These bounds are only mildly sensitive to the precise fraction of mCP DM. They do not extend to larger values of millicharge due to too strong interaction rates of mCPs in the Earth's crust. This suppresses the velocity of mCPs that, therefore, produce too-soft recoils in DD detectors.
However, these limitations can be avoided in %
ballon-borne and satellite experiments. 
We present the corresponding bounds obtained with the XQC and RRS detectors assuming that mCPs constitute a minuscule fraction of DM, $f_{\textrm{DM}} = 10^{-5}$~\cite{Mahdawi:2018euy}. 
Light mCPs coupled to massless dark photons $A^\prime$ could also enhance the $\gamma\to A^\prime$ transition rate in the early Universe and leave imprints on CMB. The relevant projected exclusion bound obtained for the proposed PIXIE mission is presented, following Ref.~\cite{Berlin:2022hmt}.

As can be seen, the FPF@FCC ionization detector could close the gap between the future projected accelerator-based bounds and dark matter DD searches for mCP masses up to several tens of GeV and extend the constraints by another order of magnitude to even larger $m_\chi$.
In Fig.~\ref{fig:mCP} we also show a simple theory target with a black dotted line. In minimal setups, the kinetic mixing arises at the one-loop level through the exchange of heavy particles with mass $M$ that are charged under both the dark $U(1)$ and SM hypercharge groups~\cite{Holdom:1985ag,Cheung:2009qd}; see Ref.~\cite{Gherghetta:2019coi} for further discussion. One then expects that $\epsilon^\prime \simeq ee^\prime/16\pi^2$. The precise value of this parameter is determined in detail by the heavy particle content of the UV-complete model, but it only logarithmically sensitive to the separation between $M$ and the renormalization scale. In the plot, we further assume that $e^\prime$ is fixed by the relic density of mCPs, which is set to be equal to the fraction $f_{\textrm{DM}} = 10^{-5}$ of DM. Here, the mCP relic abundance is driven by secluded annihilations $\chi\bar{\chi}\to A^\prime A^\prime$. Notably, the theory target line obtained in this way lies fully within the reach of the FPF@FCC.

It is interesting to consider scenarios with different fractions of mCP DM component $f_{\textrm{DM}}$. In particular, we assume that $f_{\textrm{DM}}\lesssim 0.004$. Otherwise, stringent bounds on the impact of scatterings between DM and baryons on the CMB spectrum apply, which exclude too-large fractions of mCP DM~\cite{Kovetz:2018zan,dePutter:2018xte}. Instead, lower values of $f_{\textrm{DM}}$ remain \textsl{a priori} unconstrained, although they require increasingly large coupling constants $e^\prime$, if the $\chi$ abundance has thermal origin. This also impacts the one-loop kinetic mixing target lines discussed above. Eventually, for minuscule fractions of mCP DM, one predicts the relevant target lines to lie in the region of the parameter space that is already excluded by accelerator-based or CMB constraints. Therefore, the kinetic mixing generated radiatively at the one-loop level predicts a minimal value of $f_{\textrm{DM}}$. 

We illustrate this sensitivity in the right panel of Fig.~\ref{fig:mCP}, which presents the mCP DM parameter space in the $(m_\chi,f_{\textrm{DM}})$ plane.
In the plot, we assume that mCPs are thermally produced via $\chi\bar{\chi}\to A^\prime A^\prime$ annihilations and that the kinetic mixing is given by $\epsilon^\prime = e e^\prime/16\pi^2$ in the entire plot. As discussed above, $f_{\textrm{DM}}$ is bounded from above by CMB constraints on DM-baryon interactions, and from below and for low $m_\chi$ by accelerator-based searches and $\Delta N_{\textrm{eff}}$ bounds. 
In the plot we also show DD bounds from the RRS experiment, while the XQC experiment and underground detectors do not constrain the region of the parameter space shown in the plot.
As can be seen, the mCP search in the FPF@FCC could cover almost the entire minimal $1$-loop theory target region up to $m_\chi$ of order few hundred GeV. It also extends significantly beyond future experiments proposed to operate during the HL-LHC period.

These projected bounds will be complementary to future CMB surveys that will constrain the kinetic-mixing theory target parameter space by further limiting $\Delta N_{\textrm{eff}}$ and testing possible DM-baryon interactions. Additional strong constraints can be obtained if large accumulations of mCP DM exist inside the Earth. These, however, can be washed out by efficient dark annihilations, $\chi\bar{\chi}\to A^\prime A^\prime$; see, e.g., Refs.~\cite{Pospelov:2020ktu,Budker:2021quh}.
In contrast, the direct search for mCPs at the FPF@FCC is not sensitive to such assumptions.

\section{Summary and outlook}
\label{sec:summary}

In this work we have presented an initial assessment of representative SM and BSM physics  opportunities enabled by far-forward experiments operating concurrently with an FCC-hh at $\sqrt{s}=100$ TeV.
These experiments are expected to accumulate up to $10^{9}$ electron and muon neutrino and up to $10^{7}$ tau neutrino interactions.
To highlight the physics reach of the FPF@FCC for QCD and hadron structure analyses, we have quantified the constraints on the unpolarised and polarised structure of the nucleon provided by high-energy neutrino DIS and determined their sensitivity to nuclear dynamics in the extreme small-$x$ region enabled by the neutrinos produced in proton-lead collisions.
We have also demonstrated how the FPF@FCC would make possible the measurement of the neutrino charge radius for electron and muon neutrinos down to the SM predictions.
Concerning BSM sensitivity, the FPF@FCC would provide competitive constraints on a plethora of compelling BSM scenarios such as dark Higgs bosons, relaxion-type models, quirks, and mCPs.
Indeed, the FPF@FCC detectors could discover  LLPs with masses as large as 50 GeV and couplings as small as $10^{-8}$, as well as quirks with masses up to 10 TeV.

Beyond the studies presented in this work, other topics which may deserve attention to further motivate the FPF@FCC may include testing the universality of the interactions of the three generations of neutrinos at the per mille level and constraining anomalous interactions up to energies of tens of TeV, a direct measurement of $\sin^2\theta_W$ at the highest energies ever probed, detailed multi-dimensional studies of proton and nuclear structure (sensitive to their transverse and spatial distributions) with neutrino beams, and stringent constraints on forward light particle production in hadronic collisions.
Furthermore, while not explicitly considered here, the FPF@FCC shares many connections with astroparticle  physics with its HL-LHC counterpart, from the possibility of improved modelling of extended air showers to the data-driven extraction of the prompt neutrino fluxes.

While it is impossible to predict how the HEP landscape will look like several decades from now, our analysis demonstrates the remarkable potential of far-forward experiments to extend the physics reach of the integrated FCC project (or any other high-energy hadron collider) in a cost-effective manner in several promising directions.
In this work, we have assumed the baseline FCC infrastructure and generic detectors placed in a location not far from the IP. Given the far-future nature of the FCC, no attempt has been made to optimize the accelerator infrastructure or define these detectors more precisely.  However, the results presented here show that the FPF@FCC has the potential to greatly enhance the physics case for the FCC, extending it in directions that are inaccessible at all other planned detectors.  These results therefore motivate more detailed studies of how to fully realize this physics potential.  These detailed studies include theoretical work that can establish the unique potential of the FPF@FCC to explore other topics, as noted above.

Most importantly, our findings provide additional motivation to realize the FPF at the HL-LHC, where many of the studies investigated here can be fully explored for the first time in a real experimental setting.  The experience gained at the FPF at the HL-LHC will inform decisions about the FCC accelerator complex, from its beam optics to the inclusion of muon sweeper magnets, which, with a little planning, have negligible impact on accelerator performance or high-$p_T$ physics, but which will greatly impact the ability of the FPF@FCC to fully realize its potential for both SM studies and BSM searches.  Of course, insights gained from operating experiments in the FPF at the HL-LHC will also be essential for designing optimal detectors for the FPF@FCC.

As the ongoing feasibility study of the integrated FCC project advances, with an eventual decision towards its realisation being taken as soon as 2026, it is of utmost importance that its ultimate physics potential is extensively fingerprinted.
With this motivation, our work further broadens the physics program of the FCC in both its $pp$ and heavy-ion collision modes, demonstrating how a dedicated suite of far-forward experiments would make possible realising novel studies of QCD, neutrino, and BSM physics with unprecedented sensitivity.

\subsection*{Acknowledgements}

We are grateful to Christophe Grojean and Michelangelo Mangano for encouragement to investigate the topics presented in this work.
We also thank Frank Zimmermann for providing useful input on the geometry of the insertion region at the FCC. 
We thank Saeid Foroughi-Abari for sharing the code used in part of the neutrino charge radius analysis.
We thank Stefano Forte for discussions on neutrino DIS.
We are grateful to Pasquale Di Nezza, Norihiro Doshita, and Bakur Parsamyan for information on the COMPASS and LHCspin polarised detectors.
We thank Luca Rotolli for his help to simulate forward charm production at the FCC and for useful discussions.  
We thank Tommaso Giani for providing code for {\sc\small POWHEG} simulations.
We thank Paolo Nason for useful discussions on {\sc\small POWHEG} simulations.
We are grateful to Ilkka Helenius and Leif Lönnblad for helpful discussions of heavy ion collisions in {\sc\small PYTHIA}. 
The work of R.M.A., J.L.F., and M.F.~was supported in part by U.S.~National Science Foundation Grants PHY-2111427 and PHY-2210283. 
The work of J.L.F.~was supported in part by Simons Investigator Award \#376204, Heising-Simons Foundation Grant Nos.~2019-1179 and 2020-1840, and Simons Foundation Grant No.~623683.  
The work of M.F.~was also supported by NSF Graduate Research Fellowship Award No. DGE-1839285.
J.A.~and S.T.~are supported by the National Science Centre, Poland, research grant No. 2021/42/E/ST2/00031. J.A.~is also partially supported from the STER programme - Internationalisation of Doctoral Schools by NAWA. 
The work of F.K.~was supported by the Deutsche Forschungsgemeinschaft under Germany's Excellence Strategy -- EXC 2121 Quantum Universe -- 390833306.
The work of J.L.~and J.P.~was supported by the Natural Science Foundation of Sichuan Province under grant No.~2023NSFSC1329 and the National Natural Science Foundation of China under grant Nos.~11905149 and 12247119.  
The work of J.R.~and T.R.~is partially supported by NWO, the Dutch Research Council, and by the Netherlands eScience Center (NLeSC).

\appendix
\section{Neutrino polarised structure functions}
\label{app:SFs}

In this appendix, we provide, for completeness, the LO expressions of the unpolarised and polarised neutrino structure functions on a proton target, as well as the resulting expressions for the polarised asymmetries of Eq.~(\ref{eq:asymmetry_polarised_cross-sections}) used to estimate the statistical precision of a measurement of polarised differential cross sections at the FPF@FCC in Sect.~\ref{sec:protonspin}.

For  unpolarised neutrino-proton structure functions, assuming that we work in a scheme with $n_f^{\rm (max)}=5$ active flavours and a diagonal CKM matrix, the LO expressions in terms of PDFs are
\bea
F_2^{\nu p}(x,Q^2) &=& 2x \lp \bar{u} + d +s +\bar{c} \rp (x,Q^2) \,, \nonumber\\
F_2^{\bar{\nu} p}(x,Q^2) &=& 2x \lp u + \bar{d} +\bar{s} +c\rp (x,Q^2) \,, \\
xF_3^{\nu p}(x,Q^2) &=& 2x \lp -\bar{u} + d +s -\bar{c} \rp (x,Q^2) \,, \nonumber\\
xF_3^{\bar{\nu} p}(x,Q^2) &=& 2x \lp u - \bar{d} -\bar{s} +c \rp (x,Q^2) \,. \nonumber
\eea
below the top production threshold $Q=m_t$, such that the $b\to t$ contribution is absent.
At LO, $F_2(x,Q^2)=2xF_1(x,Q^2)$, and so there are only two independent structure functions (with $F_L$ becoming non-zero from NLO onwards).
Likewise, the LO expressions for the two independent polarised structure functions $g_1$ and $g_5$ on a proton target~\cite{Forte:2001ph} are given by
\bea
xg_1^{\nu p}(x,Q^2) &=& x\lp \Delta \bar{u}
+ \Delta d + 
 \Delta s + \Delta \bar{c}  \rp (x,Q^2)\, , 
 \nonumber\\
x g_1^{\bar{\nu} p}(x,Q^2) &=& x\lp \Delta u
+ \Delta \bar{d} + \Delta \bar{s} + \Delta c
 \rp (x,Q^2)\, ,
 \label{eq:sf_pdfs}\\ 
x g_5^{\nu p}(x,Q^2) &=& x\lp \Delta \bar{u}
- \Delta d - \Delta s + \Delta \bar{c}
 \rp (x,Q^2)\, , \nonumber
\\
xg_5^{\bar{\nu} p}(x,Q^2) &=& x\lp -\Delta u
+ \Delta \bar{d} + \Delta \bar{s} - \Delta c
  \rp (x,Q^2)\, ,\nonumber 
\nonumber
\eea
where at this order we have $g_4(x,Q^2)=2xg_5(x,Q^2)$ for the same reasons as in the unpolarised case.
Expressions for any other target can be obtained by applying suitable isospin transformations. 
Recall that the structure of the DIS hadronic tensor establishes the correspondence $F_1 \to -g_5$, $F_2 \to -g_4~(=-2xg_5)$, and $F_3\to 2g_1$ between polarised and unpolarised structure functions.

At LO in the strong coupling expansion, in terms of unpolarised and polarised PDFs, the double-differential cross sections are
\bea
\label{eq:unpol_2d_xsec_app}
\nonumber 
\frac{d^2\sigma^{\nu p}(x,y,Q^2)}{dxdy}&=&
\frac{G^2_F}{2\pi (1+Q^2/m_W^2)^2}
\frac{Q^2}{xy}
\lc  \left(y-\frac{y^2}{2}\right) x F_3^{\nu p}
     + \lp 1-y+\frac{y^2}{2}\rp  F_2^{\nu p} \rc \, , \\
  &=&  \frac{G^2_F}{\pi (1+Q^2/m_W^2)^2}
\frac{Q^2}{y}  \lc \lp d + s\rp  + \lp \bar{u} + \bar{c}\rp  (1-y)^2 \rc \, , 
     \eea
\bea
\label{eq:pol_2d_xsec_app}
\frac{d^2\Delta\sigma^{\nu p}(x,y,Q^2)}{dxdy}&=&
\frac{2G^2_F}{\pi (1+Q^2/m_W^2)^2}
\frac{Q^2}{xy}
\lc  \left(y-\frac{y^2}{2}\right) xg_1^{\nu p} -\lp 1-y+\frac{y^2}{2}\rp x g_5^{\nu p}\rc \, , \nonumber \\
&=&
\frac{2G^2_F}{\pi (1+Q^2/m_W^2)^2}
\frac{Q^2}{y}
\lc  \lp \Delta d + \Delta s \rp  - \lp \Delta \bar{u}  + \Delta \bar{c}\rp  (1-y)^2\rc \, .
\eea
For antineutrinos, the corresponding expressions are
\bea
\frac{d^2\sigma^{\bar{\nu} p}(x,y,Q^2)}{dxdy}&=&
\frac{G^2_F}{2\pi (1+Q^2/m_W^2)^2}
\frac{Q^2}{xy}
\lc  -\left(y-\frac{y^2}{2}\right) x F_3^{\bar{\nu} p}
     + \lp 1-y+\frac{y^2}{2}\rp  F_2^{\bar{\nu} p} \rc \, , \nonumber \\
  &=&  \frac{G^2_F}{\pi (1+Q^2/m_W^2)^2}
\frac{Q^2}{y}  \lc \lp  \bar{d} + \bar{s} \rp  + \lp u + c\rp  (1-y)^2 \rc  \, ,
     \eea
\bea
\frac{d^2\Delta\sigma^{\bar{\nu} p}(x,y,Q^2)}{dxdy}&=&
\frac{2G^2_F}{\pi (1+Q^2/m_W^2)^2}
\frac{Q^2}{xy}
\lc  -\left(y-\frac{y^2}{2}\right) xg_1^{\bar{\nu} p} -\lp 1-y+\frac{y^2}{2}\rp x g_5^{\bar{\nu} p}\rc \, , \nonumber \\
&=&
\frac{2G^2_F}{\pi (1+Q^2/m_W^2)^2}
\frac{Q^2}{y}
\lc  -\lp \Delta \bar{d} + \Delta \bar{s} \rp  + \lp \Delta u  + \Delta c\rp  (1-y)^2\rc \, .
\eea
Expressed at LO in terms of polarised and unpolarised PDFs, the asymmetries of Eq.~(\ref{eq:asymmetry_polarised_cross-sections}) are therefore
\bea
\label{eq:polarised_asymmetry_lo}
\mathcal{R}^{\nu p}(x,Q^2,y)&=& 2\frac{\lc  \lp \Delta d+ \Delta s \rp  - \lp \Delta \bar{u}  + \Delta \bar{c}  \rp  (1-y)^2 \rc }{  \lp  d + s \rp  + \lp \bar{u} + \bar{c}\rp  (1-y)^2 } \, , \\
\label{eq:polarised_asymmetry_lo_nubar}
\mathcal{R}^{\bar{\nu} p}(x,Q^2,y)&=& -2\frac{\lc  \lp \Delta \bar{d}+ \Delta \bar{s} \rp  - \lp \Delta u  + \Delta c\rp  (1-y)^2 \rc }{  \lp  \bar{d} + \bar{s} \rp  + \lp u + c\rp  (1-y)^2 } \, ,
\eea
for neutrino and antineutrino projectiles, respectively.
Eqs.~(\ref{eq:polarised_asymmetry_lo})-(\ref{eq:polarised_asymmetry_lo_nubar}) are combined in Eq.~(\ref{eq:polarised_asy_uncertainties}) with the binned event yields to estimate the statistical uncertainties in the measurement of polarised double-differential cross sections at the FPF@FCC.

\begin{figure}[t]
\begin{center}
\includegraphics[width=0.49\textwidth]{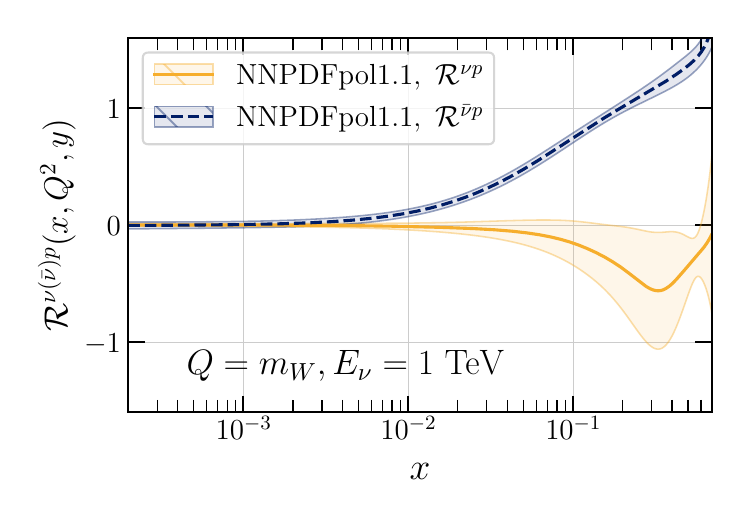}
\includegraphics[width=0.49\textwidth]{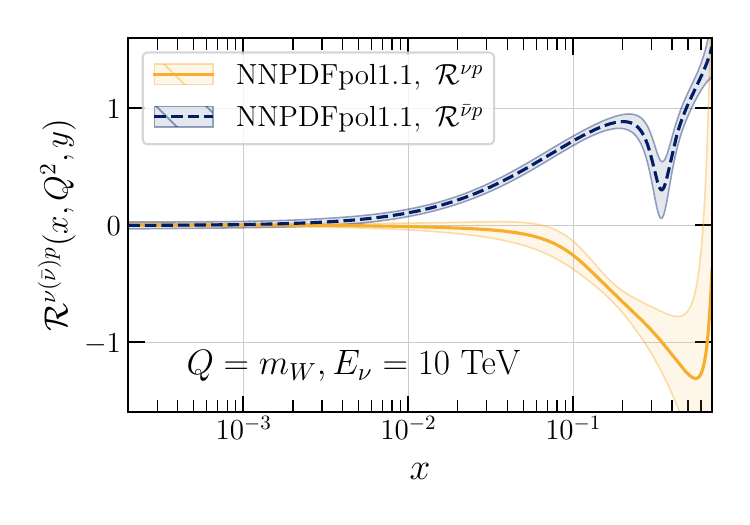}
\includegraphics[width=0.99\textwidth]{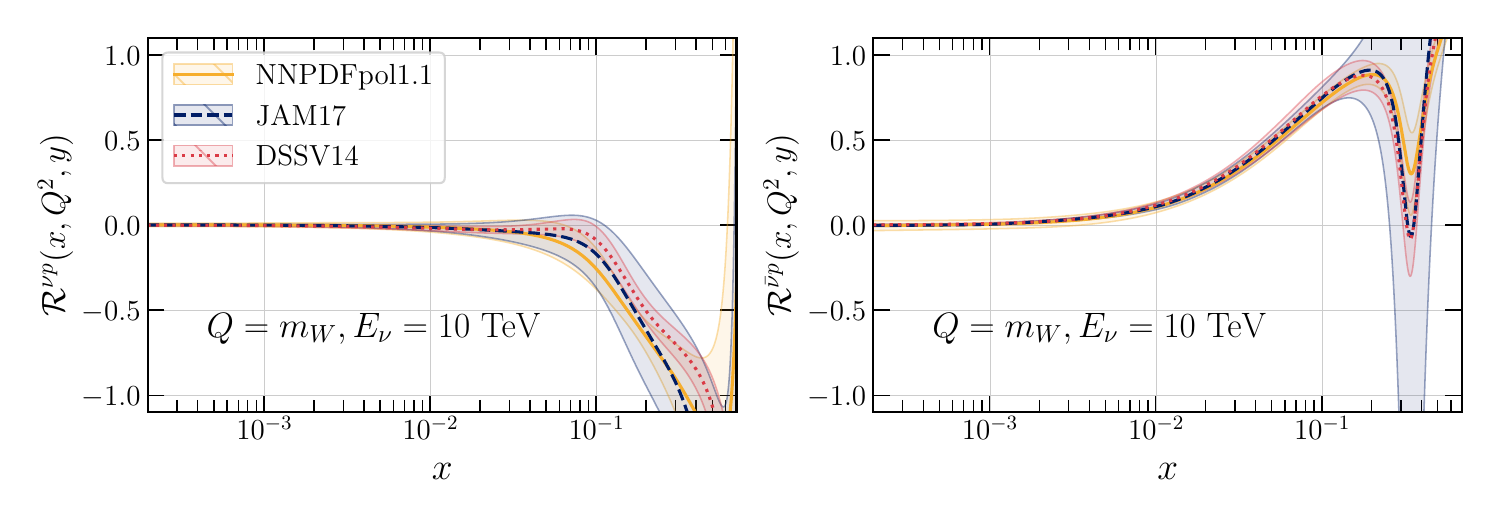}
\caption{Top: Polarised asymmetries for neutrino and antineutrino scattering evaluated at LO, Eq.~(\ref{eq:polarised_asymmetry_lo}), using the NNPDFpol1.1 and NNPDF4.0 NLO sets, at $Q=m_W$ and $E_\nu=1$ TeV (left) and 10 TeV (right panel).
Bottom: The same now comparing the NNPDFpol1.1 asymmetries with those of JAM17 and DSSV14 at $E_\nu=10$ TeV.
}
\label{fig:PolarisedPDFs-ratioasy}
\end{center}
\end{figure}

We display the numerical values of the asymmetries from Eqns.~(\ref{eq:polarised_asymmetry_lo})--(\ref{eq:polarised_asymmetry_lo_nubar}) in the top panels of Fig.~\ref{fig:PolarisedPDFs-ratioasy} for the case of the NNPDFpol1.1 and NNPDF4.0 NLO determinations for the polarised and unpolarised PDFs, respectively.
Only uncertainties associated with the polarised PDFs, which are dominant here, are displayed. 
We adopt representative values of the kinematics relevant for polarised scattering at FCC$\nu$-pol, namely $E_\nu=1~(10)$ TeV and $Q=m_W$.
At small $x$, the asymmetries $\mathcal{R}^{\nu p}$, $\mathcal{R}^{\bar{\nu} p}$ tend to zero because of the more rapid growth of unpolarised PDFs as compared to the polarised ones, and increase in magnitude moving to larger $x$ values. 
Note that the sign of the asymmetry is PDF-dependent, since polarised PDFs are not positive-definite quantities.
The bottom panels of Fig.~\ref{fig:PolarisedPDFs-ratioasy} show $R^{\nu p}$ and $R^{\bar{\nu} p}$ comparing NNPDFpol1.1 with JAM17 and DSSV14 at $E_\nu=10$ TeV, finding good qualitative agreement within PDF uncertainties.

\begin{figure}[t]
\begin{center}
\includegraphics[width=0.99\textwidth]{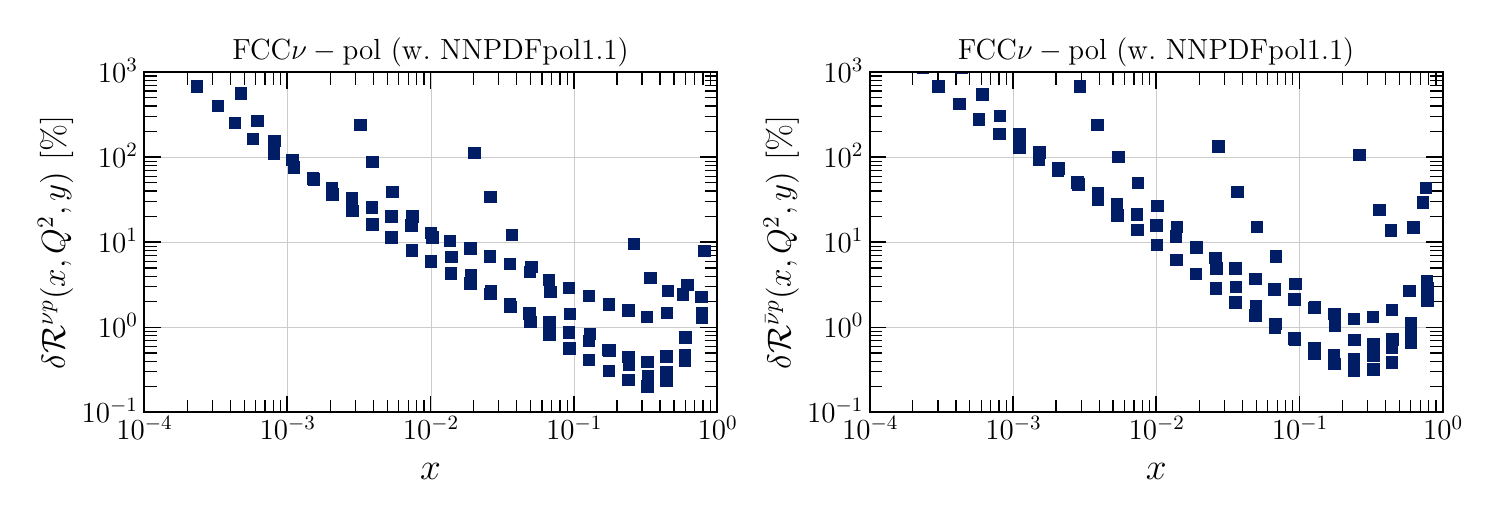}
\vspace{-0.5cm}
\caption{The projected statistical uncertainties associated with the measurement of $\mathcal{R}^{\nu p}$ (left) and $\mathcal{R}^{\bar{\nu} p}$ (right) at the kinematics accessible at the FCC$\nu$-pol detector (Fig.~\ref{fig:kinplane_COMPASS}), evaluated using Eq.~(\ref{eq:polarised_asy_uncertainties}) with NNPDFpol1.1.
}
\label{fig:nuPol-StatErrors}
\end{center}
\end{figure}

Fig.~\ref{fig:nuPol-StatErrors} displays the projected statistical uncertainties for the measurement of $\mathcal{R}^{\nu p}$ and $\mathcal{R}^{\bar{\nu} p}$ at the FCC$\nu$-pol detector evaluated using Eq.~(\ref{eq:polarised_asy_uncertainties}) with NNPDFpol1.1.
A measurement of $\mathcal{R}^{\nu p}$ with few-percent statistical precision would be possible in the $x\gsim 10^{-2}$ region.
For smaller values of $x$, statistical uncertainties become large due to the combination of lower event yields and the polarised asymmetries vanishing as $x\to 0$, as shown in Fig.~\ref{fig:PolarisedPDFs-ratioasy}.
Nevertheless, given the limited experimental constraints on the polarised PDFs for $x\lsim 10^{-2}$, measurements of $\mathcal{R}^{\nu p}$ and $\mathcal{R}^{\bar{\nu} p}$ at small-$x$ and $Q^2$, even with large uncertainties, provide valuable information on the pPDFs.

\section{Quirk searches at the FPF@FCC}
\label{app:quirk}

In this appendix, we provide additional information for the quirk search analysis presented in Sect.~\ref{subsec:quirks}, including timing distributions and the sensitivities for scalar quirk models.

\begin{figure}[tb]
\begin{center}
\includegraphics[width=0.49\textwidth]{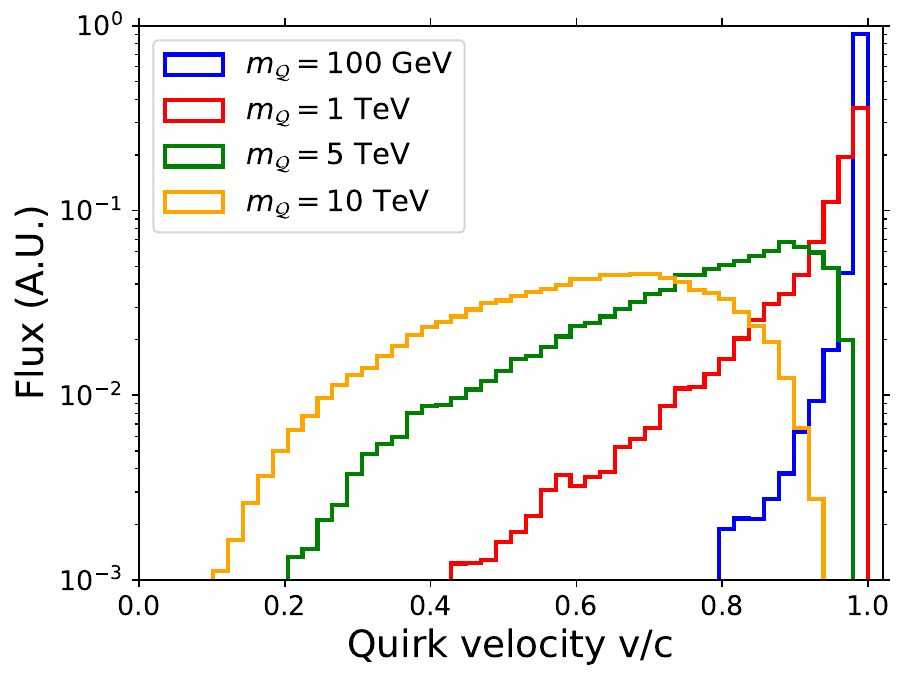}
\hfill
\includegraphics[width=0.49\textwidth]{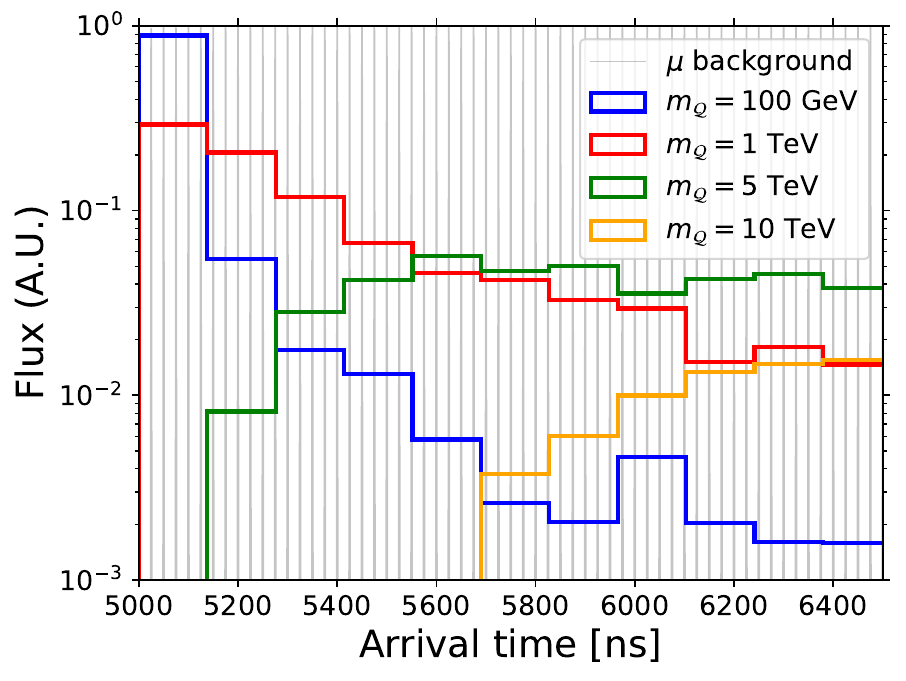}
\end{center}	
 \vspace*{-0.1in}
\caption{Left: The distribution of quirk velocities entering the detector for different values of $m_{\mathcal{Q}}$.
Right: The arrival time distributions of quirks at a distance of 1.5 km from the IP.}
\label{fig::delaydist}
\end{figure}

\begin{figure}[tb]
	\begin{center}
		\includegraphics[width=0.49\textwidth]{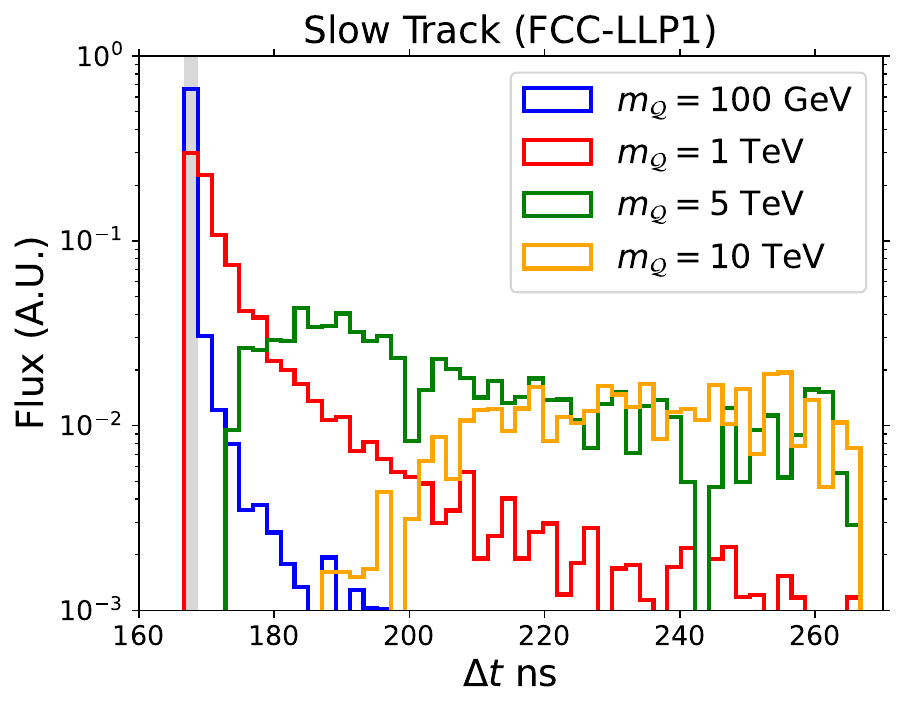}
  \hfill
		\includegraphics[width=0.49\textwidth]{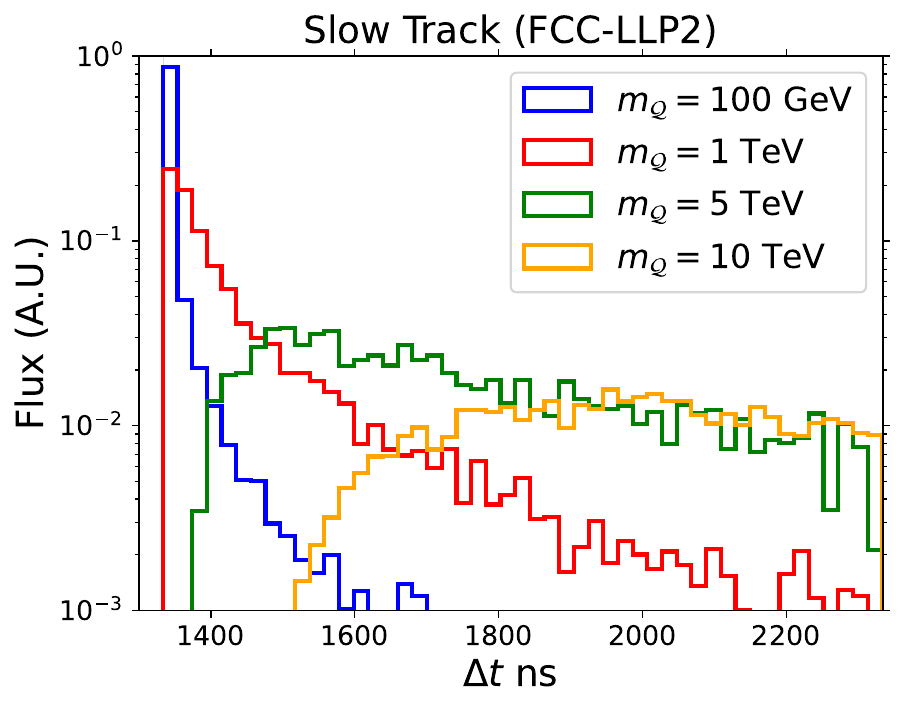}
	\end{center}
 \vspace*{-0.1in}
\caption{The time difference between quirk hits in the front and back scintillators at the FCC-LLP1 (left) and FCC-LLP2 (right) detectors. }
\label{fig::slowdist}
\end{figure}

In the left panel of Fig.~\ref{fig::delaydist}, we display the distribution of individual quirk velocities for various quirk masses.  We see that many quirks have velocities well below the maximum value of $c$, especially for heavier quirks. 
In the right panel of Fig.~\ref{fig::delaydist}, we show the distributions of arrival times for both the quirk signal and muons at a distance of 1.5 km from the IP, assuming that muons travel at the speed of light. 
The arrival time distributions for quirks are highly dependent on the quirk mass, but show little sensitivity to the quirk type and confinement scale. 
Quirks with masses exceeding 100 GeV already exhibit a significantly delayed arrival time, because the FPF@FCC detectors are so far from the IP. In contrast, energetic muons consistently arrive every 25 ns. 
Fig.~\ref{fig::slowdist} shows the distributions of time difference between quirk hits in the front and back scintillators at the FCC-LLP1 and FCC-LLP2 detectors.  
Again, a large fraction of quirks are measurably slow compared to tracks traveling at the speed of light.

\begin{figure}[t]
	\begin{center}
\includegraphics[width=0.46\textwidth]{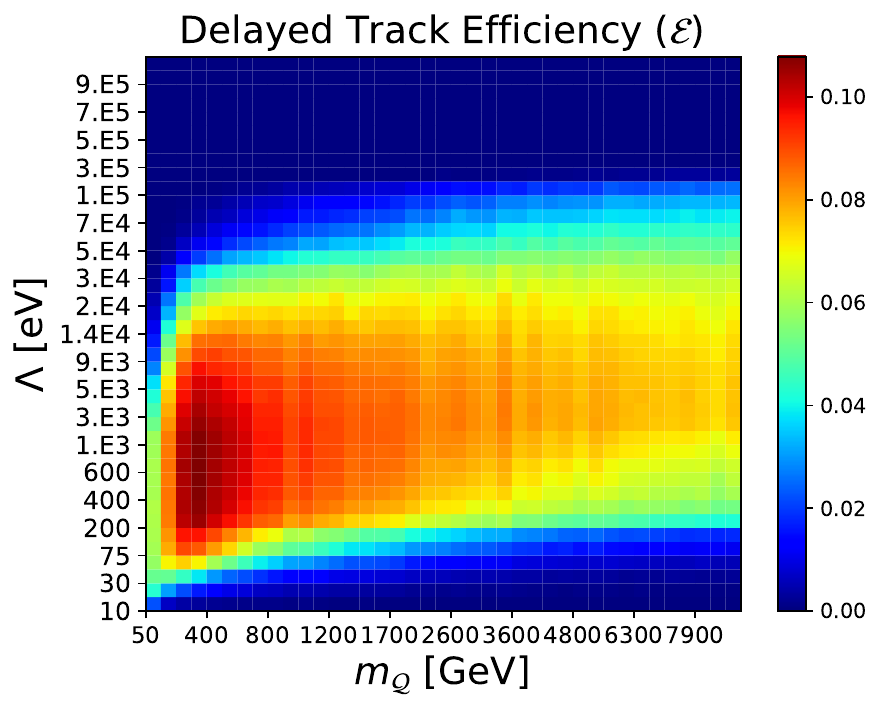}
	\includegraphics[width=0.46\textwidth]{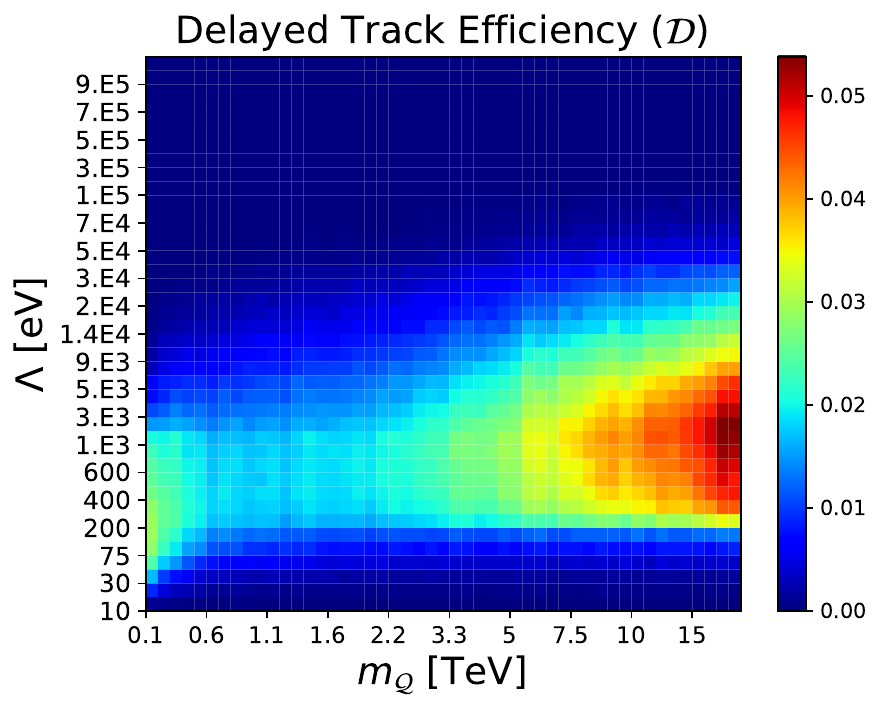}
	\includegraphics[width=0.46\textwidth]{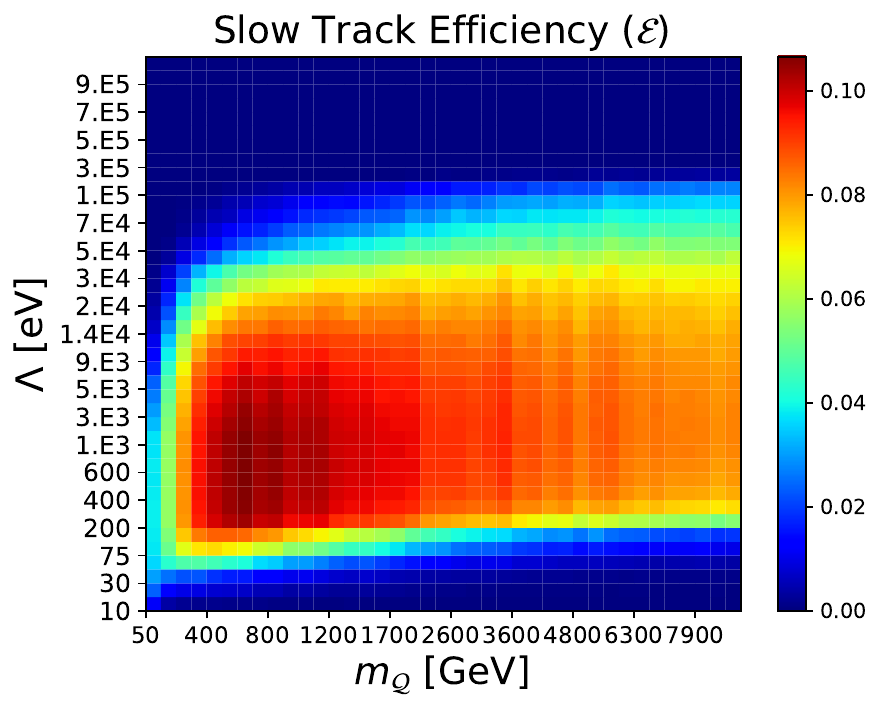}
	\includegraphics[width=0.46\textwidth]{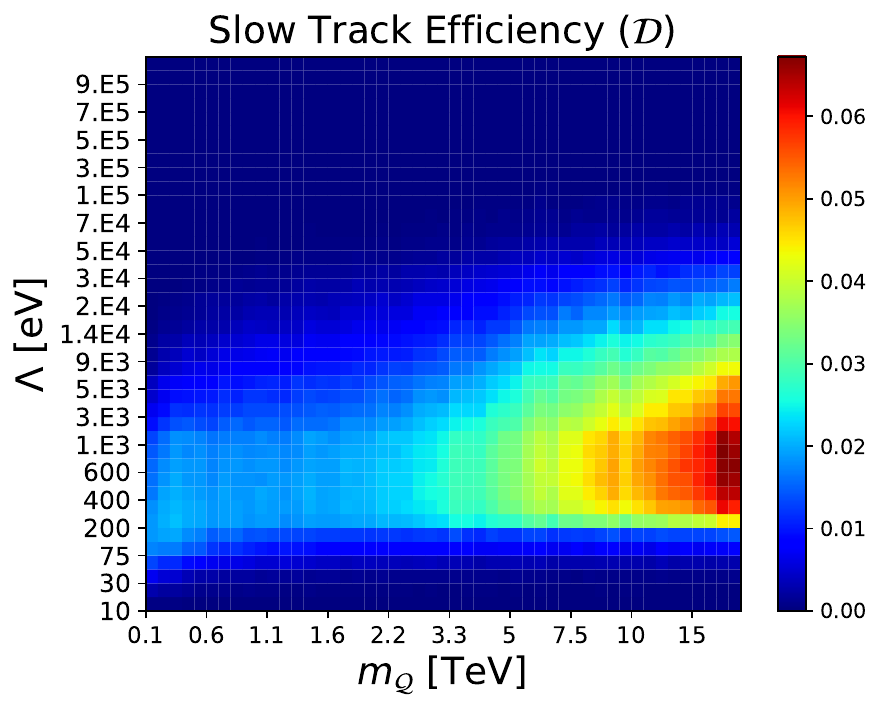}
	\end{center}
 \vspace*{-0.15in}
\caption{Efficiencies of the delayed track (DT) (upper) and slow track (ST) analyses (lower panels) at FCC-LLP2 for the fermionic quirk models indicated. The efficiencies are the fraction in all quirk events.  The energy loss radiation parameters are set to $\epsilon = 0.1$ and $\epsilon' = 0.01$.  Note that the $x$- and $y$-axes are not uniform; the values of $m_{\mathcal{Q}}$ and $\Lambda$ indicated represent the grid in parameter space that has been simulated. }
	\label{fig::effs}
\end{figure}

Fig.~\ref{fig::effs} shows the signal efficiencies of the DT and ST analyses at FCC-LLP2 in the $(m_{\mathcal{Q}}, \Lambda)$ plane for the two fermionic quirk models. The main characteristics of these distributions are similar for scalar quirk models. 
The fraction of quirk events with $\thetaq < 10/1500$, corresponding roughly to the FCC-LLP2 acceptance, is shown in the right panel of Fig.~\ref{fig::effskins}. 
The efficiencies presented in Figure~\ref{fig::effs} correspond to the fraction of quirk events that have $\thetaq < 10/1500$ and also satisfy the DT and ST criteria described in Sect.~\ref{subsec:quirks}.
For low $\Lambda \lesssim 30~\text{eV}$, the signal efficiency is reduced in both analyses. At such low $\Lambda$, the oscillation amplitude becomes large, causing most events to be excluded by the requirement that both quirks pass through the scintillators. Conversely, for high $\Lambda \gtrsim 100~\text{keV}$, the efficiency is also diminished. In this case, the oscillation amplitude is small, energy loss through radiation is rapid, and the requirement that the quirk pair survives long enough to reach the detector further reduces the efficiency. 
For $\Lambda$ values between 200 eV and 20 keV, the efficiency of both analyses is highest in the low quirk mass region for color-neutral quirks and in the high quirk mass region for colored quirks. 
This is because the fraction of forward events increases with mass for colored quirks, but decreases with mass for color-neutral quirks, as shown in Fig.~\ref{fig::effskins}.

\begin{figure}[t]
	\begin{center}
	\includegraphics[width=0.45\textwidth]{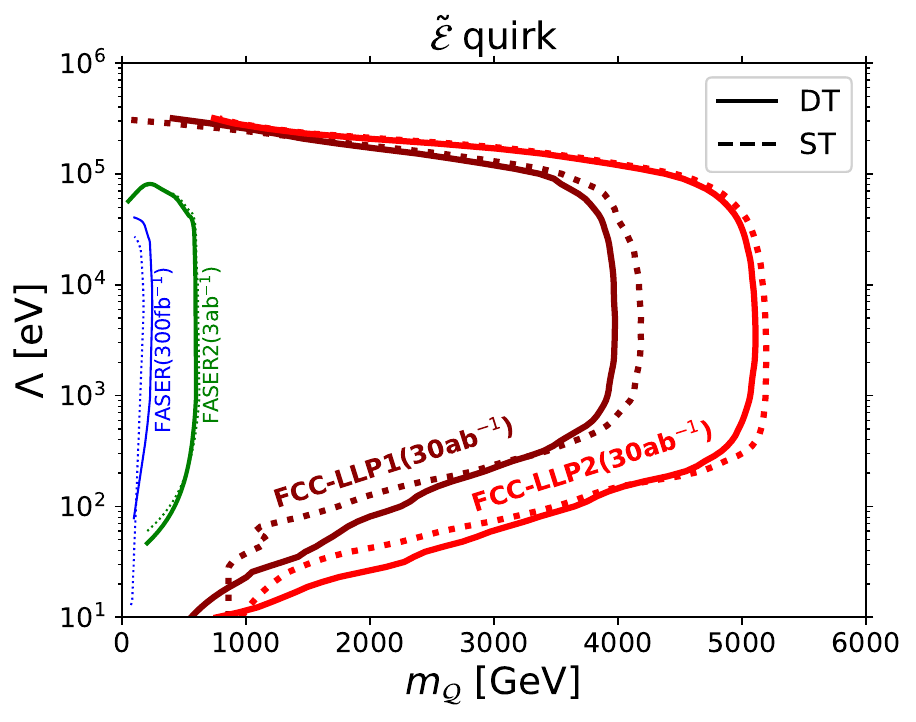}
	\includegraphics[width=0.45\textwidth]{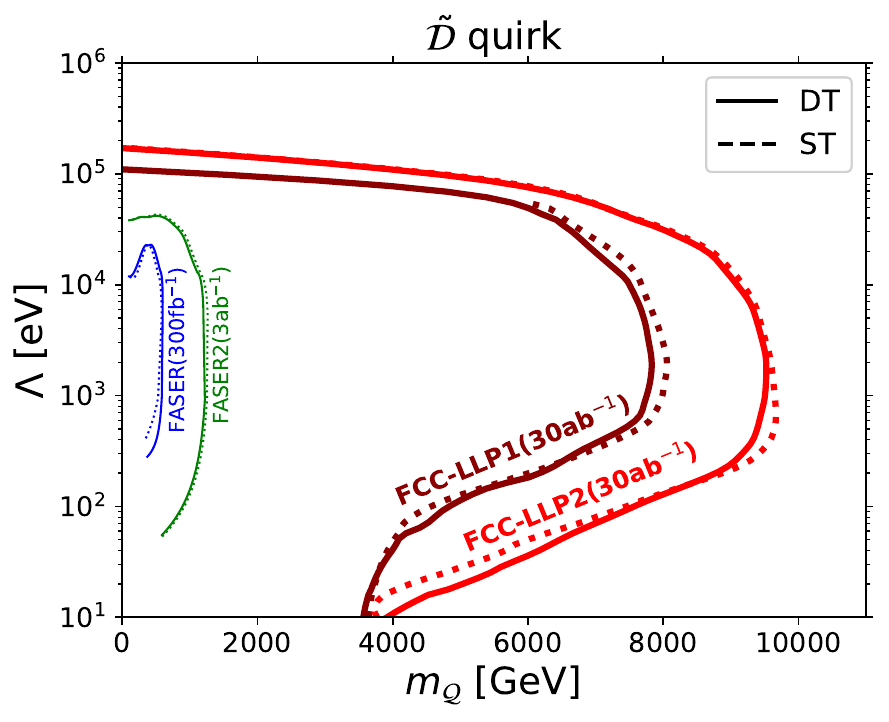}
	\end{center}
 \vspace*{-0.15in}
\caption{The $N=3$ sensitivity contours for FCC-LLP1 and FCC-LLP2 for scalar quirk models ($\tilde{\mathcal{E}}$ and $\tilde{\mathcal{D}}$) and the DT and ST analyses. There are currently no published bounds for scalar quirks, but projected sensitivities of FASER and FASER2 are given by the blue and green contours, respectively~\cite{Feng:2024zgp}.  The energy loss radiation parameters are set to $\epsilon = 0.1$ and $\epsilon' = 0.01$.}
	\label{fig::exclusionquirk2}
\end{figure}

In Fig.~\ref{fig::exclusionquirk2}, the $N=3$ signal event contours for the DT and ST analyses of scalar quirk models ($\tilde{\mathcal{E}}$ and $\tilde{\mathcal{D}}$) at FCC-LLP1 and FCC-LLP2 are displayed, where the energy loss radiation parameters are set to $\epsilon = 0.1$ and $\epsilon' = 0.01$. 
With an integrated luminosity of 30~ab$^{-1}$, FCC-LLP1 (FCC-LLP2) has the capability to detect quirks with masses up to 4.2 (5.2)~TeV in the $\tilde{\mathcal{E}}$ scenario and 8.1 (9.7)~TeV in the $\tilde{\mathcal{D}}$ scenario.

\FloatBarrier

\bibliographystyle{utphys}
\bibliography{FASERatFCCpp}

\end{document}